\documentclass[showpacs,aps,pra,floatfix,reprint,superscriptaddress,footinbib,citeautoscript,obeyspaces]{revtex4-2}
\usepackage{amssymb,amsfonts,amsmath}
\usepackage{graphicx}
\usepackage{verbatim}
\usepackage[T1]{fontenc}
\usepackage[table]{xcolor}
\usepackage{soul,colortbl}
\usepackage{multirow}
\usepackage{bm}
\usepackage{array}
\usepackage{color}
\usepackage{hyperref} \hypersetup{colorlinks=true,citecolor=blue,linkcolor=blue,urlcolor=blue}
\usepackage{mdframed}
\usepackage{longtable}
\usepackage{booktabs}
\usepackage{enumitem}
\usepackage{lmodern}
\usepackage{enumitem}
\usepackage{listings}
\usepackage{tcolorbox}
\usepackage{hhline}
\usepackage{enumitem}

\newcommand{\inlineCommand}{\texttt}
\newcommand{\inlineCommandLong}{\nolinkurl}
\newcommand{\fileName}{\texttt}
\newcommand{\fileNameLong}{\nolinkurl}
\definecolor{binarycolor}{rgb}{0,0,1}
\newcommand{\binary}[1]{\texttt{\color{binarycolor} #1}}
\newcommand{\myfbox}[1]{\hspace*{-3\fboxrule}\setlength{\fboxsep}{3\fboxrule}\framebox[1\width]{\hspace*{3\fboxrule}#1\hspace*{3\fboxrule}}\hspace*{-3\fboxrule}}
\definecolor{codegreen}{rgb}{0,0.6,0}
\definecolor{codegray}{rgb}{0.5,0.5,0.5}
\definecolor{codepurple}{rgb}{0.58,0,0.82}
\definecolor{codebgcolorCommand}{rgb}{0.95,0.95,0.92}
\definecolor{codegreen}{rgb}{0,0.6,0}
\definecolor{codegray}{rgb}{0.5,0.5,0.5}
\definecolor{codepurple}{rgb}{0.58,0,0.82}
\definecolor{codebgcolorCommand}{rgb}{0.95,0.95,0.92}
\definecolor{codebgcolorFile}{rgb}{0.92,0.92,0.95}
\lstdefinelanguage{aflowBash}{
  language=bash,
  basicstyle=\ttfamily\footnotesize,
  morekeywords={aflow,mpirun,vasp46s},
  otherkeywords={install-aflow.sh},
  frame=single,
  breaklines=true,
  backgroundcolor=\color{codebgcolorCommand},
  keywordstyle=\color{binarycolor},
  commentstyle=\it\color{codegreen},
  moredelim=[is][\it]{/*}{*/},
  postbreak=\raisebox{0ex}[0ex][0ex]{\ensuremath{\color{red}\hookrightarrow\space}},
}
\newtcolorbox{aflowBashTBox}{
  colback=codebgcolorCommand,
  sharp corners,
  size=fbox,
  top=0.2mm,
  bottom=0.2mm,
  oversize,
  on line,
  fontupper=\ttfamily\footnotesize,
  before skip balanced=0.20\baselineskip plus 0pt,
  after skip balanced=0.2\baselineskip plus 0pt,
}
\lstdefinelanguage{fileContentAFLOWIN}{
  basicstyle=\ttfamily\footnotesize,
  morekeywords={},
  otherkeywords={VASP_FORCE_OPTION,VASP_INCAR_MODE_EXPLICIT,VASP_KPOINTS_MODE_IMPLICIT,VASP_KPOINTS_FILE,VASP_POTCAR_MODE_IMPLICIT,VASP_POTCAR_FILE,VASP_POTCAR_AUID,VASP_KPOINTS_FILE,AFLOW_AEL,AFLOW_AGL,AFLOW_APL,AFLOW_QHA,AFLOW_AAPL},
  frame=single,
  breaklines=true,
  backgroundcolor=\color{codebgcolorFile},
  commentstyle=\it\color{codegreen},
  keywordstyle=\color{codegreen},
  morecomment=[l]{//},
  postbreak=\raisebox{0ex}[0ex][0ex]{\ensuremath{\color{red}\hookrightarrow\space}},
}
\lstdefinelanguage{fileContent}{
  basicstyle=\ttfamily\footnotesize,
  morekeywords={},
  otherkeywords={},
  frame=single,
  breaklines=true,
  backgroundcolor=\color{codebgcolorFile},
  commentstyle=\it\color{codegreen},
  keywordstyle=\color{codegreen},
  morecomment=[l]{//},
  postbreak=\raisebox{0ex}[0ex][0ex]{\ensuremath{\color{red}\hookrightarrow\space}},
}
\newtcolorbox{fileContentTBox}{
  colback=codebgcolorFile,
  sharp corners,
  size=fbox,
  top=0.5mm,
  bottom=0.5mm,
  oversize,
  on line,
  fontupper=\ttfamily\footnotesize,
  before skip balanced=0.5\baselineskip plus 2pt,
  after skip balanced=0.5\baselineskip plus 2pt,
}
\setlist[itemize]{noitemsep, topsep=0pt}
\newlist{myitemize}{itemize}{3}
\setlist[myitemize,1]{label=\textbullet,leftmargin=1em}
\setlist[myitemize,2]{label=--,leftmargin=1em}
\setlist[myitemize,3]{label=$\diamond$,leftmargin=1em}
\setlist[myitemize]{noitemsep, topsep=0pt}

\setlist[itemize]{noitemsep, topsep=0pt}
\newlist{myitemize_nobullet}{itemize}{3}
\setlist[myitemize_nobullet,1]{label=,leftmargin=1em}
\setlist[myitemize_nobullet,2]{label=,leftmargin=1em}
\setlist[myitemize_nobullet,3]{label=,leftmargin=1em}
\setlist[myitemize_nobullet]{noitemsep, topsep=0pt}

\definecolor{pranab_green}{rgb}{0.31,0.53,0.10}
\definecolor{pranab_red}{rgb}{0.85,0.23,0.11}
\definecolor{orcid_green}{rgb}{0.5725,0.8039,0.1137}

\newcolumntype{L}[1]{>{\raggedright\arraybackslash}p{#1}}
\newcolumntype{C}[1]{>{\centering  \arraybackslash}p{#1}}
\newcolumntype{R}[1]{>{\raggedleft \arraybackslash}p{#1}}

\def\AFLOWnopp{{\texttt{aflow}}}
\def\AFLOWpp{{\texttt{aflow++}}}
\def\AFLOW{\AFLOWpp}
\def\AFLOWorg{{\sf {a}{f}{l}{o}{w}{.}{o}{r}{g}}}
\def\AFLOWPOCC{\AFLOWnopp{\small -POCC}}
\def\AFLOWPI{{\small AFLOW$\pi$}}
\def\ACBNZero{{\small ACBN0}}
\def\AFLOWSYM{\AFLOWnopp{\small -SYM}}
\def\AFLOWCHULL{\AFLOWnopp{\small -CHULL}}
\def\AFLOWCCE{\AFLOWnopp{\small -CCE}}
\def\AFLOWAEL{\AFLOWnopp{\small -AEL}}
\def\AFLOWAGL{\AFLOWnopp{\small -AGL}}
\def\AELAGL{{\small AEL-AGL}}
\def\AFLOWAELAGL{\AFLOWnopp{\small -AEL-AGL}}

\def\AFLOWQCA{\AFLOWnopp{\small -QCA}}
\def\AFLOWQCA{\AFLOWnopp{\small -QCA}}
\def\AFLOWAPL{\AFLOWnopp{\small -APL}}
\def\AFLOWQHA{\AFLOWnopp{\small -QHA}}
\def\AFLOWAAPL{\AFLOWnopp{\small -AAPL}}
\def\AFLOWGFA{\AFLOWnopp{\small -GFA}}
\def\AFLOWAPE{\AFLOWnopp{\small -APE}}
\def\AFLOWXTALFINDER{\AFLOWnopp{\small -XtalFinder}}
\def\AFLOWXTALFINDERSHORT{{XtalFinder}}

\def\LOTO{{\small LO-TO}}
\def\GFA{{\small GFA}}
\def\AE{{\small AE}}
\def\REST{{\small REST}}
\def\RESTAPI{{\small REST-API}}
\def\API{{\small API}}
\def\MDFIVE{{\small MD5}}
\def\AFLUX{{\small AFLUX}}
\def\ICSD{{\small ICSD}}
\def\ITC{{\small ITC}}

\def\JSON{{\small JSON}}
\def\UNIX{{\small UNIX}}
\def\WSL{{\small WSL}}
\def\GNU{{\small GNU}}
\def\MPI{{\small MPI}}

\def\PDF{{\small PDF}}

\def\GIF{{\small GIF}}
\def\DFT{{\small DFT}}
\def\DFTU{{\small DFT$+U$}}
\def\DOS{{\small DOS}}
\def\EFA{{\small EFA}}

\def\SQS{{\small SQS}}

\def\KPPRA{{\small KPPRA}}
\def\VASP{{\small VASP}}
\def\VASPFOUR{{\small VASP4}}
\def\VASPFIVE{{\small VASP5}}
\def\VASPFIVEFOUR{{\small VASP5.4}}

\def\PAW{{\small PAW}}
\def\PBE{{\small PBE}}
\def\LDA{{\small LDA}}
\def\SCAN{{\small SCAN}}
\def\GGA{{\small GGA}}
\def\POSCAR{\fileName{POSCAR}}
\def\INCAR{\fileName{INCAR}}
\def\KPOINTS{\fileName{KPOINTS}}
\def\POTCAR{\fileName{POTCAR}}

\def\CIF{{\small CIF}}
\def\QUANTUMESPRESSO{\textsc{Quantum {\small ESPRESSO}}}
\def\FHIAIMS{{\small FHI-AIMS}}
\def\ABINIT{{\small ABINIT}}
\def\ELK{{\small ELK}}
\def\CALPHAD{{\small CALPHAD}}
\def\ATAT{{\small ATAT}}

\def\CCE{{\small CCE}}

\def\HTQC{{\small HTQC}}

\def\AURL{{\small AURL}}
\def\PROTOTYPEENCYCLOPEDIA{{Prototype Encyclopedia}}
\def\LFA{{\small LFA}}
\def\LIBONE{{\small LIB1}}
\def\LIBTWO{{\small LIB2}}
\def\LIBTHREE{{\small LIB3}}

\def\EOS{{\small EOS}}

\def\AGL{{\small AGL}}
\def\AEL{{\small AEL}}
\def\APL{{\small APL}}
\def\AAPL{{\small AAPL}}
\def\GIBBS{{\small GIBBS}}
\def\acoustic{{\mathrm{a}}}
\def\README{{\small README}}

\def\PAO{{\small PAO}}
\def\PAOFLOW{{\small PAOFLOW}}

\def\TB{{\small TB}}

\def\versus{vs.}
\def\ie{i.e.,}
\def\eg{e.g.,}
\def\etc{etc.}
\def\etal{et al.}

\def\Fig{Fig.}
\def\Figs{Figs.}
\def\Ref{Ref.}
\def\Refs{Refs.}
\def\Eq{Eq.}
\newcommand{\pan}[1]{(#1)}

\def\AFLOWVERSION{{3.2.12}}

\renewcommand{\vec}[1]{\textbf{#1}}

\def\APL{{\small APL}}
\def\kappap{{\kappa^\prime}}
\def\lp{{l^\prime}}
\def\IFCs{{\small IFCs}}
\def\IFCharm{{\Phi_{\alpha\beta}\left(l\kappa;\lp\kappap\right)}}

\def\Fvib{F_\text{vib}}
\def\Svib{S_\text{vib}}
\def\Uvib{U_\text{vib}}

\def\kB{{k_\text{\tiny B}}}

\def\aflowin{\fileName{aflow.in}}

\def\citeANRL{\cite{curtarolo:art121,curtarolo:art145,aflowANRL3}}
\def\onlineciteANRL{\onlinecite{curtarolo:art121,curtarolo:art145,aflowANRL3}}

\def\MAE{{\small MAE}}

\renewcommand{\vec}[1]{\textbf{#1}}
\def\QHA{{\small QHA}}
\def\EOS{{\small EOS}}
\def\AAPL{{\small AAPL}}
\def\IFCs{{\small IFCs}}
\def\lamp{{\lambda^{\prime}}}
\def\lampp{{\lambda^{\prime\prime}}}
\def\lp{{l^\prime}}
\def\lpp{{l^{\prime\prime}}}
\def\kappapp{{\kappa^{\prime\prime}}}
\def\IFCanharm{{\Phi_{\alpha\beta\gamma}\left(l\kappa;\lp\kappap,\lpp\kappapp\right)}}
\def\Felec{F_\text{elec}}
\def\fFD{f_\text{FD}}
\def\EF{E_\text{\tiny F}}
\def\Veq{V_\text{eq}}
\def\VeqT{\Veq(T)}
\def\gel{g_\text{e}\left(\epsilon, V\right)}
\newcommand{\taul}[1]{\tau_\lambda^{#1}}
\newcommand{\invtaul}[1]{\left(\taul{#1}\right)^{-1}}

\def\sPOCC{{\substack{\scalebox{0.6}{POCC}}}}
\def\APL{{\small APL}}
\def\POCC{{\small POCC}}
\def\GQCA{{\small GQCA}}
\def\POCCAPL{{\small POCC-APL}}
\def\sVRH{{\substack{\scalebox{0.6}{VRH}}}}

\def\AUID{{\small AUID}}

\def\FAIR{{\small FAIR}}

\setcitestyle{square}

\makeatletter \renewcommand\frontmatter@abstractwidth{\dimexpr\textwidth\relax} \makeatother

\begin{document}
\title{\AFLOWpp: a \texttt{C++} framework for autonomous materials design}

\author{Corey~Oses}
\affiliation{Department of Mechanical Engineering and Materials Science, Duke University, Durham, NC 27708, USA}
\affiliation{Center for Autonomous Materials Design, Duke University, Durham, NC 27708, USA}
\author{Marco~Esters}
\affiliation{Department of Mechanical Engineering and Materials Science, Duke University, Durham, NC 27708, USA}
\affiliation{Center for Autonomous Materials Design, Duke University, Durham, NC 27708, USA}
\author{David~Hicks}
\affiliation{LIFT, American Lightweight Materials Manufacturing Innovation Institute, Detroit, MI 48216, USA}
\affiliation{Department of Mechanical Engineering and Materials Science, Duke University, Durham, NC 27708, USA}
\affiliation{Center for Autonomous Materials Design, Duke University, Durham, NC 27708, USA}
\author{Simon~Divilov}
\affiliation{Department of Mechanical Engineering and Materials Science, Duke University, Durham, NC 27708, USA}
\affiliation{Center for Autonomous Materials Design, Duke University, Durham, NC 27708, USA}
\author{Hagen~Eckert}
\affiliation{Department of Mechanical Engineering and Materials Science, Duke University, Durham, NC 27708, USA}
\affiliation{Center for Autonomous Materials Design, Duke University, Durham, NC 27708, USA}
\author{Rico~Friedrich}
\affiliation{Institute of Ion Beam Physics and Materials Research, Helmholtz-Zentrum Dresden-Rossendorf, 01328 Dresden, Germany}
\affiliation{Theoretical Chemistry, Technische Universit\"{a}t Dresden, 01062 Dresden, Germany}
\affiliation{Center for Autonomous Materials Design, Duke University, Durham, NC 27708, USA}
\author{Michael~J.~Mehl}
\affiliation{Department of Mechanical Engineering and Materials Science, Duke University, Durham, NC 27708, USA}
\affiliation{Center for Autonomous Materials Design, Duke University, Durham, NC 27708, USA}
\author{Andriy~Smolyanyuk}
\affiliation{Institute of Solid State Physics, Technische Universit\"{a}t Wien, A-1040 Wien, Austria}
\affiliation{Center for Autonomous Materials Design, Duke University, Durham, NC 27708, USA}
\author{Xiomara~Campilongo}
\affiliation{Center for Autonomous Materials Design, Duke University, Durham, NC 27708, USA}
\author{Axel~van~de~Walle}
\affiliation{School of Engineering, Brown University, Providence, RI 02912, USA}
\author{Jan~Schroers}
\affiliation{Department of Mechanical Engineering and Materials Science, Yale University, New Haven, CT 06511, USA}
\author{A.~Gilad~Kusne}
\affiliation{Materials Measurement Science Division, National Institute of Standards and Technology, Gaithersburg, MD 20899, USA}
\affiliation{Department of Materials Science and Engineering, University of Maryland, College Park, MD 20742, USA}
\author{Ichiro~Takeuchi}
\affiliation{Department of Materials Science and Engineering, University of Maryland, College Park, MD 20742, USA}
\author{Eva~Zurek}
\affiliation{Department of Chemistry, State University of New York at Buffalo, Buffalo, NY 14260, USA}
\author{Marco~Buongiorno~Nardelli}
\affiliation{Department of Physics and Department of Chemistry, University of North Texas, Denton, TX 76203, USA}
\affiliation{Santa Fe Institute, Santa Fe, NM 87501, USA}
\affiliation{Center for Autonomous Materials Design, Duke University, Durham, NC 27708, USA}
\author{Marco~Fornari}
\affiliation{Department of Physics and Science of Advanced Materials Program, Central Michigan University, Mount Pleasant, MI 48859, USA}
\affiliation{Center for Autonomous Materials Design, Duke University, Durham, NC 27708, USA}
\author{Yoav~Lederer}
\affiliation{Department of Physics, NRCN, Beer-Sheva, 84190, Israel}
\affiliation{Center for Autonomous Materials Design, Duke University, Durham, NC 27708, USA}
\author{Ohad~Levy}
\affiliation{Department of Mechanical Engineering and Materials Science, Duke University, Durham, NC 27708, USA}
\affiliation{Center for Autonomous Materials Design, Duke University, Durham, NC 27708, USA}
\affiliation{Department of Physics, NRCN, P.O. Box 9001, Beer-Sheva 84190, Israel}
\author{Cormac~Toher}
\affiliation{Department of Materials Science and Engineering and Department of Chemistry and Biochemistry, University of Texas at Dallas, Richardson, Texas 75080, USA}
\affiliation{Center for Autonomous Materials Design, Duke University, Durham, NC 27708, USA}
\author{Stefano~Curtarolo}
\email[]{stefano@duke.edu}
\affiliation{Department of Mechanical Engineering and Materials Science, Duke University, Durham, NC 27708, USA}
\affiliation{Center for Autonomous Materials Design, Duke University, Durham, NC 27708, USA}

\date{\today}

\begin{abstract}
\noindent
The realization of novel technological opportunities given by computational and autonomous
materials design requires efficient and effective frameworks.
For more than two decades, \AFLOW\ (\underline{A}utomatic-\underline{Flow} Framework for
Materials Discovery) has provided an interconnected collection of
algorithms and workflows to address this
challenge.
This article contains an overview of the software and some of its most heavily-used functionalities,
including algorithmic details, standards, and examples.
Key thrusts are highlighted: the calculation of structural, electronic, thermodynamic, and thermomechanical
properties in addition to the modeling of complex materials, such as high-entropy ceramics and bulk metallic glasses.
The \AFLOW\ software prioritizes interoperability, minimizing the
number of independent parameters
and tolerances.
It ensures consistency of results across property sets ---  facilitating machine learning studies.
The software also features various validation schemes, offering
{\it real-time} quality assurance for data generated in a high-throughput fashion.
Altogether, these considerations contribute to the development of
large and reliable materials databases that
can ultimately deliver future materials systems.
\end{abstract}

\maketitle
\onecolumngrid
\clearpage
\tableofcontents
\clearpage
\twocolumngrid
\clearpage

\section{Introduction}

The \underline{A}utomatic-\underline{Flow} (\AFLOW) Framework for
Materials Discovery is an interconnected collection of
algorithms and workflows that have been developed to address the
challenge of accelerated materials' calculation and identifications.
We clarify the difference between \AFLOW\ and
\AFLOWorg.
The subject of this article, \AFLOW, is a set of codes which enables
data generation, materials discovery, analysis, identification and
optimization.
On the other hand,
\AFLOWorg\ is a web ecosystem of \FAIR\ databases,  software and tools,
including online machinery to analyze and download data,
as well as different educational resources \cite{afloworg_web_2022,aflowlib,curtarolo:art142}.
Many of the functionalities of \AFLOWorg\ rely on performing on-the-fly \AFLOW\ operations,
facilitating a
more effective, reliable, and reusable development.

The framework \AFLOW\ comprises a comprehensive suite of functionalities
that through the years has been successfully applied to the discovery
of many new systems, e. g., permanent magnets~\cite{curtarolo:art109,Sanvito_MLMag_2018},
superalloys~\cite{curtarolo:art113,ReyesTirado_ActaMat_TernarySuperalloys_2018},
high-entropy carbides~\cite{curtarolo:art140,curtarolo:art148,curtarolo:art164}, and
phase-change memory compositions~\cite{curtarolo:art166}.
The framework is written in \texttt{C++} ($\sim 650,000$ lines as of
version \AFLOWVERSION) with a growing Python environment,
and operates on \UNIX\ architectures (\GNU-Linux,  macOS).
It automates the input file generation, job submission and queue management,
error correction, analysis, storage, and dissemination of the results~\cite{afloworg_web_2022}.
Workflows are fully parallelizable, having break-points where
independent components can be run simultaneously on a cluster
accelerated with multi-threaded functionalities.
Integral to \AFLOW{}'s automation is the standardization of
input parameter sets~\cite{curtarolo:art104} and
structure prototypes~\cite{curtarolo:art170}.
The software offers direct and programmatic access to a broad range
of experimentally-observed structures~\cite{curtarolo:art121,curtarolo:art145,aflowANRL3}
with adjustable internal degrees of freedom,
enabling the construction of virtually any conceivable periodic structure.
\AFLOW\ is fully integrated to work with the
\underline{V}ienna \underline{A}b-initio \underline{S}imulation \underline{P}ackage (\VASP)~\cite{vasp2}
and provides some support for
\QUANTUMESPRESSO~\cite{Giannozzi:2017io},
\ABINIT~\cite{gonze:abinit},
the \underline{F}ritz-\underline{H}aber-\underline{I}nstitut
\underline{A}b \underline{I}nitio \underline{M}aterials \underline{S}imulation Package
(\FHIAIMS)~\cite{blum:fhi-aims},
the \ELK\ Code~\cite{elk}, and
the \underline{A}lloy \underline{T}heoretic \underline{A}utomated \underline{T}oolkit
(\ATAT)~\cite{atat4}.

Historically, \AFLOW\
started from the characterization of inorganic intermetallic crystals,
e.g., early adoption in machine-learning/data-mining \cite{curtarolo:mit_thesis,curtarolo:prl_2003_datamining} and
high-throughput \cite{monster},
in line with the use of a plane-wave basis and the
\underline{g}eneralized \underline{g}radient \underline{a}pproximation (\GGA) pseudopotentials of
\underline{P}erdew, \underline{B}urke, and \underline{E}rnzerhof (\PBE)~\cite{PBE} by default.
Recently, this scope has been extended to include
{\bf i.}~ceramics, largely enabled by the \underline{c}oordination \underline{c}orrected \underline{e}nthalpies
(\AFLOWCCE) method~\cite{curtarolo:art150,curtarolo:art172} and
{\bf ii.}~structurally and chemically disordered systems,
facilitated by the creation of the
thermodynamic \underline{d}ensity \underline{o}f \underline{s}tates (\DOS) descriptor~\cite{curtarolo:art112,curtarolo:art142} and
the \underline{G}lass-\underline{F}orming-\underline{A}bility (\AFLOWGFA)~\cite{curtarolo:art112,curtarolo:art154},
\underline{P}artial \underline{Occ}upation (\AFLOWPOCC)~\cite{curtarolo:art110}, and
\underline{Q}uasi-\underline{C}hemical \underline{A}pproximation (\AFLOWQCA)~\cite{curtarolo:art139} modules.
Descriptors are feasibly-calculated quantities based on microscopic features
that offer predictive power of macroscopic properties of the material~\cite{curtarolo:art81}.
Their development and application
remain at the heart of \AFLOW,
particularly for the prediction of
thermodynamic stability/synthesizability~\cite{curtarolo:art144},
electronic~\cite{curtarolo:art58,curtarolo:art133,curtarolo:art181}, and
thermomechanical~\cite{curtarolo:art96,curtarolo:art115,curtarolo:art100,curtarolo:art180,curtarolo:art114,curtarolo:art146,curtarolo:art125}
properties.

In this article, we highlight functionality and workflows that have been developed since
the original \AFLOW\ report ~\cite{curtarolo:art65} and
demonstrate their interoperability within the overall environment.
Examples of their application for the discovery of new materials are presented,
providing a practical guide for future materials informatics investigations.
Through ongoing innovation and implementation of robust descriptors and workflows,
\AFLOW\ continues to deliver valuable solutions~\cite{MGI} as well as
playing a role in accelerating the pace of automation in the materials community.

\onecolumngrid

{\small
  \begin{table*}[h!]
    \caption{Current \AFLOW\ tools in the standard distribution
      (version \AFLOWVERSION, Fall 2022).
      Non-\texttt{C++} modules include
      \AFLOWPI:
      medium-throughput framework for \QUANTUMESPRESSO\ and the \ACBNZero\ function~\cite{curtarolo:art127,curtarolo:art93}, and
      \PAOFLOW:
      procedure for projecting the full plane-wave solution on a reduced space of pseudoatomic orbitals~\cite{curtarolo:art133,curtarolo:art181}
      as described in Section~\ref{sec:paoflow}.
    }\label{tableacronyms}
    \begin{tabular}{@{}L{0.2\textwidth}@{}|@{}L{0.15\textwidth}@{}|@{}L{0.45\textwidth}@{}|@{}L{0.2\textwidth}@{}}
      \toprule
      ~{\bf Acronym} & ~{\bf Section} & ~{\bf Module or Library} & ~{\bf Refs.}
                                                                   \tabularnewline \midrule
                                                                   ~\AFLOWAAPL\ & ~\ref{sec:aapl} & ~\underline{A}utomatic \underline{A}nharmonic \underline{P}honon \underline{L}ibrary & ~\cite{curtarolo:art125}\\
      ~\AFLOWAEL\ & ~\ref{sec:aelagl} & ~\underline{A}utomatic \underline{E}lasticity \underline{L}ibrary & ~\cite{curtarolo:art115}\\
      ~\AFLOWAGL\ & ~\ref{sec:aelagl} & ~\underline{A}utomatic {\small \underline{G}IBBS} \underline{L}ibrary & ~\cite{curtarolo:art96}\\
      ~\AFLOWAPE\ & ~\ref{sec:ape} & ~{\small \underline{A}FLOW} \underline{P}ython \underline{E}nvironment & ~\cite{curtarolo:art135,curtarolo:art144,curtarolo:art170,curtarolo:art172}\\
      ~\AFLOWAPL\ & ~\ref{sec:apl} & ~\underline{A}utomatic \underline{P}honon \underline{L}ibrary & ~\cite{curtarolo:art180}\\
      ~\AFLOWCCE\ & ~\ref{sec:cce} & ~\underline{C}oordination \underline{C}orrected \underline{E}nthalpies Module & ~\cite{curtarolo:art150,curtarolo:art172}\\
      ~\AFLOWCHULL\ & ~\ref{sec:chull} & ~\underline{C}onvex \underline{Hull} Module & ~\cite{curtarolo:art144}\\
      ~\AFLOWGFA\ & ~\ref{sec:gfa} & ~\underline{G}lass-\underline{F}orming-\underline{A}bility Module & ~\cite{curtarolo:art112,curtarolo:art154}\\
      ~\AFLOWPOCC\ & ~\ref{sec:pocc} & ~\underline{P}artial \underline{Occ}upation Module & ~\cite{curtarolo:art110}\\
      ~\AFLOWQCA\ & ~\ref{sec:qca} & ~\underline{Q}uasi-\underline{C}hemical \underline{A}pproximation Module & ~\cite{curtarolo:art139}\\
      ~\AFLOWQHA\ & ~\ref{sec:qha} & ~\underline{Q}uasi-\underline{H}armonic \underline{A}pproximation Library & ~\cite{curtarolo:art114,curtarolo:art146,aflow_qhpocc_2022}\\
      ~\AFLOWSYM\ & ~\ref{sec:sym} & ~\underline{Sym}metry Module & ~\cite{curtarolo:art135}\\
      ~\AFLOWXTALFINDER\ & ~\ref{sec:xtalfinder} & ~Crys\underline{tal} \underline{Finder} Module & ~\cite{curtarolo:art170}\\
      \bottomrule
   \end{tabular}
  \end{table*}
}
\clearpage
\twocolumngrid

\clearpage
\section{Installation and Basic Usage}

\noindent{\bf Installation.}
\AFLOW\ was built to run on \UNIX-based platforms and thus natively runs on Linux and macOS. It can
be compiled from its monolithic source using the provided Makefile and without having to configure other libraries.
The location
of the executable can then be added to the \inlineCommand{\textit{\$PATH}} variable
to make it available everywhere for the
user. Windows, on the other hand, cannot directly run \AFLOW\ because of its different
architecture and system APIs. Short of setting up a virtual machine running Linux, Windows users
can compile \AFLOW\ on the \underline{W}indows \underline{S}ubsystem for \underline{L}inux~(\WSL),
which provides a \GNU/Linux environment
for multiple popular distributions, including Ubuntu and
Debian~\cite{wsl}
\footnote{The WSL is available for
Windows~10 and newer --- users of older Windows versions can use Cygwin instead.}.

To increase portability across platforms, compiling from source is a viable option \footnote{Both the WSL
and Cygwin have only few tools to compile \texttt{C++} code pre-installed.}.
Users with limited command-line experience, can benefit from an automated installation
script (\binary{install-aflow.sh}). It supports many Linux
distributions~(including Ubuntu and Debian for the \WSL),  macOS, and
Cygwin~\cite{cygwin}.
It can be downloaded at
\href{https://aflow.org/install-aflow/}{aflow.org/install-aflow}.
\footnote{The website contains
further instructions on how to prepare systems for the execution of the script, including on how to
set up and install the WSL and Cygwin.}
The script automatically installs the dependencies needed to compile and run all
features of \AFLOW.
This includes external packages required for graphics.
It then compiles the \AFLOW\ source code and adds it to the \inlineCommand{\textit{\$PATH}} variable so that
the executable \binary{aflow} can be used after the terminal is restarted. It can also install
\AFLOW's Python wrappers inside a virtual environment using the \inlineCommand{-{}-venv} option.

Some features of the installation can be customized.
By default, dependencies for all \AFLOW\ features are installed,
which includes \binary{pdflatex} used in some modules.
These plotting packages require considerable disk space, and often cannot be
installed by individual users on a shared system. They can be skipped entirely by executing the script as:
\begin{lstlisting}[language=aflowBash,morekeywords={install-aflow.sh}]
install-aflow.sh --slim
\end{lstlisting}
Other opportunities to customize the installation
include selecting the location of the \AFLOW\ binary by using the option \inlineCommand{-{}-ULB=\textit{path}},
where \inlineCommand{\textit{path}} is the desired path for the \AFLOW\ binary. Similarly, the location of
the \AFLOW\ source directory and the installed virtual environment can be set by using
\inlineCommand{-{}-AWD=\textit{path}} and \inlineCommand{-{}-venv=\textit{path}}, respectively.

\noindent{\bf The \fileName{aflow.rc} configuration file.}
The \fileName{aflow.rc} is the configuration file defining the default settings for running
within the \AFLOW\ environment.
The file emulates the \fileName{bashrc} script that runs when initializing an interactive environment in \binary{bash}
(\underline{B}ourne \underline{a}gain \underline{sh}ell).
The file is read upon running the \binary{aflow} binary, with settings that can be overridden by
flags passed through the command line or those set in the \aflowin\ file,
allowing for calculation-specific customization.
Upon running \binary{aflow}, a fresh \fileName{aflow.rc} file is created in \inlineCommand{\textit{\$HOME}} if one is not already present.
Just like the \fileName{bashrc} file, the \fileName{aflow.rc} file is hidden (\ie\ \fileName{\textit{\$HOME}/.aflow.rc}).

Tunable settings include the compression algorithm, output file names, \MPI\ settings,
paths for binaries, databases, module-specific settings, and machine settings.
Users of \VASP\ might consider configuring the following settings for their machine:
\begin{lstlisting}[language=fileContent]
// DEFAULT GENERIC MPI
MPI_START_DEFAULT="ulimit -s unlimited"
MPI_STOP_DEFAULT=""
MPI_COMMAND_DEFAULT="mpirun -np"
MPI_NCPUS_DEFAULT=4
MPI_NCPUS_MAX=4

// DEFAULTS BINARY
DEFAULT_VASP_BIN="vasp46s"
DEFAULT_VASP_MPI_BIN="mpivasp46s"
DEFAULT_VASP5_BIN="vasp_std"
DEFAULT_VASP5_MPI_BIN="vasp_std"

// DEFAULTS POTCARS
DEFAULT_VASP_POTCAR_DIRECTORIES="~/src/vasp,/home/Tools/src/vasp"
DEFAULT_VASP_POTCAR_DATE="current"
DEFAULT_VASP_POTCAR_SUFFIX="/POTCAR"
DEFAULT_VASP_POTCAR_DIR_POT_LDA="pot_LDA"
DEFAULT_VASP_POTCAR_DIR_POT_GGA="pot_GGA"
DEFAULT_VASP_POTCAR_DIR_POT_PBE="pot_PBE"
DEFAULT_VASP_POTCAR_DIR_POTPAW_LDA="potpaw_LDA"
DEFAULT_VASP_POTCAR_DIR_POTPAW_GGA="potpaw_GGA"
DEFAULT_VASP_POTCAR_DIR_POTPAW_PBE="potpaw_PBE"
DEFAULT_VASP_POTCAR_DIR_POTPAW_LDA_KIN="potpaw_LDA.54"
DEFAULT_VASP_POTCAR_DIR_POTPAW_PBE_KIN="potpaw_PBE.54"
\end{lstlisting}
\inlineCommand{MPI\_START\_DEFAULT} and \inlineCommand{MPI\_STOP\_DEFAULT}
are commands to run before and after the \MPI\ program.
For example, running \inlineCommand{{\color{binarycolor} ulimit} -s unlimited} allows \VASP\ to
access as much stack memory as it needs.
\inlineCommand{MPI\_COMMAND\_DEFAULT} together with
\inlineCommand{MPI\_NCPUS\_DEFAULT} and \inlineCommand{MPI\_NCPUS\_MAX}
define the \MPI\ portion of the \VASP\ command, \eg\ \inlineCommand{{\color{binarycolor} mpirun} -np 4}.
\inlineCommand{MPI\_COMMAND\_DEFAULT} is overridden in the
\aflowin\ with the following setting:
\inlineCommand{[{\color{codegreen} AFLOW\_MODE\_MPI\_MODE}]COMMAND=``mpirun -np''}.
\inlineCommand{MPI\_NCPUS\_DEFAULT} is chosen if no overriding
\inlineCommand{NCPUS} option is provided on either the command line
(\inlineCommand{{\color{binarycolor} aflow} -{}-run -{}-ncpus=8}) or in the \aflowin\
(\inlineCommand{[{\color{codegreen} AFLOW\_MODE\_MPI\_MODE}]NCPUS=8}), and
\inlineCommand{MPI\_NCPUS\_MAX} is chosen if the maximum number of cores are
requested via the command line (\inlineCommand{-{}-ncpus=max}) or
in the \aflowin\ (\inlineCommand{[{\color{codegreen} AFLOW\_MODE\_MPI\_MODE}]NCPUS=MAX}).

The \VASP\ binary name is
specified with the \inlineCommand{DEFAULT\_VASP\_BIN}
variable, with corresponding serial and \MPI\ variants (\inlineCommand{DEFAULT\_VASP\_MPI\_BIN}).
The serial and \MPI\ variants are overridden in the \aflowin\ with
\inlineCommand{[{\color{codegreen} AFLOW\_MODE\_BINARY}=vasp46s]}
and \inlineCommand{[{\color{codegreen} AFLOW\_MODE\_MPI\_MODE}]BINARY=``mpivasp46s''}, respectively.
The default can be either a \VASPFOUR\ or \VASPFIVE\ binary.
There are some cases where the \VASPFIVE\ binary is required by the workflow
(\eg\ calculation of force constants using linear-response or to
determine longitudinal optic - transverse optical (\LOTO) lattice
vibration splitting).
\AFLOW\ tries to determine the \VASP\ version automatically and, for these cases,
changes the settings and formats to run with \VASPFIVE\ (\inlineCommand{DEFAULT\_VASP5\_BIN} serial and \MPI\ variants).
These binaries should be made accessible in the environment of the compute-node
through the \inlineCommand{\textit{\$PATH}}.
Together with the \MPI\ settings, these variables define the full \MPI\ \VASP\ command to be executed by
\AFLOW:
\begin{lstlisting}[language=aflowBash]
mpirun -np 4 mpivasp46s > vasp.out
\end{lstlisting}
This can be checked in the calculation's \fileName{LOCK} file, looking for the line that starts with:
\begin{fileContentTBox}
00000  MESSAGE Executing: \textit{command}
\end{fileContentTBox}

Proper organization of the pseudopotential files is critical for enabling \AFLOW\ to employ them automatically.
\AFLOW\ breaks the subdirectory structure into different variables.
For example, the following path
\fileNameLong{/home/Tools/src/vasp/potpaw\_PBE/current/Mn\_pv/POTCAR}
can be constructed
to find the manganese pseudopotential file calculated
with the \PBE\ functional~\cite{PBE}
using the \underline{p}rojector-\underline{a}ugmented-\underline{w}ave formalism (\PAW)~\cite{PAW}.
\inlineCommand{DEFAULT\_VASP\_POTCAR\_DIRECTORIES} is a comma-separated list of paths
to check for the pseudopotential files.
It resembles the \inlineCommand{\textit{\$PATH}} environment variable in that
it will check each path sequential until a match is found.
The level of theory, formalization, and components are specified in the \aflowin:
\begin{fileContentTBox}
[{\color{codegreen} VASP\_FORCE\_OPTION}]AUTO\_PSEUDOPOTENTIALS=\textit{mode}\tcbbreak
[{\color{codegreen} VASP\_POTCAR\_MODE\_IMPLICIT}]\tcbbreak
[{\color{codegreen} VASP\_POTCAR\_FILE}]Mn\tcbbreak
\end{fileContentTBox}
\noindent
\inlineCommand{\textit{mode}} can be any of the following:
\inlineCommand{pot\_LDA}, \inlineCommand{pot\_GGA}, \inlineCommand{potpaw\_LDA}, \inlineCommand{potpaw\_GGA}, \inlineCommand{potpaw\_PBE},
\inlineCommand{potpaw\_LDA\_KIN}, \inlineCommand{potpaw\_PBE\_KIN},
where \inlineCommand{LDA} refers to the \underline{l}ocal \underline{d}ensity \underline{a}pproximation,
and the \inlineCommand{KIN} variants refer to the pseudopotentials released with
\VASPFIVEFOUR\ that include information on the \underline{kin}etic energy density of the core-electrons.
The \inlineCommand{\textit{mode}} chosen corresponds to one of the pseudopotential subdirectories:
\inlineCommand{DEFAULT\_VASP\_POTCAR\_DIR\_\textit{MODE}}.
The \inlineCommand{pv} suffix in \inlineCommand{Mn\_pv} refers to the treatment of $p$ semi-core states as valence states,
as specified in the \VASP~Wiki: \href{https://www.vasp.at/wiki/index.php/Available_PAW_potentials}{\nolinkurl{vasp.at/wiki/index.php/Available\_PAW\_potentials}}.
The selection of the right pseudopotential file can be checked in the calculation's \fileName{LOCK} file,
looking for the line that starts with:
\begin{fileContentTBox}
00000  MESSAGE POTCAR  FILE: Found potcar FilePotcar=\textit{path}
\end{fileContentTBox}
\noindent
To keep track of different pseudopotentials released with new versions of \VASP,
\AFLOW\ identifies each reference with its \MDFIVE\ hash~\cite{RFC1321,curtarolo:art104} and appends this information to the bottom
of the \aflowin:
\begin{lstlisting}[language=fileContentAFLOWIN]
[VASP_POTCAR_AUID]99be850476e2dfb3
\end{lstlisting}

\noindent{\bf Generating geometry files.}
Geometry files for crystalline materials
can be automatically generated with \AFLOW\ by decorating prototype structures with different elements.
The command-line syntax to generate prototypes is:
\begin{lstlisting}[language=aflowBash]
aflow --proto=/*label*/./*ordering*/:/*elements*/ --params=/*parameters*/
\end{lstlisting}
where \inlineCommand{\textit{label}} is the prototype designation (or alias) and
\inlineCommand{\textit{parameters}} are the comma-separated degrees of freedom for the prototype.
\inlineCommand{\textit{elements}} is a colon-separated list of elements in alphabetical order decorating the structure
(\eg\ \inlineCommand{Ag:C:Cu}).
By default, the structure is decorated with fictitious atoms (\ie\ \inlineCommand{A}, \inlineCommand{B}, \inlineCommand{C}, \inlineCommand{D}, $\ldots$).
\inlineCommand{\textit{ordering}} specifies the site decoration (\eg\ \inlineCommand{ABC} vs. \inlineCommand{BAC}),
where the placement of the letter corresponds to the site on the prototype and
the letter corresponds to the alphabetically-ordered \inlineCommand{\textit{elements}}.

There are two prototype libraries in \AFLOW:
the \underline{H}igh-\underline{T}hroughput \underline{Q}uantum \underline{C}omputing (\HTQC) library~\cite{curtarolo:art65} and the
\AFLOW\ Prototype Encyclopedia~\cite{curtarolo:art121,curtarolo:art145,aflowANRL3}.
\HTQC\ prototypes are hard-coded structures that do not require any degrees of freedom to be specified,
and whose labels are largely ad hoc \eg\ \inlineCommand{201} and \inlineCommand{T0001} denote the rocksalt and Heusler structures, respectively.
An example command to generate an \HTQC\ structure is:
\begin{lstlisting}[language=aflowBash]
aflow --proto=201:Cl:Na
\end{lstlisting}
For Prototype Encyclopedia prototypes~\cite{curtarolo:art121,curtarolo:art145,aflowANRL3},
the label --- or \AFLOW\ prototype label --- is an underscore-delimited string describing the symmetry of
a crystal structure.
For example, corundum has the prototype label \inlineCommand{A2B3\_hR10\_167\_c\_e-001}, where:
\begin{myitemize}
\item the first field indicates the reduced stoichiometry ($A_{2}B_{3}$),
\item the second field indicates the Pearson symbol (hR10),
\item the third field indicates the space group number (167),
\item the fourth field indicates the Wyckoff letters associated with species $A$ ($c$),
\item the fifth field indicates the Wyckoff letters associated with species $B$ ($e$), and
\item the sixth field indicates the alias for the corundum parameters-set
  extracted and generalized by \AFLOWXTALFINDER\ (001).
\end{myitemize}
In addition to the prototype label, \AFLOWXTALFINDER\ determines the lattice and Wyckoff parameters that are not fixed
by symmetry and returns their values for the particular geometry.
For corundum, these degrees of freedom are lattice parameters $a$ and $c/a$ and
Wyckoff coordinates $x_{1}$ and $x_{2}$ in direct/fractional space:
4.7607, 2.7296, 0.3522, and 0.5561, respectively, for Al$_{2}$O$_{3}$.
This structure can be generated with the parameters provided explicitly:
\begin{lstlisting}[language=aflowBash]
aflow --proto=A2B3_hR10_167_c_e:Al:O --params=4.7607,2.7296,0.3522,0.5561
\end{lstlisting}
or using the \inlineCommand{001} alias:
\begin{lstlisting}[language=aflowBash]
aflow --proto=A2B3_hR10_167_c_e-001:Al:O
\end{lstlisting}
Options specific to the Prototype Encyclopedia prototypes include:
\begin{myitemize}
  \item \inlineCommand{-{}-add\_equations} :
    The symbolic version of the geometry file (in terms of the variable degrees of freedom)
    is printed after the numeric geometry file.
  \item \inlineCommand{-{}-equations\_only} :
    Only prints the symbolic version of the geometry file (in terms of the variable degrees of freedom).
\end{myitemize}

\AFLOW\ can read and generate geometry files (\inlineCommand{\textit{GEOM\_FILE}}) in different file formats:
\VASP~\cite{vasp},
\QUANTUMESPRESSO~\cite{quantum_espresso_2009},
\FHIAIMS~\cite{blum:fhi-aims},
\ABINIT~\cite{gonze:abinit},
\ELK~\cite{elk}, and
the \underline{C}rystallographic \underline{I}nformation \underline{F}ile
(\CIF)~\cite{Hall_CIF_1991}.
In the following sections, the variable \inlineCommand{\textit{GEOM\_FILE}}
can be replaced with files of any name having
any of the aforementioned formats;
\AFLOW\ will detect their type and process them automatically.
To convert between formats, use the following commands, respectively:
\inlineCommand{-{}-vasp},
\inlineCommand{-{}-qe},
\inlineCommand{-{}-aims},
\inlineCommand{-{}-abinit},
\inlineCommand{-{}-elk}, and
\inlineCommand{-{}-cif}.
For example:
\begin{lstlisting}[language=aflowBash]
aflow --qe < POSCAR
\end{lstlisting}
Note, \AFLOW\ functions are generally overloaded to read inputs from the input stream,
enabling commands to be compounded via \binary{bash}'s pipe.
For example:
\begin{lstlisting}[language=aflowBash]
aflow --proto=A2B3_hR10_167_c_e-001:Al:O | aflow --sconv | aflow --aflowSG
\end{lstlisting}
will create the geometry file for corundum,
convert it to the \AFLOW\ Standard Conventional representation~\cite{curtarolo:art58},
and calculate its space group~\cite{curtarolo:art135}.

\phantomsection
\noindent{\bf The \aflowin\ input file.}\label{sec:aflowin}
Calculations performed by \AFLOW\ are controlled via the \aflowin\ file,
containing directives to create and run ensembles of \underline{d}ensity \underline{f}unctional \underline{t}heory (\DFT)
calculations with \VASP\
for the analysis of materials' structural, electronic, thermal, and elastic properties.
For a given calculation workflow, options can be specified to control:
\textbf{i.}~the symmetry analyses of the input geometry,
\textbf{ii.}~type and order of \VASP\ calculations,
\textbf{iii.}~schemes fixing \VASP\ errors (and subsequent calculation resubmission), and
\textbf{iv.}~results analysis.
Depending on the calculation,
\VASP\ input files are generated and organized automatically by \AFLOW, namely
the \POSCAR\ (lattice vectors and atomic positions),
\INCAR\ (\VASP\ settings),
\KPOINTS\ ({\bf k}-point grid information), and
\POTCAR\ (pseudopotential information).
The \aflowin\ file enables high-throughput calculations of material properties in a consistent
and repeatable manner, expanding the breadth of materials in the \AFLOWorg\ repositories.

In general, the \aflowin\ file includes the
required machine/compute settings and \VASP\ setup for the calculations.
A summary of the specific content in the \aflowin\ file is as follows.
First, the system name is given at the top of the file, usually consisting of a string containing the
elements, associated pseudopotential designation, and prototype structure.
Next, settings to perform the \VASP\ calculation are given, namely the \VASP\ binary name/location, number of computing resources
(compute cores/nodes), \underline{m}essage \underline{p}assing \underline{i}nterface (\MPI) settings for calculation parallelization, and
commands to launch the \VASP\ application (\eg\ \binary{mpirun} and \binary{aprun}).
These machine settings are followed by the options for the various \AFLOW\ submodules to calculate
different material properties, including:
\begin{myitemize}
\item crystallographic symmetry (\AFLOWSYM),
\item phonons via the harmonic approximation (\AFLOWAPL),
\item phonons via the quasi-harmonic approximation (\AFLOWQHA),
\item anharmonic phonons (\AFLOWAAPL), and
\item thermomechanical properties (\AFLOWAELAGL).
\end{myitemize}

\aflowin\ files can be created automatically, generally by converting
the aforementioned \inlineCommand{-{}-proto} commands into
\inlineCommand{-{}-aflow\_proto} commands.
For example:
\begin{lstlisting}[language=aflowBash]
aflow --aflow_proto=A2B3_hR10_167_c_e-001:Al:O
\end{lstlisting}
will generate an \aflowin\ within the following directory structure (to avoid writing collisions):
\inlineCommandLong{AFLOWDATA/AlO/A2B3\_hR10\_167\_c\_e-001.AB}
The first layer, \inlineCommand{AFLOWDATA}, is the general container for automatically generated \aflowin\ files.
The second layer specifies the chemistry (species), and the third specifies the structure (prototype).
The default parameters written inside this automatically-generated \aflowin\ are controlled
by the \fileName{aflow.rc} and command-line options.
More information can be found under the \inlineCommand{-{}-aflow\_proto} command in the following \README:
\begin{lstlisting}[language=aflowBash]
aflow --readme=aconvasp
\end{lstlisting}
\section{Ab-Initio Calculations}

\noindent
{Ab-initio} structure-energy calculations remain by far the most time- and resource-intensive
component of \AFLOW's workflows.
As such, substantial efforts have been devoted to integrating with \VASP,
the default ab-initio software employed by \AFLOW.
\VASP\ offers well-tuned default settings, especially for their pseudopotentials~\cite{kresse_vasp_paw},
that ensure fast convergence and high accuracy of results without much need for additional customization.
\AFLOW\ also provides support for other ab-initio software and frameworks used in the community,
especially for structure characterization and manipulation.
Beyond this dedicated functionality, \AFLOW's \inlineCommand{alien} mode enables the execution of any binary
in high-throughput fashion~\cite{curtarolo:art65}.

The full documentation for running automated ab-initio calculations with \AFLOW,
including parameter-tuning and error-handling,
can be found in the \AFLOW\ \README:
\begin{aflowBashTBox}
{\color{binarycolor} aflow} -{}-readme=aflow
\end{aflowBashTBox}

\noindent{\bf Standard calculation types and protocols.}
\AFLOW\ offers three basic run schemes for ab-initio calculations with \VASP:
\inlineCommand{RELAX}, \inlineCommand{STATIC}, and \inlineCommand{BANDS}.
By default, two relaxations are performed to ensure structural convergence,
which is specified in the \aflowin\ with
\inlineCommand{[{\color{codegreen} VASP\_RUN}]RELAX=2}.
To incorporate a \inlineCommand{STATIC} and \inlineCommand{BANDS} run into the workflow,
the setting should be modified to
\inlineCommand{[{\color{codegreen} VASP\_RUN}]RELAX\_STATIC\_BANDS=2}.
The run schemes are described below.
\begin{myitemize}
  \item A \inlineCommand{RELAX} run optimizes the geometry of the structure ---
either by minimizing the energy (default setting)
or the forces (changed in the \aflowin\ with \inlineCommand{[{\color{codegreen} VASP\_FORCE\_OPTION}]RELAX\_MODE=FORCES}) ---
while trying to converge the electronic charge density at each structure-snapshot.
Convergence of the electronic charge density is facilitated by smearing techniques
(\inlineCommand{ISMEAR} and \inlineCommand{SIGMA} in \VASP~\cite{VASPwiki})
allowing partial occupancy of orbitals at the Fermi edge, controlled in the \aflowin\
by \inlineCommand{[{\color{codegreen} VASP\_FORCE\_OPTION}]TYPE=DEFAULT},
which also takes values of \inlineCommand{METAL} and \inlineCommand{INSULATOR}.
For high-throughput calculations, the material assumes the parameters of a metal as the default:
using the method of Methfessel-Paxton with first order corrections~\cite{Methfessel_prb_1989}
and a width of 0.1~eV.
This sets the following parameters in the \fileName{INCAR}:
\begin{lstlisting}[language=fileContent,xrightmargin=\leftmargin]
ISMEAR=1
SIGMA=0.1
\end{lstlisting}
Within the \AFLOW\ workflow, the components of the stress tensor are checked after the final relaxation;
if any are in excess of 10~kB, the calculations are automatically rerun with
increased precision and cutoffs for the plane-wave basis set~\cite{curtarolo:art128}.
As a standard, \AFLOW\ runs spin-polarized calculations with initial magnetic moments for all atoms set to 1.0~$\mu_{\text{B}}$/atom;
a good default if the magnetic properties of the system are unknown.
Spin-polarization is turned off to reduce computational resources if the magnetization resulting from the second relaxation is found to be
below 0.05~$\mu_{\text{B}}$/atom.
These settings are found in the \aflowin:
\begin{lstlisting}[language=fileContentAFLOWIN,xrightmargin=\leftmargin]
[VASP_FORCE_OPTION]SPIN=ON,REMOVE_RELAX_2
#[VASP_FORCE_OPTION]AUTO_MAGMOM=ON
\end{lstlisting}
\inlineCommand{[{\color{codegreen} VASP\_FORCE\_OPTION}]AUTO\_MAGMOM=ON}
would change the initial magnetic moments from 1.0 to 5.0~$\mu_{\text{B}}$/atom,
and is turned off (commented out) by default.
To change how these settings are written with the automatic generation of \aflowin\ files,
the following variables should be tuned in the \fileName{aflow.rc}:
\begin{lstlisting}[language=fileContent,xrightmargin=\leftmargin]
DEFAULT_VASP_FORCE_OPTION_SPIN=1
DEFAULT_VASP_FORCE_OPTION_SPIN_REMOVE_RELAX_1=0
DEFAULT_VASP_FORCE_OPTION_SPIN_REMOVE_RELAX_2=1
DEFAULT_VASP_SPIN_REMOVE_CUTOFF=0.05
DEFAULT_VASP_FORCE_OPTION_AUTO_MAGMOM=0
\end{lstlisting}
\item A \inlineCommand{STATIC} run converges the electronic charge density of a fixed geometry
with settings that ensure a high-resolution calculation of the total energy and electronic \DOS.
This is accomplished by running with a high {\bf k}-point density and
performing Brillouin-zone integrations with the tetrahedron method
with Bl\"{o}chl corrections~\cite{Bloechl1994a} as a default.
The {\bf k}-point density is controlled by the
\underline{\bf k}-\underline{p}oints \underline{p}er \underline{r}eciprocal \underline{a}tom (\KPPRA)~\cite{curtarolo:art104}
parameter, defining a grid size that scales inversely with the number of atoms.
The parameter can be adjusted in the \aflowin:
\begin{lstlisting}[language=fileContentAFLOWIN,xrightmargin=\leftmargin]
[VASP_KPOINTS_FILE]STATIC_KPPRA=10000
\end{lstlisting}
\item A \inlineCommand{BANDS} run uses the well-converged electronic charge density from the \inlineCommand{STATIC}
run and calculates the energy levels along the \AFLOW\ standard {\bf k}-paths
to resolve the full electronic band structure and associated properties, such as the band gap.
Standard semi-local \DFT\ (with, \eg\ the \PBE\ functional) tends to underestimate the band gap~\cite{Perdew_IJQC_1985,curtarolo:art93},
an issue that is addressed in \AFLOW\ with the \DFTU\ approach based on the formulations developed by
Liechtenstein~\cite{Liechtenstein1995} and
Duradev~\cite{Dudarev1998}.
The relevant \aflowin\ parameters are:
\begin{lstlisting}[language=fileContentAFLOWIN,xrightmargin=\leftmargin]
[VASP_FORCE_OPTION]LDAU2=ON
[VASP_FORCE_OPTION]LDAU_PARAMETERS=Ga,Sb;2,-1;3.9,0;0,0
\end{lstlisting}
where \inlineCommand{[{\color{codegreen} VASP\_FORCE\_OPTION}]LDAU\textit{i}=ON}
chooses the formulation as developed by Liechtenstein (\inlineCommand{\textit{i}=1})
or Duradev (\inlineCommand{\textit{i}=2}; default).
\inlineCommand{[{\color{codegreen} VASP\_FORCE\_OPTION}]LDAU\_PARAMETERS}
is a semicolon-separated string of \DFTU\ on-site interaction parameters for each species (comma-separated),
namely
the $l$-quantum number
and the strengths of the effective Coulomb ($U$) and exchange ($J$) interactions~\cite{VASPwiki}.
Note the values for Sb for which no \DFTU\ corrections will be applied: \inlineCommand{-1}, \inlineCommand{0}, \inlineCommand{0}.
The aforementioned \aflowin\ parameters would generate the following lines in the \fileName{INCAR}:
\begin{lstlisting}[language=fileContent,xrightmargin=\leftmargin]
LDAU=.TRUE.
#LDAU_SPECIES=Ga Sb
LDAUL=2 0
LDAUU=3.9 0
LDAUJ=0 0
LDAUTYPE=2
LMAXMIX=4
\end{lstlisting}
The parameters for several systems have been defined as part of the \AFLOW\ standard~\cite{curtarolo:art104,curtarolo:art58}.
For alloys having \DFTU\ parameters, an \aflowin\ will automatically be generated
with \inlineCommand{[{\color{codegreen} VASP\_FORCE\_OPTION}]LDAU2=ON}.
To turn off this behavior, add \inlineCommand{-{}-noldau} to the \inlineCommand{-{}-aflow\_proto} command.
\end{myitemize}
The default parameters for these run schemes, including convergence tolerances, grid densities, and
pseudopotential choices, have been defined as part of the \AFLOW\ standard detailed in \Ref~\onlinecite{curtarolo:art104}.
These include:
{\bf i.}~a \VASP\ precision set to the highest pre-defined setting~\cite{VASPwiki}
(\inlineCommand{[{\color{codegreen} VASP\_FORCE\_OPTION}]PREC=ACCURATE} in the \aflowin,
\inlineCommand{DEFAULT\_VASP\_FORCE\_OPTION\_PREC\_SCHEME=ACCURATE} in the \fileName{aflow.rc}),
{\bf ii.}~a plane-wave basis cutoff increased by a factor of 1.4 above that set by \VASP~\cite{VASPwiki}
(\inlineCommand{[{\color{codegreen} VASP\_FORCE\_OPTION}]ENMAX\_MULTIPLY=1.4} in the \aflowin,
\inlineCommand{DEFAULT\_VASP\_PREC\_ENMAX\_ACCURATE=1.4} in the \fileName{aflow.rc}), and
{\bf iii.}~the stable Davidson blocked scheme for diagonalizing the Hamiltonian~\cite{Liu_rep_1978,Davidson_1983}
(\inlineCommand{[{\color{codegreen} VASP\_FORCE\_OPTION}]ALGO=NORMAL} in the \aflowin,
\inlineCommand{DEFAULT\_VASP\_FORCE\_OPTION\_ALGO\_SCHEME=NORMAL} in the \fileName{aflow.rc}).
As demonstrated, settings are highly tunable through a combination of the \aflowin\ and \fileName{aflow.rc} files.
In addition to the \inlineCommand{[{\color{codegreen} VASP\_FORCE\_OPTION}]} keys,
the \aflowin\ offers explicit and implicit control of \VASP\ input files.
For example,
\begin{lstlisting}[language=fileContentAFLOWIN]
[VASP_INCAR_MODE_EXPLICIT]START
SYSTEM=Ga_hGe_h.11
PSTRESS=000 # for hand modification
#NBANDS=XX  # for hand modification
#IALGO=48   # for hand modification
[VASP_INCAR_MODE_EXPLICIT]STOP
[AFLOW] ***************************
[VASP_KPOINTS_MODE_IMPLICIT]
[VASP_KPOINTS_FILE]KSCHEME=M
[VASP_KPOINTS_FILE]KPPRA=6000
[VASP_KPOINTS_FILE]STATIC_KSCHEME=M
[VASP_KPOINTS_FILE]STATIC_KPPRA=10000
[VASP_KPOINTS_FILE]BANDS_LATTICE=AUTO
[VASP_KPOINTS_FILE]BANDS_GRID=20
\end{lstlisting}
where \inlineCommand{[{\color{codegreen} VASP\_INCAR\_MODE\_EXPLICIT}]}
allows direct injection of content into the \fileName{INCAR},
and \inlineCommand{[{\color{codegreen} VASP\_KPOINTS\_MODE\_IMPLICIT}]}
defines a namespace of keys that control the creation of the \fileName{KPOINTS} files.
These settings are generally overridden by \inlineCommand{[{\color{codegreen} VASP\_FORCE\_OPTION}]}
settings, unless \inlineCommand{[{\color{codegreen} VASP\_FORCE\_OPTION}]NOTUNE} is provided (and uncommented).

\noindent{\bf Error-handling.}
\AFLOW\ offers automatic \VASP\ error detection and correction,
with various treatment routes accessible depending on the errors and the order in which they are encountered.
\AFLOW\ is efficient in its application of the corrections,
only applying the ones that have not been tried before or would conflict with previous ones.
Treatments include modifications of the diagonalization algorithm, precision, {\bf k}-points grid and scheme,
and rescaling of the atomic distances during relaxations.
Errors are detected in the \fileName{vasp.out} file (\VASP{}'s \fileName{standard output})
and corrections are documented in the \fileName{LOCK} file.
The treatments and order in which they are applied have been heavily tested and optimized to minimize the
need of human intervention.
In the event that \AFLOW\ over-corrects, the following command can be added to the \aflowin:
\begin{lstlisting}[language=fileContentAFLOWIN]
[VASP_FORCE_OPTION]IGNORE_AFIX=ERROR:DENTET,FIX:ALGO=FAST
\end{lstlisting}
which will ignore \VASP{}'s \inlineCommand{DENTET} error and not apply
any treatment which includes changing the algorithm to the \inlineCommand{FAST} preset~\cite{VASPwiki}.
The full list of errors detected and treatments available can be found under the
\inlineCommand{IGNORE\_AFIX} section of the \AFLOW\ \README:
\begin{aflowBashTBox}
{\color{binarycolor} aflow} -{}-readme=aflow
\end{aflowBashTBox}
\section{Structure Analysis}
\subsection{\AFLOW\ Standard Cell Representations}

\noindent
Determination of a standard cell representation is essential to an autonomous workflow, and is usually the first step.
\AFLOW\ employs the standard primitive and standard conventional cells as defined in
\Ref~\onlinecite{curtarolo:art58}.

\noindent{\bf Standard primitive cell.}
The \AFLOW\ standard primitive representations for all Bravais lattices have been constructed to have
corresponding Minkowski-reduced lattices in the reciprocal space,
ensuring both speed and convergence
of electronic structure calculations using plane-wave bases~\cite{Nguyen2009_Minkowsky,Nguyen_Minkowski_greedy_Gaussian_algorithm_2004}.
Such a lattice is guaranteed to be composed of the
three smallest linearly-independent vectors --- and is thus maximally compact ---
and have a bounded orthogonality defect,
where an orthogonal basis has a defect of zero.
The orthogonality defect is associated with the loss of completeness of a truncated
plane-wave basis, needing more plane-wave terms (and computational resources) to reach the required accuracy.
Note that in the search for the most primitive lattice,
vectors defined by the atomic basis are considered,
so the shape of the lattice may change in such a way that the symmetry of the crystal is still preserved~\cite{curtarolo:art135}.
A structure can be converted to the \AFLOW\ standard primitive representation with the following command:

\begin{lstlisting}[language=aflowBash]
aflow --sprim < /*GEOM_FILE*/
\end{lstlisting}

\noindent{\bf Standard conventional cell.}
The \AFLOW\ standard conventional representations for all Bravais lattices have been constructed to highlight
symmetry properties of the lattices (\eg\ defining lattice vectors along important
symmetry directions).
A structure can be converted to the \AFLOW\ conventional primitive representation with the following command:
\begin{lstlisting}[language=aflowBash]
aflow --sconv < /*GEOM_FILE*/
\end{lstlisting}
Many \AFLOW\ standard conventional representations match with those defined
in the \underline{I}nternational \underline{T}ables for \underline{C}rystallography (\ITC)~\cite{tables_crystallography},
with others making use of different, equally-valid choices;
such is the case for the monoclinic system.
Since the Wyckoff positions are standardized with respect to the \ITC\ conventional cells,
\AFLOW\ is also able to generate structures in the \ITC\ representation with the following command:
\begin{lstlisting}[language=aflowBash]
aflow --itc < /*GEOM_FILE*/
\end{lstlisting}
where the \inlineCommand{-{}-itc} flag can be appended with other output formats,
\eg\
\begin{lstlisting}[language=aflowBash]
aflow --itc --qe < /*GEOM_FILE*/
\end{lstlisting}
to convert the structure to the \QUANTUMESPRESSO\ geometry format.

By default, \AFLOW\ will convert structures to the \AFLOW\ standard primitive representation
before running an ab-initio calculation.
This setting is controlled with the following line in the \texttt{aflow.in}:
\begin{lstlisting}[language=fileContentAFLOWIN]
[VASP_FORCE_OPTION]CONVERT_UNIT_CELL=SPRIM
\end{lstlisting}
which also takes \inlineCommand{SCONV} (\AFLOW\ conventional representation),
\inlineCommand{NIGGLI} (Niggli standard form~\cite{Gruber_Niggli_ActaCristA_1973,Niggli1928}),
\inlineCommand{MINK} (Minkowski-reduced lattice),
\inlineCommand{INCELL} (moving atoms inside the inequivalent unit cell),
\inlineCommand{COMPACT} (moving atoms to reduce distance between them and expose bonds),
\inlineCommand{WS} (Wigner-Seitz cell),
\inlineCommand{CART}/\inlineCommand{FRAC} (Cartesian/direct coordinates),
\inlineCommand{PRES} (no modification of input structure).
The \AFLOW\ standard conventional representation has been useful for phonon calculations (via finite-displacement),
achieving more spherical supercells that include
more full coordination shells while keeping cell sizes as small as possible (see Section~\ref{sec:apl})~\cite{curtarolo:art180}.

\def\FINDSYM{{{\small FINDSYM}}}
\def\PLATON{{{\small PLATON}}}
\def\SPGLIB{{{\small Spglib}}}

\subsection{\AFLOWSYM: The Crystal Symmetry Module}\label{sec:sym}

\noindent{\textbf{Identifying crystallographic symmetries.} To identify the isometries of a crystal structure,
candidate symmetries are applied to the atomic positions in the unit cell ($\{\mathbf{x}\}$).
A structure exhibits that symmetry if all transformed atomic positions ($\{\mathbf{x}_\mathrm{transformed}\}$)
map one-to-one with the original positions ($\{\mathbf{x}_\mathrm{original}\}$).
In general, the transformed and original atomic positions will:
\textbf{i.}~match exactly (ideal mapping),
\textbf{ii.}~significantly differ (no mapping), or
\textbf{iii.}~slightly differ (possible mapping)
(depicted in \Fig~\ref{fig:cpp_aflow_sym}\pan{a}).
To determine whether the transformed and original atoms map,
a threshold, $\epsilon_{\mathrm{sym}}$, is employed:
  \[
||\mathbf{x}_{\mathrm{orig}} - \mathbf{x}_{\mathrm{transformed}}|| \le \epsilon_{\mathrm{sym}}, \forall \mathbf{x} \in \{\mathbf{x}\}.
\]

\begin{figure*}
  \includegraphics[width=\textwidth]{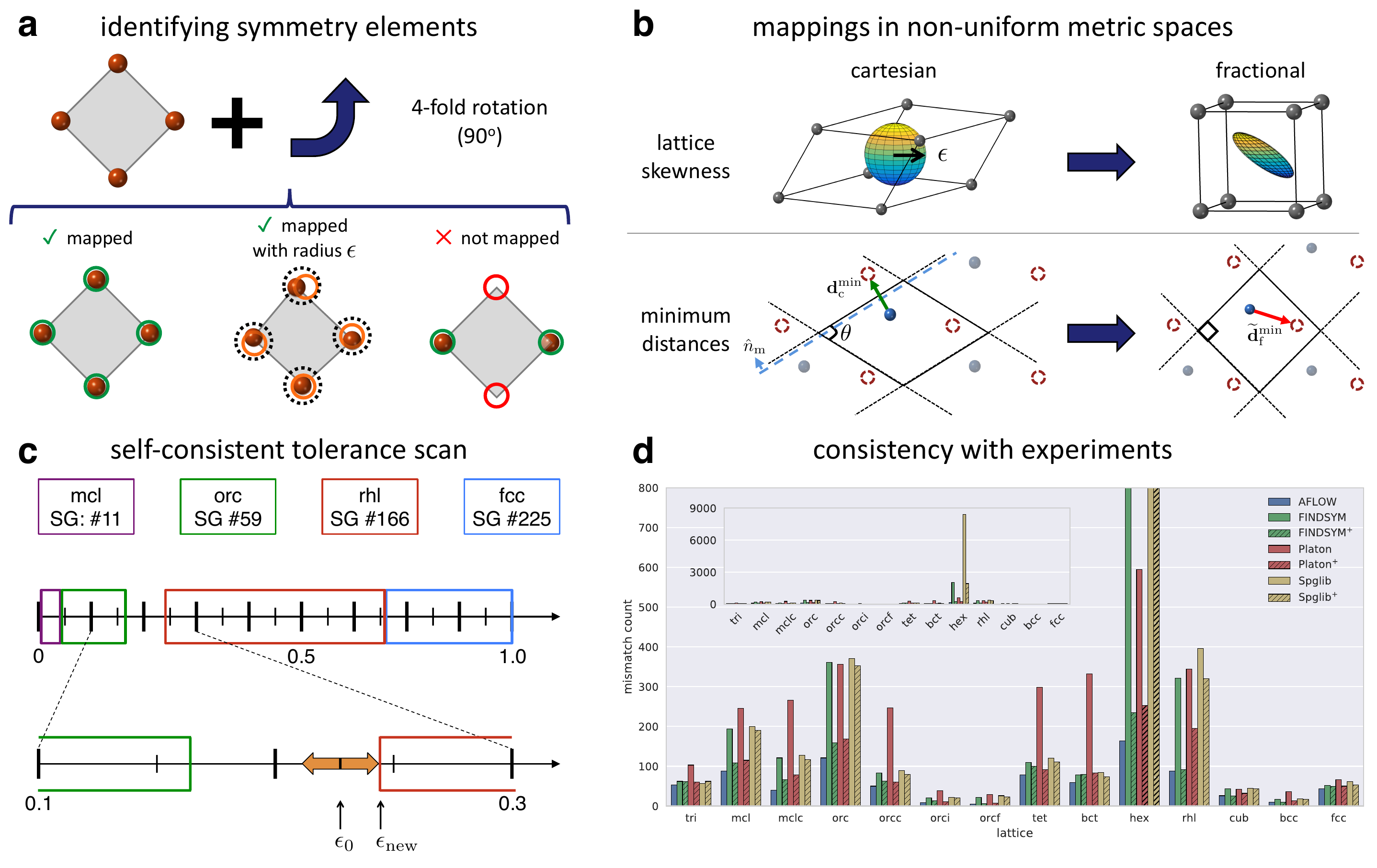}
  \caption{\textbf{Overview of \AFLOWSYM\ functionality and example of high-accuracy results.}
  ({\bf a})~A schematic highlighting how a symmetry element is tested on an arrangement of atoms.
  Outcomes of mapping the original atomic positions (solid circles) to their transformed counterparts (hollow circles)
  are grouped into the following categories: atom positions exactly map (left),
  atoms positions map within a tolerance $\epsilon$ (middle), or
  some/all of the atomic positions are not mapped (right).
  ({\bf b})~Examples are showing how distances and tolerances change between Cartesian and fractional coordinates.
  \AFLOWSYM\ considers lattice skewness in its mapping procedure to ensure it is correct regardless of the
  coordinate system.
  ({\bf c})~An illustration of the different space groups found during a tolerance scan.
  ({\bf d})~Results from a benchmark study, highlighting the accuracy of space groups calculated with different software packages compared to
  experiment (information provided by the Inorganic Crystal Structure Database).
  \AFLOWSYM\ has the fewest mismatches with experiment (best accuracy) across all 14 Bravais lattice types.
  }
  \label{fig:cpp_aflow_sym}
\end{figure*}

For crystals, atom mappings in periodic systems are considered via the method of images~\cite{Hloucha_minimumimage_1998}.
However, determination of the minimum distance
--- required to identify the correct one-to-one mappings ---
is often computationally expensive in Cartesian (Euclidean) space.
Algorithms to minimize distances in fractional (direct, non-Euclidean) space
are generally faster; however, incorrect minimum distances (mappings)
are possible since the metric tensor in this space is not uniform.
The problem is exacerbated in systems with skewed lattices; \eg\
a triclinic lattice ($\alpha\ne\beta\ne\gamma\ne90^{\circ}$) in Cartesian coordinates
will be converted
to a unit cube in fractional coordinates, warping the space (as shown in \Fig~\ref{fig:cpp_aflow_sym}\pan{b}).
To determine the impact of the warping,
the \AFLOWSYM\ module
compares the lattice skewness and minimum interatomic distance in Cartesian space $d^{\mathrm{nn,min}}_{\mathrm{c}}$ to the symmetry tolerance:
  \[
\epsilon_\mathrm{max} \equiv \left[1-\mathrm{max}\left(\mathrm{cos}\alpha, \mathrm{cos}\beta, \mathrm{cos}\gamma\right)\right]d^{\mathrm{nn,min}}_{\mathrm{c}}.
\]
If $\epsilon_{\mathrm{sym}}$ is below $\epsilon_\mathrm{max}$,
the minimum distance algorithm in fractional space (faster) should yield the correct mappings.
Above this value, the warping may yield incorrect mappings and
the slower, but more reliable minimization in Cartesian space is required.
This heuristic was validated for the more than 3.5 million entries in the \AFLOWorg\ repositories at the time of writing.

\noindent{\textbf{Symmetry tolerance.}
To alleviate the burden of identifying suitable symmetry tolerances,
\AFLOW\ offers two preset values:
$\epsilon_{\mathrm{sym}}^{\mathrm{tight}} = d^{\mathrm{nn,min}}_{\mathrm{c}}/100$ (default)
and
$\epsilon_{\mathrm{sym}}^{\mathrm{loose}} = d^{\mathrm{nn,min}}_{\mathrm{c}}/10$.
These presets can be used in any of the symmetry commands by appending
\inlineCommand{=tight} or \inlineCommand{=loose} to any of the symmetry commands, respectively
(see the \AFLOWSYM\ \README\ for examples).
Since the tolerance thresholds are based on the minimum interatomic distance in the crystal,
they are system-specific and generally more consistent with experimental space group determinations
(see the \hyperref[sec:sym_validation]{Validation with experiments} subsection).
Alternatively, users can input their own numerical tolerance values,
as long as they are below the minimum interatomic distance in the crystal.

Additionally, to ensure consistent symmetry descriptors for a wide range of tolerance values,
\AFLOWSYM\ employs an adaptive tolerance scan (\Fig~\ref{fig:cpp_aflow_sym}\pan{c}).
Namely, \AFLOWSYM\ checks that descriptors are commensurate with
group theory and crystallographic conventions (see \Ref~\onlinecite{curtarolo:art135} for details).
If the checks fail at a particular tolerance value $\left(\epsilon_{0}\right)$,
\AFLOWSYM\ will perform a radial tolerance scan (\ie\ in $\pm$ increments around the initial value) and
recalculate the symmetries at new values $\left(\epsilon_{\mathrm{new}}\right)$.
The process continues until consistent symmetry is found at a new tolerance value.
This procedure reduces errors between symmetry descriptions and mitigates the need for users to
tune tolerances to obtain accurate results.

\noindent{\textbf{Symmetry calculator.} \AFLOWSYM\ identifies the
entire symmetry profile of a crystal in any arbitrary unit cell representation.
These routines determine all crystallographic symmetry groups afforded by group theory in both real and reciprocal spaces.
In particular, the following groups are calculated:
point groups (real lattice, reciprocal lattice, Patterson, and atom-centered),
factor group representatives (unit cell), and
space groups.
The different symmetry groups and their \AFLOW\ aliases (in parentheses) include:
\begin{myitemize}
\item Lattice point group (\texttt{pgroup}):
Describes the point group symmetry (rotations, inversion, and roto-inversions) of the lattice points.
\item Reciprocal lattice point group (\texttt{pgroupk}):
Describes the point group symmetry (rotations, inversion, and roto-inversions)
of the reciprocal lattice (\ie\ Brillouin zone).
\item Crystallographic point group (\texttt{pgroup\_xtal}):
Describes the point group symmetry of the lattice faces~\cite{tables_crystallography,Nespolo_pointgroups_2009}.
\item Dual of the crystallographic point group (\texttt{pgroupk\_xtal}):
Describes the point group symmetry of the dual of the crystallographic point group (\ie\ irreducible Brillouin zone).
\item Patterson point group (\texttt{pgroupk\_Patterson}):
Describes the Patterson point group symmetry in reciprocal space, \ie\ symmetry of the inter-atomic vectors.
\item Factor group representative (\texttt{fgroup}):
Describes the rotations, inversion, roto-inversions, screws, and glides of the unit cell.
Note, the factor group representative itself is not a group:
the closure axiom is violated since the lattice translations are not present in the group~\cite{DeAngelis_AM_1972}.
\item Space group (\texttt{sgroup}):
Describes the rotations, inversion, roto-inversion, screws, and glides of the entire periodic crystal. Since the space group represents
the symmetry of an infinite object, a finite number of space group operations are calculated in \AFLOWSYM\ (see \Ref~\onlinecite{curtarolo:art135} for details).
\item Atom-site point group (\texttt{agroup}):
Describes the point group symmetry centered on each atom in the unit cell.
To make the calculation more efficient,
\AFLOWSYM\ only calculates the atom point group operations
for symmetrically-inequivalent atoms and transforms them for the other atoms.
The symmetrically-equivalent atoms are grouped into an \texttt{iatoms} object
(analogous to Wyckoff positions).
\end{myitemize}
Note that the Patterson point group is a new addition featured after the original publication of \Ref~\onlinecite{curtarolo:art135}.
All the aforementioned symmetry groups are guaranteed to be commensurate with crystallographic conventions
due to a variety of consistency checks implemented into the routines.
Any discrepancies initiate the adaptive tolerance scheme to change the symmetry tolerance ($\epsilon_{\mathrm{sym}}$)
and recalculate all symmetry groups until consistency is achieved.

\noindent{\textbf{Symmetry representations.}
All crystallographic symmetry representations are categorized as either a
\textbf{i.}~translation,
\textbf{ii.}~fixed-point (rotations, inversion, and roto-inversions), or
\textbf{iii.}~fixed-point free (screws and glides) operations.
For each of these symmetry elements, \AFLOWSYM\ provides multiple representations to cater
to different applications.
Translations are represented as $3\times1$ vectors,
fixed-point operations are represented as $3\times3$ matrices, and
fixed-point free elements are a combination of the two.
All of these symmetry elements are returned in both Cartesian and fractional (direct) coordinate systems.

\AFLOWSYM\ provides additional representations for pure rotations, comprising the $SO(3)$ Lie group~\cite{Gilmore_LieGroups_2008}, including the
axis-angle representation,
matrix generator ($so(3)$, Lie algebra),
quaternion ($4\times4$ matrix and $4\times1$ matrix),
$SU(2)$ Lie group ($2\times2$ complex matrix), and
$su(2)$ Lie algebra ($2\times2$ complex matrix).

\noindent{\textbf{Consistency with the \ITC.} In addition to determining the symmetry of any arbitrary unit cell,
\AFLOWSYM\ determines the space group symmetry and Wyckoff positions,
commensurate with the \ITC~\cite{tables_crystallography}.
From this analysis, \AFLOWSYM\ determines the
space group number,
International (Hermann-Mauguin) designation,
Schoenflies designation, and
Hall symbol.
Furthermore, \AFLOWSYM\ identifies the symmetrically-equivalent atomic positions --- or Wyckoff positions --- and
returns the corresponding Wyckoff letter designation, multiplicity, site symmetry, and representative Wyckoff coordinate.
Space groups can also be converted into special space group settings or unit cell choices for
monoclinic, rhombohedral, and centrosymmetric space groups.
Namely, conventional cells and Wyckoff positions can be represented via
\textbf{i.}~unique axis $b$ or $c$ for monoclinic space groups,
\textbf{ii.}~rhombohedral or hexagonal unit cells, or
\textbf{iii.}~origins centered on inversion sites or other high-symmetry sites for centrosymmetric space groups.
Lastly, since the choice of Wyckoff positions can differ with lattice and/or origin choices,
\AFLOWSYM\ prefers the Wyckoff sequence with the smallest enumerated Wyckoff lettering.

\noindent{\textbf{Crystal-spin symmetry.}} The magnetic moment (spin) associated with each atom in a crystal
can impact the properties of a crystal.
Thus, \AFLOWSYM\ calculates the crystal-spin symmetry by incorporating the magnetic moment into the symmetry analysis,
acting as a degree of freedom that can break (lower) symmetry.
This is analogous to how decorating a lattice with different atomic species lowers the symmetry.
Thus, in general, the crystal-spin symmetry forms a subgroup with respect to the crystal symmetry.
The analysis distinguishes the symmetry between varying spin configurations
(\ie\ ferromagnetic, ferrimagnetic, and antiferromagnetic).
This type of symmetry analysis is relevant to ab-initio codes --- such as \VASP\ --- that break orbital
symmetry based on spin considerations.
The \AFLOWSYM\ crystal-spin symmetry routines are designed for both collinear and non-collinear systems.

\phantomsection
\noindent{\textbf{Validation with experiments.}}\label{sec:sym_validation}
Other software solutions to calculate symmetry are available,
including \FINDSYM~\cite{findsym}, \PLATON~\cite{platon_2003}, and \SPGLIB~\cite{spglib}, each catering to different symmetry objectives.
Compared to space groups determined by experimental methods, \AFLOWSYM\ is the most consistent
(\Fig~\ref{fig:cpp_aflow_sym}\pan{d}).
\AFLOWSYM's high-fidelity results are attributed to its
\textbf{i.}~robust mapping scheme for skewed lattices,
\textbf{ii.}~default tolerance values, and
\textbf{iii.}~adaptive tolerance scheme with integrated consistency checks.

\noindent{\textbf{Applications.}}
The symmetry routines discussed herein are used throughout the \AFLOW\ codebase to categorize crystallographic structures
and reduce the cost of simulations.
For example, the point group symmetries are used to determine the high-symmetry {\bf k}-paths for electronic and
phonon band structure calculations.
Furthermore, in phonon simulations, the factor group (\texttt{fgroup}) and atom-site point group (\texttt{agroup})
are used to identify the symmetrically-inequivalent atoms and distortion directions, respectively, to reduce the simulation cost.

To enable adoption into user workflows, \AFLOWSYM\ features a Python module to call the major symmetry functions
in a Python environment.
Furthermore, symmetry results can be printed in either human-readable text or
\underline{J}ava\underline{S}cript \underline{O}bject \underline{N}otation (\JSON)
for easy manipulation and extension to other environments.

\noindent{\textbf{Command-line options.}}
There are three main functions that provide all symmetry information for a given input structure.
These functions allow an optional tolerance value (\inlineCommand{\textit{tol}}) to be specified,
accepting a number (double) or the
strings \inlineCommand{tight} and \inlineCommand{loose} corresponding to $\epsilon_{\mathrm{tight}}$ and $\epsilon_{\mathrm{loose}}$, respectively.
To perform the symmetry analysis of a crystal,
the functions are called with the following commands:
\begin{lstlisting}[language=aflowBash]
aflow --aflowSYM < /*GEOM_FILE*/
\end{lstlisting}
calculates and returns the symmetry operations for the lattice point group, reciprocal lattice point group,
factor group representatives, crystal point group, dual of the crystal point group, Patterson symmetry, site symmetry, and space group.
It also returns the unique and equivalent sets of atoms.
The tolerance can be appended to the \inlineCommand{aflowSYM} option: \inlineCommand{-{}-aflowSYM=\textit{tol}}.
The isometries of the different symmetry groups are saved to their own files:
\inlineCommand{aflow.\textit{group}.out} or \inlineCommand{aflow.\textit{group}.json}.
The \inlineCommand{\textit{group}} labels are as follows: \inlineCommand{pgroup} (lattice point group),
\inlineCommand{pgroupk} (reciprocal lattice point group), \inlineCommand{fgroup} (factor group representatives),
\inlineCommand{pgroup\_xtal} (crystal point group), \inlineCommand{pgroupk\_xtal}, (dual of the crystal point group),
\inlineCommand{pgroupk\_Patterson} (Patterson point group),
\inlineCommand{agroup} (site symmetry), and \inlineCommand{sgroup} (space group).
\begin{lstlisting}[language=aflowBash]
aflow --edata < /*GEOM_FILE*/
\end{lstlisting}
calculates and returns the extended crystallographic symmetry data (crystal, lattice,
reciprocal lattice, and superlattice symmetry), while incorporating the full set of checks
for robust symmetry determination.
The tolerance can be appended to the \inlineCommand{edata} option: \inlineCommand{-{}-edata=\textit{tol}}.
\begin{lstlisting}[language=aflowBash]
aflow --sgdata < /*GEOM_FILE*/
\end{lstlisting}
calculates and returns the space group symmetry of the crystal, while only validating the symmetry descriptions matching with the \ITC\ conventions.
The tolerance can be appended to the \inlineCommand{sgdata} option: \inlineCommand{-{}-sgdata=\textit{tol}}.
The \inlineCommand{-{}-print} option
specifies the output format --- accepting \inlineCommand{txt} (default) or \inlineCommand{json} ---
and can be appended to
the \inlineCommand{aflowSYM}, \inlineCommand{edata}, and \inlineCommand{sgdata} commands: \eg\ \inlineCommand{-{}-print=json}.

\subsection{\AFLOWXTALFINDER: The Crystal Prototypes Module}\label{sec:xtalfinder}

\noindent{\textbf{Autonomous prototype finder.}} To identify the prototype of a given
crystallographic compound, \AFLOWXTALFINDER\ computes the structure's
Pearson symbol, space group, and Wyckoff positions via \AFLOWSYM\ routines~\cite{curtarolo:art135} (\Fig~\ref{fig:cpp_xtalfinder}\pan{a}).
With the underlying \AFLOWSYM\ adaptive tolerance mechanism,
prototype designations are guaranteed to be consistent, automatically changing the symmetry
tolerance otherwise.
The default tolerance for the symmetry analysis is $\epsilon_\mathrm{sym}=d^{\mathrm{min}}_{\mathrm{nn}}/100$, where
$d^{\mathrm{min}}_{\mathrm{nn}}$ is the minimum interatomic distance within the crystal.
Based on benchmarks performed in \Ref~\onlinecite{curtarolo:art135}, this value is consistent with experimentally-resolved
space group symmetries.

\begin{figure*}
  \includegraphics[width=\textwidth]{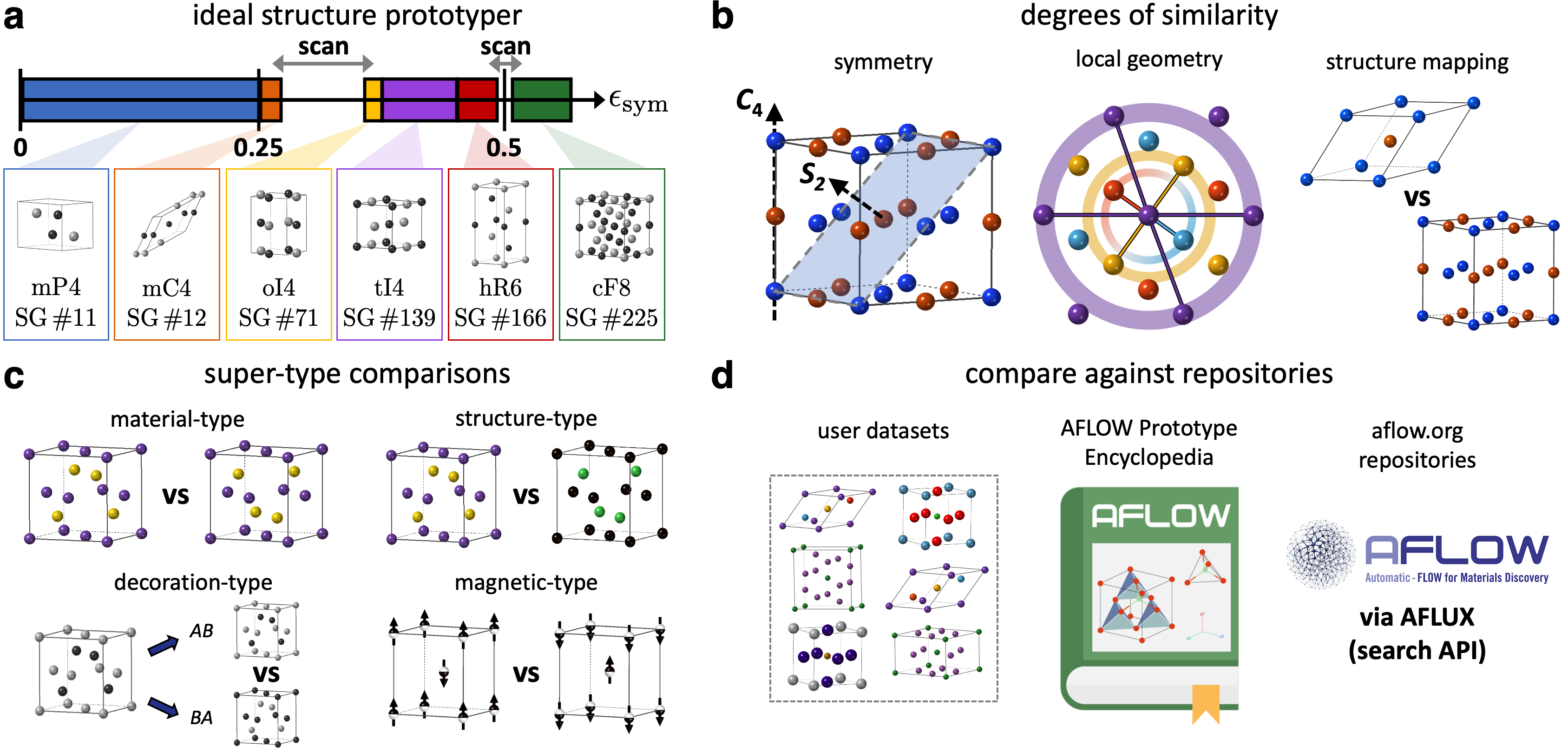}
  \caption{\textbf{Ideal prototyper and structure comparison tools in \AFLOWXTALFINDER.}
  ({\bf a})~The ideal prototyper determines a structure's prototype label and degrees of freedom based on the \AFLOWSYM\ symmetry analyses, employing a tolerance scan scheme if incommensurate descriptions are found.
  ({\bf b})~Structures are compared to varying degrees of similarity via symmetry, local geometry, and geometric structure mapping.
  ({\bf c})~\AFLOWXTALFINDER\ enables different structure comparison modes:
  material-type (map like atoms),
  structure-type (map any atom types of equal stoichiometry),
  decoration-type (generate and compare different atom colorings on a structure), and
  magnetic-type (map alike atoms and magnetic moments).
  ({\bf d})~Input structures can be automatically compared to user datasets,
  the \AFLOW\ Prototype Encyclopedia,
  or the \AFLOWorg\ repositories.
  }
  \label{fig:cpp_xtalfinder}
\end{figure*}

\AFLOWXTALFINDER\ casts these structures into their ideal prototype label and degrees of freedom based on the
\AFLOW\ Prototype Encyclopedia representation~\cite{curtarolo:art121,curtarolo:art145,aflowANRL3}.
To determine the prototype label, \AFLOWXTALFINDER\ calculates the
reduced stoichiometry, lattice, and Wyckoff positions
(via \AFLOWSYM\ routines)~\cite{curtarolo:art135}.
Once the symmetry is calculated, \AFLOWXTALFINDER\ identifies which
lattice (\ie\ $a$, $b/a$, $c/a$, $\alpha$, $\beta$, and $\gamma$) and
Wyckoff parameters (\ie\ $x$, $y$, and $z$ for all Wyckoff positions)
are not fixed by symmetry and returns their values for the particular geometry.
For example, the wurtzite structure has the prototype label \inlineCommand{AB\_hP4\_186\_b\_b} and
its corresponding degrees of freedom ($a$, $c/a$, $z_{1}$ and $z_{2}$) are 3.82, 1.63, 0.3748, and 0, respectively.

This particular representation of a prototype is advantageous because:
\textbf{i.}~it is consistent with the \ITC\ (the {\it{de facto}} standard)~\cite{tables_crystallography},
\textbf{ii.}~it allows users to tune the degrees of freedom while preserving the underlying symmetry, and
\textbf{iii.}~geometry files for any arbitrary structure in this designation can be generated using \AFLOW's symbolic
prototype generator.
Furthermore, this representation has been fruitful in performing symmetry-constrained relaxations~\cite{Lenz_SymConstrain_2019}.

\noindent{\textbf{Degrees of similarity.}}
\AFLOWXTALFINDER\ performs symmetry, local atomic geometry, and complete geometric comparisons
to analyze structural similarity to varying degrees (\Fig~\ref{fig:cpp_xtalfinder}\pan{b}).
Symmetry comparisons are performed to identify structures that are isopointal (same symmetry).
This is done by calculating the space group symmetries and
Wyckoff positions of the relevant structures with \AFLOWSYM~\cite{curtarolo:art135}.
Structures are considered isopointal if
\textbf{i.}~their space groups are the same or form an enantiomorphic pair (mirror image) and
\textbf{ii.}~their Wyckoff sequences are similar (related via an automorphism of the space group)~\cite{tables_crystallography,Boyle_WyckoffOrigin_1973,Koch_Automorphisms_1975}.
\AFLOWXTALFINDER\ tests Wyckoff similarity by comparing the Wyckoff multiplicities and permuting the site symmetry designation.

\noindent\textbf{Isoconfigurational snapshots: comparing local geometries.}
Beyond isopointal analyses, structures are further compared by inspecting arrangements of atoms,
\ie\ local atomic geometries.
Routines to quickly identify local geometries are employed here
to analyze structures beyond symmetry considerations
(\ie\ isoconfigurational {\it{versus}} isopointal~\cite{Lima-De-Faria_StructureTypes_1990}).

Rather than determine the complete local atomic geometry for each atom,
\AFLOWXTALFINDERSHORT\ builds a reduced representation:
neighborhoods comprised of only the \underline{l}east \underline{f}requently occurring \underline{a}tom (\LFA) types.
The local \LFA\ geometry analysis provides the connectivity for a subset of atoms (\ie\ \LFA\ type) to discern
if patterns are present in both structures, regardless of cell choice and crystal orientation.
This description is preferred over the full local geometry because it is
\textbf{i.}~computationally less expensive to calculate and
\textbf{ii.}~generally less sensitive to coordination cutoff tolerances.
The latter is attributed to the fact that \LFA\ geometries are more sparse.

A local \LFA\ \underline{a}tomic \underline{g}eometry ($AG$) is
a set of vectors connecting a central atom ($c$) to its closest neighbors:
  \[
    AG_{c}\equiv\{\mathbf{d}^{\mathrm{min}}_{ic}\} \, \, \forall i \, | \, {\mathrm{atom}}_{i} \in \{\mathrm{LFAs}\},
  \]
where $\mathbf{d}^{\mathrm{min}}_{ic}$ is the minimum distance vector to the $i$-atom --- restricted to \LFA\ types only ---
and is calculated via the method of images for periodic systems~\cite{Hloucha_minimumimage_1998}:
  \[
    d^{\mathrm{min}}_{ic}=\min_{i}\left[\min_{n_{a},n_{b},n_{c}}||(\mathbf{x}_{i} - \mathbf{x}_{c} + n_{a}\mathbf{a} + n_{b}\mathbf{b} + n_{c}\mathbf{c} )||\right].
  \]
Here, $n_{a}$, $n_{b}$, and $n_{c}$ are the lattice dimensions along the lattice vectors $\mathbf{a}$, $\mathbf{b}$, and $\mathbf{c}$;
and $\mathbf{x}_{i}$ and $\mathbf{x}_{c}$ are the Cartesian coordinates of the $i$ and $c$ (center) atoms, respectively.
A coordination shell with a thickness of $d^{\mathrm{min}}_{ic}/10$ captures other atoms of the same type
to control numerical noise in the atomic coordinates (a similar tolerance metric is defined
in \AFLOWSYM, \ie\ \inlineCommand{loose} preset tolerance value~\cite{curtarolo:art135}).
This cutoff value yields expected coordination numbers for well-known systems and is
comparable to results provided by other atomic environment calculators~\cite{curtarolo:art112,Zimmermann_LocalStructure_2019}.
If there is only one \LFA\ type --- \eg\ Si in $\alpha$-cristobalite (SiO$_{2}$,
\href{http://www.aflow.org/prototype-encyclopedia/A2B_tP12_92_b_a}{A2B\_tP12\_92\_b\_a})~\cite{curtarolo:art121,curtarolo:art145,aflowANRL3} ---
then the distance to the closest neighbor of that \LFA\ type is calculated.
If there are multiple \LFA\ types --- \eg\ four for the quaternary Heusler ---
then the minimum distances to each \LFA\ type are computed.
The local atomic geometry is calculated for each atom of the \LFA\ types in the unit cell, resulting in a list of
atomic geometries ($\{AG_{c}\}$).
Therefore, $\alpha$-cristobalite has a set of four Si \LFA\ geometries (one for each Si in the unit cell:
\{$AG_{\mathrm{Si,1}}$, $AG_{\mathrm{Si,2}}$, $AG_{\mathrm{Si,3}}$, $AG_{\mathrm{Si,4}}$\}) and
the quaternary Heusler has a set of four \LFA\ geometries (one for each element type:
\{$AG_{\mathrm{Au}}$, $AG_{\mathrm{Li}}$, $AG_{\mathrm{Mg}}$, $AG_{\mathrm{Sn}}$\}, respectively).

To investigate structural compatibility, local atomic geometry lists for compounds are compared.
In general, the local geometry comparisons err on the side of caution.
For instance, comparing the cardinality of the coordination is often too strict.
Despite a sparser geometry space, slight deviations in position can move atoms
outside the coordination shell threshold, changing the atom cardinality and overlooking potential matches.
Local atomic geometries are thus compatible if
\textbf{i.}~the central atoms are of comparable types (\ie\ same element and/or stoichiometric ratio in the crystal),
\textbf{ii.}~the neighborhoods of surrounding atoms have distances that match within
20\% after normalizing with respect to $\max(AG_{c})$ (\ie\ the largest distance in the local geometry cluster), and
\textbf{iii.}~the angles formed by two atoms and the center atom match within 10 degrees.
To further alleviate the coordination problem, an exact geometry match is not required,
\ie\ some distances and angles need not match exactly at this stage.
This more lenient method favors labeling structures as near-isoconfigurational
to mitigate false negatives in subsequent geometric comparisons.

\noindent\textbf{Isoconfigurational structures: comparing full geometry.}
To resolve a commensurate representation between two structures for
geometric comparison,
one structure --- the reference $\mathbb{X}_\mathrm{ref}$ --- remains fixed and
the other structure --- the potential duplicate $\mathbb{X}_\mathrm{test}$ --- is expanded into a supercell.
Lattice vectors are identified within the supercell and compared against the reference structure.
For any similar lattices to $\mathbb{X}_\mathrm{ref}$,
$\mathbb{X}_\mathrm{test}$ is transformed into the new lattice representation ($\widetilde{\mathbb{X}}_\mathrm{test}$).
Origin shifts for this cell are then explored in an attempt to match atoms.
If one-to-one atom mappings exist between the two structures, then the similarity
is quantified with the crystal misfit method (see \hyperref[sec:misfit_criteria]{Quantitative similarity measure} subsection)~\cite{Burzlaff_ActaCrystA_1997}.
Misfit values below a given threshold indicate that structures match and the
search terminates.
Alternatively, misfit values larger than the threshold are
disregarded and the search continues until all lattices and origin shifts are exhausted.

\phantomsection
\noindent\textbf{Quantitative similarity measure.} \label{sec:misfit_criteria}
To compare two crystals in a given representation, a method proposed by Burzlaff and Malinovsky is employed~\cite{Burzlaff_ActaCrystA_1997}.
The similarity between structures is quantified by a misfit value~\cite{Burzlaff_ActaCrystA_1997}:
  \[
    \epsilon \equiv 1.0 - \left(1.0 - \epsilon_{\mathrm{latt}}\right) \left(1.0 - \epsilon_{\mathrm{coord}}\right) \left(1.0 -\epsilon_{\mathrm{fail}}\right),
  \]
which incorporates differences between lattice vectors $\left(\epsilon_{\mathrm{latt}}\right)$ and
atomic positions $\left(\epsilon_{\mathrm{coord}}\text{ and }\epsilon_{\mathrm{fail}}\right)$ and are defined below.
The misfit quantity is bound between zero and one, where zero indicates a perfect match.
Special misfit ranges defined by Burzlaff and Malinovsky are adopted~\cite{Burzlaff_ActaCrystA_1997}:
\begin{align*}
  0 < \epsilon \leq \epsilon_{\mathrm{match}} &\colon \mbox{match}, \nonumber \\
  \epsilon_{\mathrm{match}} < \epsilon \leq \epsilon_{\mathrm{family}} &\colon \mbox{same family, and} \\
  \epsilon_{\mathrm{family}} < \epsilon \leq 1 &\colon \mbox{no match} \nonumber.
\end{align*}
The ``same family'' designation generally corresponds to crystals with common symmetry subgroups.
Burzlaff and Malinovsky recommend $\epsilon_{\mathrm{match}}=0.1$ and $\epsilon_{\mathrm{family}}=0.2$
based on definitions from Pearson~\cite{Pearson_ChemPhysMetals_1972} and Parth{\'e}~\cite{Parthe_StructuralChemistry_1990}.
In the \AFLOWXTALFINDERSHORT\ article, heuristic misfit thresholds are identified based on the
allowed maximum enthalpy differences between similar structures (see \Ref~\onlinecite{curtarolo:art170} for details).

The deviation of the lattices $\left(\epsilon_{\mathrm{latt}}\right)$ captures the difference between the lattice face diagonals of
$\widetilde{\mathbb{X}}_\mathrm{test}$ and $\mathbb{X}_\mathrm{ref}$~\cite{Burzlaff_ActaCrystA_1997}:
\begin{align*}
  \epsilon_{\mathrm{latt}} &\equiv 1 - (1 - D_{12})(1 - D_{23})(1 - D_{31}), \\
    D_{kl} &\equiv \frac{  ||\widetilde{\mathbf{d}}_{kl}^{\mathrm{test}} - \mathbf{d}_{kl}^{\mathrm{ref}}||
                    + ||\widetilde{\mathbf{f}}_{kl}^{\mathrm{test}} - \mathbf{f}_{kl}^{\mathrm{ref}}|| }
                 {||\mathbf{{d}}_{kl}^{\mathrm{ref}} - \mathbf{f}_{kl}^{\mathrm{ref}}||},
\end{align*}
where $\mathbf{f}_{kl}$ and $\mathbf{d}_{kl}$ (the diagonals on the $kl$ lattice face) are calculated by adding and subtracting, respectively, the $k$ and $l$ lattice vectors.
In the lattice search algorithm, $\Delta l$, $\Delta \theta$, and $\Delta V$ tolerances
are coupled to $\epsilon_{\mathrm{latt}}$, and are tuned to ensure $\epsilon_{\mathrm{latt}}\le\epsilon_{\mathrm{family}}$.

The coordinate deviation --- measuring the disparity between atomic positions in the two structures ---
is based on the mapped atom distances
($d^{\mathrm{map}}_{i}$ or $d^{\mathrm{map}}_{j}$)
and the atoms' nearest neighbor distances in the respective structures $\left(d_{\mathrm{nn}}\right)$~\cite{Burzlaff_ActaCrystA_1997}:
\begin{align*}
    \epsilon_{\mathrm{coord}} &\equiv \frac{\sum_{i}^{{\widetilde{N}}^{\mathrm{test}}} \left(1 - \widetilde{n}_{i}^{\mathrm{test}}\right) d_{i}^{\mathrm{map}}
                    + \sum_{j}^{{N}^{\mathrm{ref}}} \left(1 - n_{j}^{\mathrm{ref}}\right) d_{j}^{\mathrm{map}}}
                   {\sum_{i}^{{\widetilde{N}}^{\mathrm{test}}} \left(1 - \widetilde{n}_{i}^{\mathrm{test}}\right) d_{\mathrm{nn},i}^{\mathrm{test}}
                    + \sum_{j}^{{N}^{\mathrm{ref}}} \left(1 - n_{j}^{\mathrm{ref}}\right) d_{\mathrm{nn},j}^{\mathrm{ref}}}.
\end{align*}
$\widetilde{N}^{\mathrm{test}}$ and $N^{\mathrm{ref}}$ are the number of atoms in the two crystals.
If $d^{\mathrm{map}} < d_{\mathrm{nn}}/2$, then a ``switch'' variable $n$ is set to zero and the
mapped atom distance is included in $\epsilon_{\mathrm{coord}}$.
Otherwise, $n$ is set to one, signifying that the mapped atoms are far apart and not considered in $\epsilon_{\mathrm{coord}}$.
These atoms are counted in the figure of failure~\cite{Burzlaff_ActaCrystA_1997}:
\begin{align*}
    \epsilon_{\mathrm{fail}} &\equiv \frac{\sum_{i}^{\widetilde{N}^{\mathrm{test}}} \widetilde{n}_{i}^{\mathrm{test}} +\sum_{j}^{N^{\mathrm{ref}}} n_{j}^{\mathrm{ref}}}
                              {\widetilde{N}^{\mathrm{test}} + N^{\mathrm{ref}}}.
\end{align*}

\noindent{\textbf{Super-type comparisons.}}
\AFLOWXTALFINDER\ offers four modes of comparing crystallographic structures by changing the mapping criteria.
These modes are material-type, structure-type, decoration-type, and magnetic-type comparisons (\Fig~\ref{fig:cpp_xtalfinder}\pan{c}).
Material-type comparisons map atoms with matching elements (\eg\ Cl$\to$Cl and Na$\to$Na in sodium chloride),
revealing duplicate compounds.
Structure-type comparisons relax this criteria and attempt to map atoms of any element type as long as they have the same stoichiometric ratio
(\eg\ Cl$\to$Cl, Cl$\to$Na, or vice versa in sodium chloride).
This comparison-type identifies compounds that share the same prototype
structure; \eg\ periclase (MgO) and sodium chloride (ClNa) are both rocksalt structures.
Decoration-type comparisons identify the atom types (or colorings) of a single structure that
leave the compound invariant (\eg\ swapping Cl and Na in sodium chloride leaves the crystal
unchanged).
In this analysis, all possible decorations of an $n$-species structure ($n!$) are created and compared if sites have
similar Wyckoff positions.
Since this analysis determines the unique atom colorings, prototypes can be
decorated such that degenerate compounds are omitted.
Lastly, magnetic-type comparisons incorporate the magnetic moment of each atom into the analysis, determining distinct spin
configurations (\eg\ $\uparrow\to\uparrow$ and $\downarrow\to\downarrow$).
Comparison of systems with collinear and non-collinear magnetic moments are supported.

\noindent{\textbf{Automatic grouping.}}
Automatically comparing structures is necessary for high-throughput classification of unique/duplicate compounds and structure-types.
In \AFLOWXTALFINDERSHORT,
compounds are first grouped into isopointal sets by analyzing and comparing the symmetries of the structures,
aggregating them by stoichiometry, space groups, and Wyckoff sets (calculated via \AFLOWSYM~\cite{curtarolo:art135}).
Next, compounds are further partitioned into near-isoconfigurational sets by determining and comparing the
local \LFA\ geometries in each structure.
Within each near-isoconfigurational group, one representative structure --- generally the first in the set ---
is compared to the other structures via geometric comparisons and the misfit values are stored.
Once the comparisons finish, any unmatched structures
(\ie\ misfit values greater than $\epsilon_{\mathrm{match}}$) are reorganized
into new comparison sets.
The process is recursive, repeating until all structures have been assembled into matching groups or all comparison pairs have been exhausted.
The three comparison analyses are performed in this order for two reasons:
\textbf{i.}~to categorize structural similarity to varying degrees
(isopointal, near-isoconfigurational, and isoconfigurational) and
\textbf{ii.}~to efficiently group compounds to reduce the computational cost of the geometric structure comparison.
This procedure is the same for material-, structure-, decoration-, and magnetic-type comparisons; however, different
atom mapping restrictions are applied depending on the comparison mode.

To enhance calculation speed, multithreading capabilities can be employed.
The three computationally intensive procedures
--- calculating the symmetry, constructing the local \LFA\ geometry, and performing geometric comparisons ---
are partitioned onto allocated threads, offering significant speed increases for large collections of structures.

\noindent{\textbf{Comparison against established repositories.}}
There are three built-in functions to compare multiple structures automatically (\Fig~\ref{fig:cpp_xtalfinder}\pan{d}):
\textbf{i.}~compare structures provided by a user,
\textbf{ii.}~compare an input structure to prototypes in \AFLOW~\citeANRL, and
\textbf{iii.}~compare an input structure to entries in the \AFLOWorg\ repositories.

\noindent\textbf{Compare user datasets.}
Users can load crystal geometries and compare them automatically with \AFLOWXTALFINDERSHORT.
Options to perform both material-type and structure-type comparisons are available to identify unique compounds and prototypes, respectively.
For structure-type comparisons, the unique atom decorations for each representative structure are determined.
Once the analysis is complete, \AFLOWXTALFINDERSHORT\ groups compatible structures together and returns the corresponding misfit values.

\noindent\textbf{Compare to \AFLOW\ prototypes libraries.}
Given an input structure, this routine returns similar \AFLOW\ prototypes along with their misfit values.
\AFLOW\ contains structural prototypes that can be rapidly decorated for high-throughput materials discovery:
1,100 in the \PROTOTYPEENCYCLOPEDIA~\citeANRL{} and
1,492 in the \HTQC~\cite{curtarolo:art65}.
In this method, \AFLOW\ prototypes are extracted --- based on similar stoichiometry, space group, and Wyckoff positions to the input --- and compared to the user's structure.
Since only matches to the input are relevant, the procedure terminates before regrouping any unmatched prototypes.
The attributes of matched prototypes are also returned, including the prototype label,
mineral name, {\it{Strukturbericht}} designation, and links to the corresponding \PROTOTYPEENCYCLOPEDIA\ webpage.
The scheme identifies common structure-types with the \AFLOW\ prototype libraries or --- if no matches are found --- reveals new prototypes.
Absent prototypes can be characterized automatically in the \AFLOW\ standard designation with \AFLOWXTALFINDERSHORT{}'s
prototyping tool.

\noindent\textbf{Compare to \AFLOWorg\ repositories.}
Compounds are compared to entries in the \AFLOWorg\ repositories using the \AFLOW\ \REST- and \AFLUX\ Search-\API{s}~\cite{curtarolo:art92,curtarolo:art128}.
An \AFLUX\ query (\ie\ matchbook and directives) is generated internally
and returns database compounds similar to the input structure based on species, stoichiometry, space group, and Wyckoff positions.
With the \AURL\ from the \AFLUX\ response, structures for the entry are retrieved via the \RESTAPI.
The most relaxed structure is extracted by default; however, options are available to obtain structures at different
{{ab-initio}} relaxation steps.
The set of entries from the database is then compared to the input structure.
Similar to the \AFLOW\ prototype comparisons, candidate entries are only compared against the input structure,
\ie\ the procedure terminates without regrouping unmatched entries.

With the underlying \AFLUX\ functionality, material properties can also be extracted,
highlighting the structure-property relationship amongst similar materials.
For instance, the enthalpy per atom ($H_{\mathrm{atom}}$) for matching database entries are printed by including the
\inlineCommand{enthalpy\_atom} \API\ keyword in the query.
Any number or combination of properties can be queried; available \API\ keywords are located in
\Ref~\onlinecite{afloworg_web_2022}.

This routine reveals equivalent \AFLOWorg\ compounds if similar materials exist in
the database.
As such, it can estimate structural properties {\it{a priori}}; before performing any calculations.
The estimation is based on the following assumptions:
\textbf{i.}~the matching \AFLOW\ material resides at a local minimum in the energy landscape and
\textbf{ii.}~the input structure relaxes to the same geometry as the matching \AFLOW\ compound, given comparable
calculation parameters.
The functionality can explore properties that are not calculated for a given entry, but are calculated for an equivalent entry.
For example, compounds in \AFLOW's prototype catalogs (\LIBONE, \LIBTWO, \LIBTHREE, \etc) do not usually have band structure data; however,
corresponding \ICSD\ entries can be found which do provide band structure information.
Finally, the method can identify compounds that are absent from the database and prioritize them for future calculation,
enhancing the diversity of the \AFLOWorg\ repositories.

\noindent{\textbf{New features.}} Since the publication of \Ref~\onlinecite{curtarolo:art170}, additional functionality has been
added to the \AFLOWXTALFINDERSHORT\ module.
The transformations for mapped structures can be returned to users by appending the \inlineCommand{-{}-print\_mapping} option to
a comparison command (\eg\ \inlineCommand{-{}-compare\_materials} or \inlineCommand{-{}-compare\_structures}).
The transformation information includes
\textbf{i.}~the basis transformation ($3\times 3$ matrix),
\textbf{ii.}~the rotation of the coordinate system ($3\times 3$ matrix),
\textbf{iii.}~the origin shift ($3\times 1$ vector), and
\textbf{iv.}~the volume scaling factor (scalar) between the two structures.
Along with changes to the lattice vectors, the basis transformation accounts for changes in the unit cell and can
describe mappings between smaller and larger cells.
Furthermore, the atom mapping information is included, indicating which atoms are mapped and their relative distances
between the two structures.

\label{xtalfinder:command_line}
\noindent\textbf{Command-line interface.}
The \AFLOWXTALFINDERSHORT\ command-line calls are detailed below.
Function descriptions and options are provided following each command.

\noindent\textbf{Prototype commands.}
A structure (\inlineCommand{\textit{GEOM\_FILE}}) is converted into its standard \AFLOW\ prototype label
with the command
\begin{lstlisting}[language=aflowBash]
aflow --prototype < /*GEOM_FILE*/
\end{lstlisting}
The parameter variables (degrees of freedom) and corresponding values are also listed.
Information about the label and parameters are described in the \Refs~\onlineciteANRL. \\
Options for this command include:
\begin{myitemize}
  \item \inlineCommand{-{}-setting=\textit{SETTING}} :
    Specify the space group setting for the conventional cell/Wyckoff positions.
    Possible values for \textit{SETTING} include: \inlineCommand{1}, \inlineCommand{2}, or \inlineCommand{aflow}.
    Setting values \inlineCommand{1} and \inlineCommand{2} generally correspond to the first and second choice listed in the \ITC, respectively.
    The \inlineCommand{aflow} setting follows the choices of the \PROTOTYPEENCYCLOPEDIA:
    axis-$b$ for monoclinic space groups,
    rhombohedral setting for rhombohedral space groups,
    and origin centered on the inversion site for centrosymmetric space groups (default: \inlineCommand{aflow}).
\end{myitemize}
\noindent\textbf{Comparison commands.}
The command
\begin{lstlisting}[language=aflowBash]
aflow --compare_materials=/*GEOM_FILES*/
\end{lstlisting}
compares a comma-separated list of geometry files of compounds comprised of the same elements and with
commensurate stoichiometric ratios, \ie\ material-type comparison,
returning their level of similarity (misfit value).
This method identifies unique and duplicate materials.
The command
\begin{lstlisting}[language=aflowBash]
aflow --compare_structures=/*GEOM_FILES*/
\end{lstlisting}
compares a comma-separated list of geometry files of compounds with commensurate stoichiometric ratios with no
requirement of the element type,
\ie\ structure-type comparison, and returns their level of similarity (misfit value).
This method identifies unique and duplicate prototypes.
For the material- and structure-type comparisons, there are three input types (examples for \inlineCommand{-{}-compare\_material} are shown below):
\begin{myitemize}
  \item \inlineCommand{aflow -{}-compare\_materials=\textit{GEOM\_FILES}} : \inlineCommand{\textit{GEOM\_FILES}} is a comma-separated list of geometry files to compare.
  \item \inlineCommand{aflow -{}-compare\_materials -D \textit{path}} : Path to directory (\inlineCommand{\textit{path}}) containing geometry files to compare.
  \item \inlineCommand{aflow -{}-compare\_materials -F=\textit{filename}} : File (\inlineCommand{\textit{filename}}) containing compounds between delimiters
    \inlineCommand{[{\color{codegreen} VASP\_POSCAR\_MODE\_EXPLICIT}]START} and \inlineCommand{[{\color{codegreen} VASP\_POSCAR\_MODE\_EXPLICIT}]STOP}.
\end{myitemize}
To do the same for structure-type comparisons, swap \inlineCommand{-{}-compare\_material}
with \inlineCommand{-{}-compare\_structures} in the commands above.

The command
\begin{lstlisting}[language=aflowBash]
aflow --compare2database < /*GEOM_FILE*/
\end{lstlisting}
compares a structure (\inlineCommand{\textit{GEOM\_FILE}}) to \AFLOWorg\ repositories entries, returning similar
compounds and quantifying their levels of similarity (misfit values).
Material properties can be extracted from the database (via \AFLUX) and printed, highlighting structure-property relationships.
This function can perform either material-type comparisons or structure-type comparisons (by adding the \inlineCommand{-{}-structure\_comparison} option).
Options specific to this command include:
\begin{myitemize}
\item \inlineCommand{-{}-properties=\textit{keywords}} : Specifies the comma-separated properties via their \API\ keyword to print the corresponding values with the comparison results.
\item \inlineCommand{-{}-catalog=\textit{string}} : Restricts the database entries to a specific catalog/library (\eg\ \inlineCommand{lib1}, \inlineCommand{lib2}, \inlineCommand{lib3}, and \inlineCommand{icsd}).
\item{\inlineCommand{-{}-geometry\_file=\textit{string}} : Compares geometries from a particular \DFT\ relaxation step (\eg\ \fileName{POSCAR.relax1}, \fileName{POSCAR.relax2}, and \fileName{POSCAR.static}).}
\end{myitemize}
The command
\begin{lstlisting}[language=aflowBash]
aflow --compare2prototypes < /*GEOM_FILE*/
\end{lstlisting}
compares a structure (\inlineCommand{\textit{GEOM\_FILE}}) against the \AFLOW\ prototype libraries, returning similar
structures and quantifying their levels of similarity (misfit values).
Options specific to this command include:
\begin{myitemize}
  \item \inlineCommand{-{}-catalog=\textit{string}} : Restricts the prototypes to a specific catalog/library (\eg\ \inlineCommand{aflow} or \inlineCommand{htqc}).
\end{myitemize}
The command
\begin{lstlisting}[language=aflowBash]
aflow --isopointal_prototypes < /*GEOM_FILE*/
\end{lstlisting}
returns prototype labels that are isopointal (\ie\ similar space group and Wyckoff positions)
to the input structure (\inlineCommand{\textit{GEOM\_FILE}}).
Options specific to this command include:
\begin{myitemize}
  \item \inlineCommand{-{}-catalog=\textit{string}} : Restricts the prototypes to a specific catalog/library (\eg\ \inlineCommand{aflow} or \inlineCommand{htqc}).
\end{myitemize}
The command
\begin{lstlisting}[language=aflowBash]
aflow --unique_atom_decorations < /*GEOM_FILE*/
\end{lstlisting}
determines the unique and duplicate atom decorations for a given structure.

A full list of the possible commands and options is available in the \AFLOWXTALFINDERSHORT\ \README,
which is printed with the command
\begin{lstlisting}[language=aflowBash]
aflow --readme=xtalfinder
\end{lstlisting}

\section{\PAOFLOW\ Electronic Analysis}\label{sec:paoflow}
\begin{figure*}
  \includegraphics[width=\textwidth]{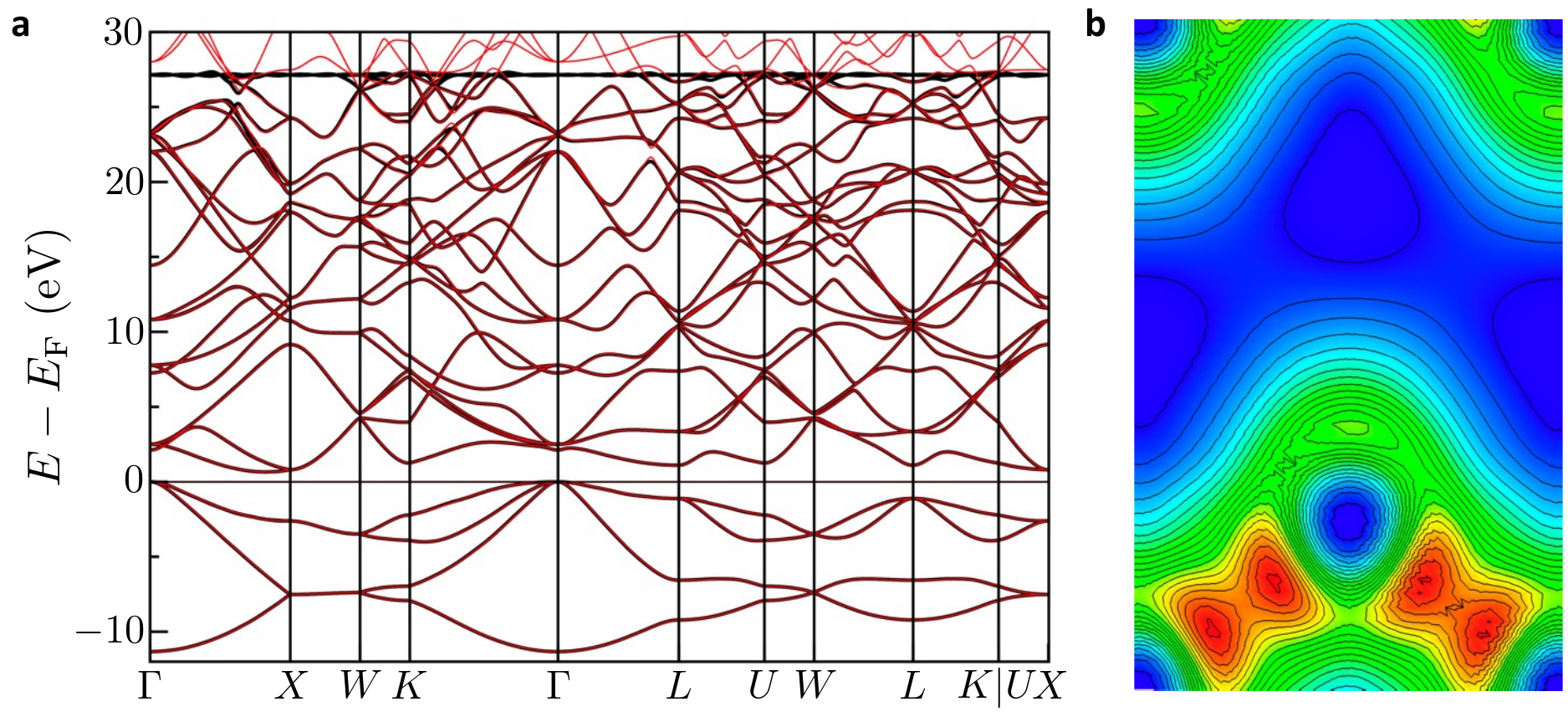}
  \caption{\textbf{Electronic band structure and density calculated with \PAOFLOW.}
  ({\bf a})~Silicon band structure calculated with \PAOFLOW\ (black) projecting
  on the basis set comprising the $3s$, $3p$, $3d$, $4s$, $4p$, $4d$, and $4f$ orbitals, compared to the \DFT\ band structure ({\bf \color{red} red}).
  ({\bf b})~Silicon electron density reconstructed with \PAOFLOW\ on the same basis set (contour plot on the $\left[1,0,-1\right]$ plane).
  }
  \label{fig:cpp_paoflow}
\end{figure*}

\noindent
\PAOFLOW~\cite{curtarolo:art133,curtarolo:art181} is a software tool to efficiently post-process standard first-principles electronic structure
plane-wave pseudopotential calculations.
From interpolated band structures and \DOS,
it promptly computes several quantities that provide insight into transport, optical, magnetic and topological properties
like anomalous and spin Hall conductivities, magnetic circular dichroism, spin circular dichroism,
and topological invariants.
The methodology is based on the \underline{p}rojection of the electronic wavefunctions
of the system on a local \underline{a}tomic \underline{o}rbitals basis (\PAO)~\cite{curtarolo:art86,curtarolo:art108,curtarolo:art111}
and is part of the \AFLOW\ software infrastructure~\cite{curtarolo:art65,curtarolo:art75}.
Currently, \PAOFLOW\ is interfaced with \QUANTUMESPRESSO~\cite{quantum_espresso_2009,Giannozzi:2017io} and the proposed procedure
is completely general and can be implemented with any \DFT\ electronic structure engine.
Accurate \PAO\ Hamiltonian matrices can be built from the direct projection of the Kohn-Sham Bloch states
$\left|\psi_{n\mathbf{k}}\right\rangle$
onto a chosen basis set of fixed localized functions, as we discussed extensively in \Refs~\onlinecite{curtarolo:art86,curtarolo:art108,curtarolo:art111}.
The real space Hamiltonians $\text{H}{\left(\mathbf{R}\right)}$ ($\mathbf{R}$ is a lattice vector)
can be directly calculated using atomic orbitals or pseudo atomic orbitals from the pseudopotential of any given element~\cite{curtarolo:art86,curtarolo:art108}.
The crucial quantities that measure the accuracy of the basis set are the
projectabilities
$p_{n\mathbf{k}} = \left\langle \psi_{n\mathbf{k}} \right| P \left| \psi_{n\mathbf{k}} \right\rangle \geq 0$
($P$ is the operator that projects onto the space of the \PAO\ basis set, as defined in \Ref~\onlinecite{curtarolo:art108}),
which indicate the representability of a Bloch state $\left|\psi_{n\mathbf{k}}\right\rangle$
on the chosen \PAO\ set.
Maximum projectability, $p_{n\mathbf{k}} = 1$, indicates that the Bloch state can be
perfectly represented in the chosen \PAO\ set; inversely, $p_{n\mathbf{k}} \approx 0$ indicates that the
\PAO\ set is insufficient and should be augmented.
Once the Bloch states with good projectabilities have been identified, the \PAO\ Hamiltonian is constructed as
\[
  \text{H}{\left(\mathbf{k}\right)}=\text{A}\text{E}\text{A}^{\dagger} + \kappa\left[\text{I}-\text{A}{\left(\text{A}^{\dagger}\text{A}\right)}^{-1}\text{A}^{\dagger}\right],
\]
as in \Ref~\onlinecite{curtarolo:art108}.
Here $E$ is the diagonal matrix of Kohn-Sham eigenenergies and $A$ is the matrix of coefficients obtained from projecting the
Bloch wavefunctions onto the \PAO\ set~\cite{curtarolo:art108}.
The expression above can be understood as a filtering procedure on the \PAO\ Hamiltonian:
the states with good projectability are kept while all others are
relegated to an orthogonal null space (the second term in the equation
above).
The null eigenvalues can be moved out from the physically relevant energy spectrum via a shifting parameter $\kappa$.
An example of this is illustrated in \Fig~\ref{fig:cpp_paoflow}\pan{a}, where the null space eigenvalues appear as a straight line at 27.5 eV.
This procedure provides an accurate real space representation of the ab-initio Hamiltonian $\text{H}{\left(\mathbf{r}\right)}$
as a tight-binding (\TB) matrix of very small dimension,
an advantage for the calculation of any physical property requiring precise integration in the reciprocal space.

By exploiting the \PAO\ projection scheme described above, we can
easily Fourier transform, $\mathcal{F} [\cdot]$, the \PAO\ real space representation
and interpolate to arbitrary precision; \ie\ $\text{H}{\left(\mathbf{r}_{\alpha}\right)} \to \text{H}{\left(\mathbf{k}\right)}$.
This procedure is computationally inexpensive because of the small dimension of the \TB\ Hamiltonian,
and it is performed using a zero-padding algorithm that operates
globally on the \PAO\ Hamiltonian with a Fourier transform.
Simply from the knowledge of the \PAO\ $\text{H}{\left(\mathbf{k}\right)}$,
one can directly construct the real-space localized Hamiltonian as
\[
\text{H}{\left(\mathbf{R}\right)} = \mathcal{F}^{-1} \left[\text{H}{\left(\mathbf{k}\right)}\right].
\]
$\text{H}{\left(\mathbf{R}\right)}$ is then zero-padded to
{\bf i.}~increase the resolution in ${\bf k}$-space upon inverse $\mathcal{F} [\cdot]$ and
{\bf ii.}~obtain the interpolated \PAO\ Hamiltonian for any arbitrary {\bf k}-vector
mesh with the same accuracy defined by the projectability number.

From here, it is a simple task to evaluate the expectation value of the momentum operator ---
the main quantity in the definition of many property descriptors.
The momentum operator is defined as~\cite{Mostofi_CPC_2014}:
\begin{align*}
  p_{nm}\left(\mathbf{k}\right) &= \left\langle \psi_{n}\left(\mathbf{k}\right)\right| p \left| \psi_{m}\left(\mathbf{k}\right)\right\rangle  \\
  &= \left\langle u_{n}\left(\mathbf{k}\right) \right| \left(m_{0}/\hbar\right) \nabla_{\mathbf{k}}
  \text{H}{\left(\mathbf{k}\right)} \left| u_{m}\left(\mathbf{k}\right)\right\rangle,
\end{align*}
where
\[
\nabla_{\mathbf{k}}\text{H}{\left(\mathbf{k}\right)} = \sum_{i} \mathbf{R} \exp \left(i \mathbf{k}\cdot\mathbf{R}\right) \text{H}{\left(\mathbf{R}\right)},
\]
$\text{H}{\left(\mathbf{R}\right)}$ being the real space \PAO\ matrix, and
$\left|\psi_{n}\left(\mathbf{k}\right)\right\rangle = \exp\left(-i \mathbf{k}\cdot\mathbf{r}\right) \left|u_{n}\left(\mathbf{k}\right)\right\rangle$
the Bloch's functions~\cite{curtarolo:art116}.
This procedure can be applied multiple times to evaluate higher order derivatives (effective masses, \etc)~\cite{Jayaraj_SciRep_2022}.

In the original formulation of the \PAOFLOW\ method, the atomic orbital basis was built from the radial pseudowavefunctions
of the pseudopotential used in the \DFT\ calculation.
This ``minimal basis set'' approach has proved satisfactory in achieving accurate \TB\ matrices for periodic systems.
However, if more unoccupied bands are needed for a particular application, it can be achieved by
progressively increasing the size of the atomic orbital basis set, effectively increasing the number of states
with high projectability and thus the spectrum of
$\text{H}{\left(\mathbf{k}\right)}$.

We have recently developed an alternative approach that achieves this while maintaining the high accuracy of the minimal basis set.
Our approach is entirely independent of the choice (or the availability) of the pseudopotential's radial functions.
We generate the basis function by solving the all-electron atomic problem and building
the basis set from the atomic radial functions, consistent with the valence states present in the pseudopotential~\cite{curtarolo:art181}.
In this way, we can increase the size of the basis set and construct Hamiltonians
that reproduce exactly the electronic states for energies high in the conduction band.
As an example, we show in \Fig~\ref{fig:cpp_paoflow}\pan{a} the band structure of silicon generated by a
\PAO\ Hamiltonian with a basis set comprising the $3s$, $3p$, $3d$, $4s$, $4p$, $4d$, and $4f$ orbitals.
The accuracy of the representation, measured as the average difference between the original
\DFT\ bands and the \PAOFLOW\ ones across the whole first Brillouin zone, is of the order of $10^{-3}$ for energies up to 20~eV.
Moreover, the introduction of an explicit basis set promotes the \PAOFLOW\ method beyond a simple tight-binding representation.
We are now able to  reconstruct the true electronic wavefunctions fully
and thus the electronic density of the system (see \Fig~\ref{fig:cpp_paoflow}\pan{b}) ---
the essential quantity to evaluate a plethora of properties in their real space representation.
\section{Thermodynamics}

\begin{figure*}
  \includegraphics[width=\textwidth]{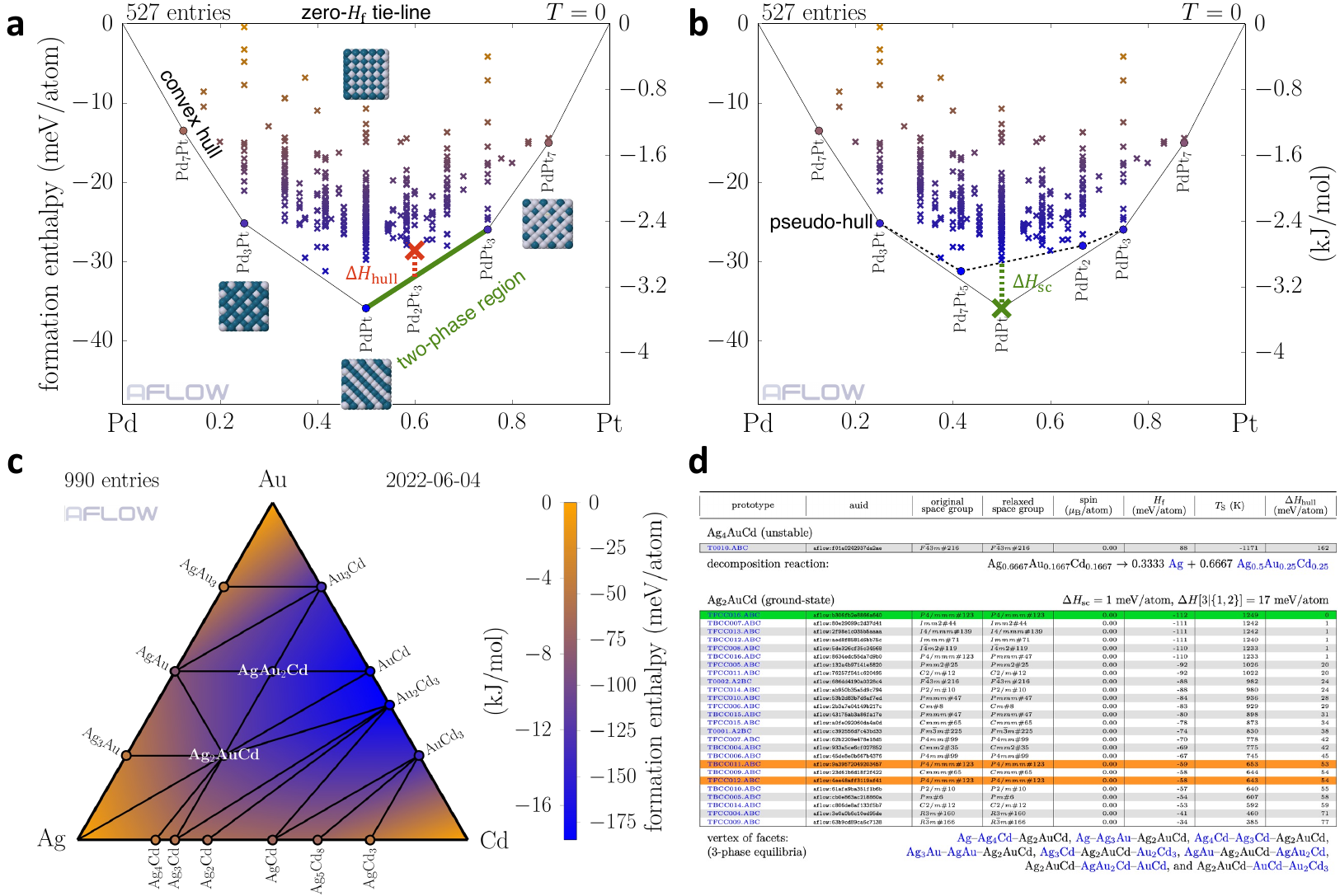}
  \caption{\textbf{Results and output from the \AFLOW\ Convex Hull Module.}
  ({\bf a})~Illustration of the PdPt convex hull, highlighting the various compounds sampled from
  the \AFLOWorg\ repositories having different structures and concentrations, the zero-formation-enthalpy tie-line,
  a two-phase region (hull facet), and the decomposition of Pd$_{2}$Pt$_{3}$.
  ({\bf b})~Illustration of the stability criterion analysis of PdPt, highlighting the construction of the
  pseudo-hull (dotted-line) from which the distance is measured.
  ({\bf c})~Illustration of the AgAuCd convex hull, where the colors (orange to blue) indicate depth (lower formation enthalpy),
  as indicated by the color bar on the right.
  Only stable compounds (on the hull) are shown.
  ({\bf d})~An excerpt from the full AgAuCd \PDF\ report, which
  organizes all the data used to construct the convex hull
  and presents the results of the analysis.
  }
  \label{fig:cpp_chull_1}
\end{figure*}

\begin{figure*}
  \includegraphics[width=\textwidth]{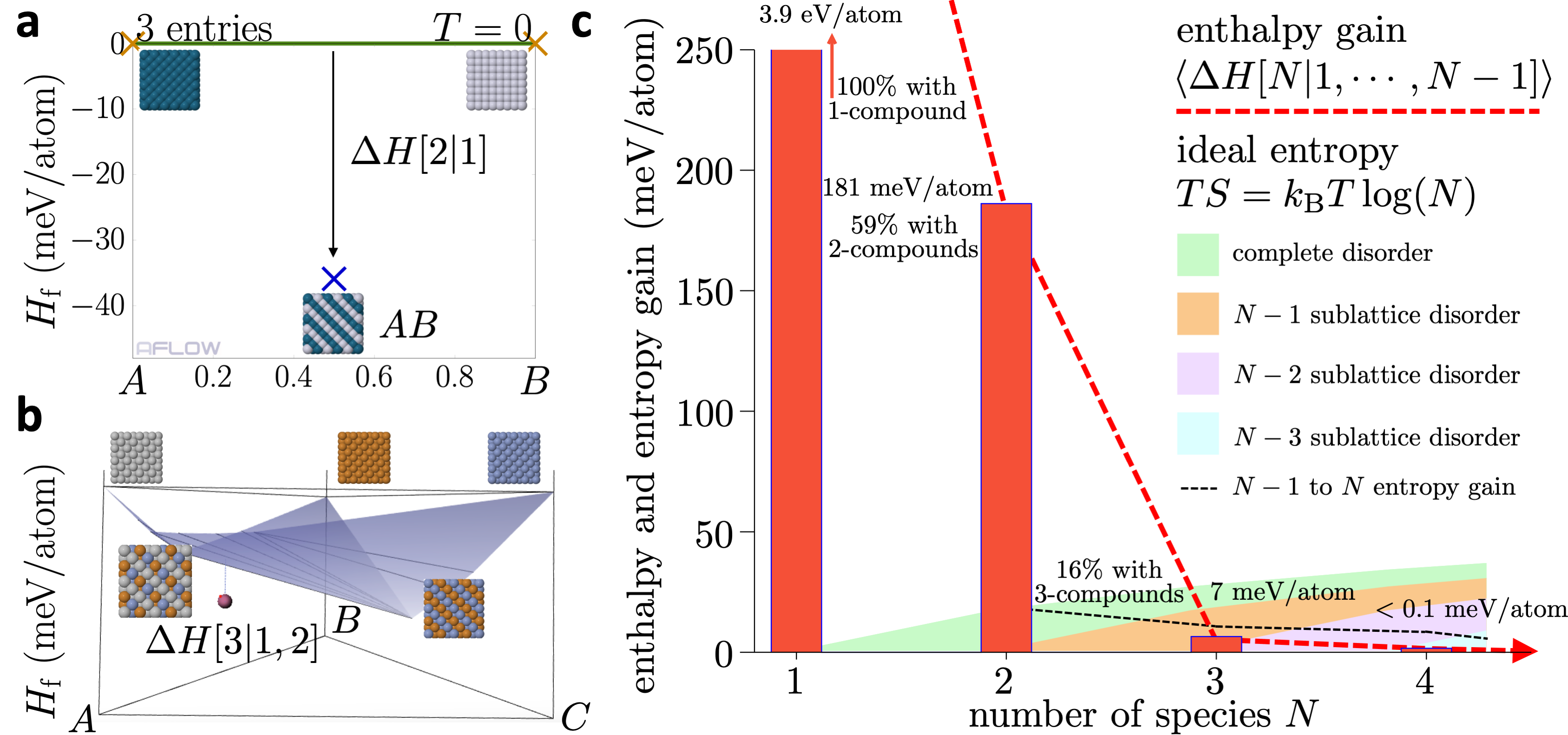}
  \caption{\textbf{The $N+1$ enthalpy gain descriptor.}
  ({\bf a})~For binary compounds, the $N+1$ enthalpy gain descriptor is
  the distance of the compound from the zero-formation-enthalpy tie-line (highlighted in green),
  which is trivially the formation enthalpy.
  ({\bf b})~For ternary compounds, the $N+1$ enthalpy gain descriptor is
  the distance of the compound from the convex hull constructed of
  unary (reference) and binary compounds (highlighted in blue).
  ({\bf c})~A plot comparing the enthalpy gains averaged over $N$-compounds having metallic components in the
  \AFLOWorg\ repositories with the configurational entropy gains for
  increasing numbers of species.
  }
  \label{fig:cpp_chull_2}
\end{figure*}

\noindent
The energetics of the formation and decomposition reactions of a material
influence whether it can be made and
its stability.
Ideally, the full reaction pathway should be considered (kinetics),
including intermediate products which may be energetically inaccessible,
rendering the reaction pathway unfeasible.
These considerations are highly dependent on the experimental method and conditions,
and are thus difficult to generalize.
Instead, formation and stability analyses first focus only on the energy difference
between the reaction endpoints (thermodynamics), which
assumes the system can freely explore all possible outcomes to realize
the minimum energy one (ergodicity).

\subsection{\AFLOWCHULL: The Convex Hull Module}\label{sec:chull}

\noindent
The \AFLOW\ Convex Hull Module (\AFLOWCHULL)~\cite{curtarolo:art144} can be used to construct
ground-state ($T=0$) ab-initio phase diagrams offering a wealth of thermodynamic data.
Identification of stable phases involves the calculation of the convex hull
in the free-energy-concentration space.
Compounds having different structures (\eg\ rocksalt \versus\ wurtzite)
and concentrations $\left(A_{x}B_{1-x}\right)$ are sampled from the
\AFLOWorg\ repositories.
These configurations have been generated through a combination of
structure prototyping of naturally occurring compounds~\cite{curtarolo:art121,curtarolo:art145,aflowANRL3}
and structure enumeration algorithms~\cite{curtarolo:art170}.
Stability $\left(\Delta G<0\right)$ is achieved by minimizing the enthalpy and,
in multi-phase regions,
can involve changing concentrations (phase-separation and tie-line construction),
as dictated by the minimum energy surface.
The collection of stable phases and the tie-lines connecting them is determined by the convex hull:
the set of outer-most points outlining
the smallest convex shape enclosing the data, as illustrated in \Fig~\ref{fig:cpp_chull_1}\pan{a}.
The convex hull defines stability and thermodynamic pathways to it,
and its construction is based solely on the geometry (positions) of the data.
The energy axis is the formation enthalpy $\left(H_{\text{f}}\right)$, which is defined as:
\[
H_{\text{f},A_{x}B_{1-x}}=H_{A_{x}B_{1-x}}-\left[xH_{A}+\left(1-x\right)H_{B}\right],
\]
where $H_{A_{x}B_{1-x}}$ is the enthalpy of a binary compound, and
$H_{A}$ is the enthalpy of the $A$-reference-state.
This compositionally-weighted shift of the raw \DFT-enthalpies
fixes the reference-states $\left(x=[0,1]\right)$ to zero,
so only the lower convex hull needs to be calculated.

Compounds below the zero-formation-enthalpy tie-line $\left(H_{\text{f}}<0\right)$ are only stable with respect to
decomposition to the reference states.
The phases on the convex hull are globally stable (will not decompose) and thus are
expected to form under similar experimental conditions~\cite{curtarolo:art49,curtarolo:art51,curtarolo:art55,curtarolo:art63,curtarolo:art67,curtarolo:art70,curtarolo:art87,curtarolo:art106,curtarolo:art126}.
Compounds above the hull will decompose into a linear combination
of the stable phases defining the tie-line directly below it.
For example, the decomposition reaction of Pd$_{2}$Pt$_{3}$ (highlighted in \Fig~\ref{fig:cpp_chull_1}\pan{a}) is:
\[
\text{Pd}_{0.4}\text{Pt}_{0.6} \xrightarrow[]{\genfrac{}{}{0pt}{}{-3}{\text{meV/atom}}} \frac{3}{5}~\text{Pd}_{0.5}\text{Pt}_{0.5} + \frac{2}{5}~\text{Pd}_{0.25}\text{Pt}_{0.75},
\]
where the distance to the hull $\left(H_{\text{hull}}\right)$ is the energy difference between the products and the reactants,
and the products define the tie-line directly below Pd$_{2}$Pt$_{3}$.
$H_{\text{hull}}$ can be used as a measure of metastability:
compounds close to the hull may stabilize at finite temperatures (room temperature: ${\sim}25~\text{meV}$).

Analogous to the distance to the hull, the stability criterion~\cite{curtarolo:art144,curtarolo:art109} $\left(\Delta H_{\text{sc}}\right)$
was devised to gauge the relative stability of compounds on the hull.
This descriptor is defined as the distance of the compound from the
pseudo-hull constructed without it (\Fig~\ref{fig:cpp_chull_1}\pan{b}),
quantifying the effect of the phase on the convex hull
and its susceptibility to
destabilization by a new phase that has yet to be explored.
The descriptor helped guide the design of two Heusler magnetic compounds,
the first magnets discovered by computational approaches~\cite{curtarolo:art109}.

\AFLOWCHULL\ can construct and analyze convex hulls for arbitrary numbers of components:
\eg\ \Ref~\onlinecite{curtarolo:art140} presents distances to the hull and decomposition reactions
for 5-metal high-entropy carbides (6D hulls).
The module offers illustrations of the hull for binary and ternary systems (\Fig~\ref{fig:cpp_chull_1}\pan{c}),
as well as a \PDF\ report summarizing the data used to construct the hull and the results of the analysis
(\Fig~\ref{fig:cpp_chull_1}\pan{d}).
Entries are organized by arity (ternaries first, then binaries) and concentration.
The report presents `unstable' \versus\ `ground-state' designations, distances to the hull,
decomposition reactions, and stability criteria.
For each stable phase, the report also provides the set of phases with which it is in equilibrium,
\ie\ the set of vertices for all the facets the stable phase defines.
For example, Ag$_{2}$AuCd in \Fig~\ref{fig:cpp_chull_1}\pan{c} (lower left) is a vertex for
eight facets, which
are enumerated in \Fig~\ref{fig:cpp_chull_1}\pan{d} under `vertex of facets'.
This information was used to discover two cobalt-based superalloys,
where candidate impurity-phase compositions potentially form during age-hardening
were screened for not being in two-phase equilibrium with the fcc host matrix~\cite{curtarolo:art113,ReyesTirado_ActaMat_TernarySuperalloys_2018}.

The $N+1$ enthalpy gain $\left(\Delta H\left[N|1,\ldots,N-1\right]\right)$
has also been implemented within \AFLOWCHULL~\cite{curtarolo:art152}.
The descriptor quantifies an $N$-compound's distance from the hull constructed only of
$\{1,\ldots,N-1\}$-compounds (\Fig~\ref{fig:cpp_chull_2}),
where, \eg\ binaries are 2-compounds and ternaries are 3-compounds.
The $N+1$ enthalpy gain for $1$-compounds is the cohesive energy,
and for $2$-compounds is the formation enthalpy (\Fig~\ref{fig:cpp_chull_2}\pan{a}).
An analysis of the \AFLOWorg\ repositories
for metal compositions reveals that with an increasing number of
species there is a diminishing enthalpy gain,
which can be rapidly overcome by the configurational entropy gain ($N\geq4)$,
see \Fig~\ref{fig:cpp_chull_2}\pan{c}.
This validates the unavoidability of disorder in multi-component systems~\cite{curtarolo:art152}.

\noindent{\textbf{Command-line options.}}
There are several ways to interact with the \AFLOWCHULL\ module, including through the web~\cite{afloworg_web_2022},
with the \AFLOW\ binary installed locally, and through Python/Jupyter wrappers (calling a local install of the \AFLOW\ binary).
The binary offers full access to functionality and options, as well as
various output formats, including plain text, \JSON, \PDF, and Jupyter notebooks.
The \PDF\ output requires the \LaTeX\ package. See the Supporting Information of \Ref~\onlinecite{curtarolo:art144}
for version and package details.
The primary \AFLOWCHULL\ command
\begin{lstlisting}[language=aflowBash]
aflow --chull --alloy=MnPdPt
\end{lstlisting}
queries the \AFLOWorg\ repositories for entries containing $\{\text{Mn},\text{Pd},\text{Pt}\}$,
calculates the convex hull, and returns the information as a \PDF\ (default, see \inlineCommand{-{}-print}).}
The flags and options include:
\begin{myitemize}
  \item \inlineCommand{-{}-chull} : Necessary flag for entering the convex-hull module.
  \item \inlineCommand{-{}-alloy=\textit{alloy}} : Necessary argument, specifies the system. This code is not dimension limited,
    \ie\ any $N$-ary system can be calculated.
    There are two input modes:  raw (comma-separated) and combinatorial (colon- and comma-separated). \\
    Raw input:  \inlineCommand{-{}-alloy=MnPdPt,AlCuZn}. \\
    Combinatorial input:  \inlineCommand{-{}-alloy=Ag,Au:Mn}.  This is interpreted as \inlineCommand{-{}-alloy=AgMn,AuMn}.
  \item \inlineCommand{-{}-np=\textit{ncpus}} : Number of threads for calculation of multiple convex hulls.
    Default is \inlineCommand{-{}-np=1} (serial).
  \item \inlineCommand{-{}-print=\textit{format}} :
    Selects the output format, options include: \inlineCommand{pdf}, \inlineCommand{png}, \inlineCommand{json}, \inlineCommand{txt}, \inlineCommand{jupyter2}, and \inlineCommand{jupyter3}.
    \inlineCommand{json} and \inlineCommand{txt} have the following extensions:  \fileName{.json} and \fileName{.txt}.
    \inlineCommand{jupyter2}/\inlineCommand{jupyter3} create a Python2/Python3 Jupyter notebook \JSON\ file that
    plots a convex hull for the specified alloy.
    Default is \inlineCommand{pdf}.
  \item \inlineCommandLong{--dist2hull=aflow:bb0d45ab555bc208} :
    Returns the distance from the hull for entry \inlineCommand{aflow:bb0d45ab555bc208},
    specified by the \AUID.
  \item \inlineCommandLong{--scriterion=aflow:bb0d45ab555bc208} :
    Returns the stability criterion for entry \inlineCommand{aflow:bb0d45ab555bc208},
    specified by the \AUID. The entry must be a ground-state structure; a warning will be issued otherwise.
    \AFLOWCHULL\ removes the point from the hull, calculates the pseudo-hull, and determines the distance of this point from
    below the pseudo-hull.
  \item \inlineCommandLong{--nplus1=aflow:bb0d45ab555bc208} :
    Returns the $N+1$ enthalpy gain for entry \inlineCommand{aflow:bb0d45ab555bc208},
    specified by the \AUID. The entry must be a ground-state structure; a warning will be issued otherwise.
    \AFLOWCHULL\ removes all points having the same dimensionality as the input entry from the hull,
    calculates the pseudo-hull, and determines the distance of this point from below the pseudo-hull.
  \item \inlineCommand{-{}-hull\_enthalpy=0.25,0.25} : Returns the value of the convex hull surface at the specified coordinate/concentration.
    Users should provide the composition in reduced form, \eg\ the Mn$_{2}$PdPt composition is specified by
    \inlineCommand{-{}-hull\_enthalpy=0.5,0.25}, where the last component is implicitly $1-\text{sum}(0.5+0.25)$.
\end{myitemize}
For the full set of options and additional information, see the \AFLOWCHULL\ \README:
\begin{lstlisting}[language=aflowBash]
aflow --readme=chull
\end{lstlisting}

\subsection{\AFLOWCCE: The Coordination Corrected Enthalpies Module}\label{sec:cce}

\noindent
While there have been significant advances in calculating finite temperature effects from first principles~\cite{PhysRevB.79.134106,Grabowski_NPJCM_2019,curtarolo:art125},
the computational modeling of formation enthalpies ---
the enthalpy difference between the material and its elemental references --- still poses a fundamental challenge.
Standard (semi-)local and even currently available advanced ab-initio approaches yield inaccurate predictions~\cite{Wang_Ceder_GGAU_PRB_2006,Lany_FERE_2008,Jain_GGAU_PRB_2011,Lany_Zunger_FERE_2012,Zhang_NPJCM_2018,Isaacs_PRM_2018,Yan_formation_PRB_2013,Jauho_PRB_2015},
with errors of several hundred meV/atom
in particular for ionic systems, which inhibits materials design.
The problem is intimately connected to the fact that computing reliable formation enthalpies ab-initio eventually requires accurate total energies for all systems involved~\cite{Lany_Zunger_FERE_2012,curtarolo:art150,curtarolo:art172}.
This is generally not possible within a \mbox{(semi-)local} approximation.
To date, it even remains unknown what level of \DFT-based
theory would be needed to achieve satisfactory accuracy for formation enthalpies, given that exact Quantum Monte Carlo results are only available for a few special systems, such as MgH$_2$~\cite{Pozzo_PRB_2008,Mao_QMC_2011}.

Physically motivated empirical correction schemes parameterizing \mbox{(semi-)local} \DFT\ errors with respect to measured values are hence the only feasible option to enable materials design.
Several correction methods based solely on the composition of the materials were established~\cite{Wang_Ceder_GGAU_PRB_2006,Jain_GGAU_PRB_2011,Wolverton_DFTUenthalpies_prb_2014,Lany_FERE_2008,Lany_Zunger_FERE_2012}.
These approaches were a major step forward, but their accuracy is limited and the relative stability of polymorphs --- sometimes erroneously predicted by \DFT~\cite{Zhang_NPJCM_2018} --- cannot be corrected.
Moreover, correction methods based only on composition can lead to incorrect thermodynamic behavior when considering activity \versus\ concentration
\cite{curtarolo:art150}.

\begin{figure*}
\centering
\includegraphics[width=\textwidth]{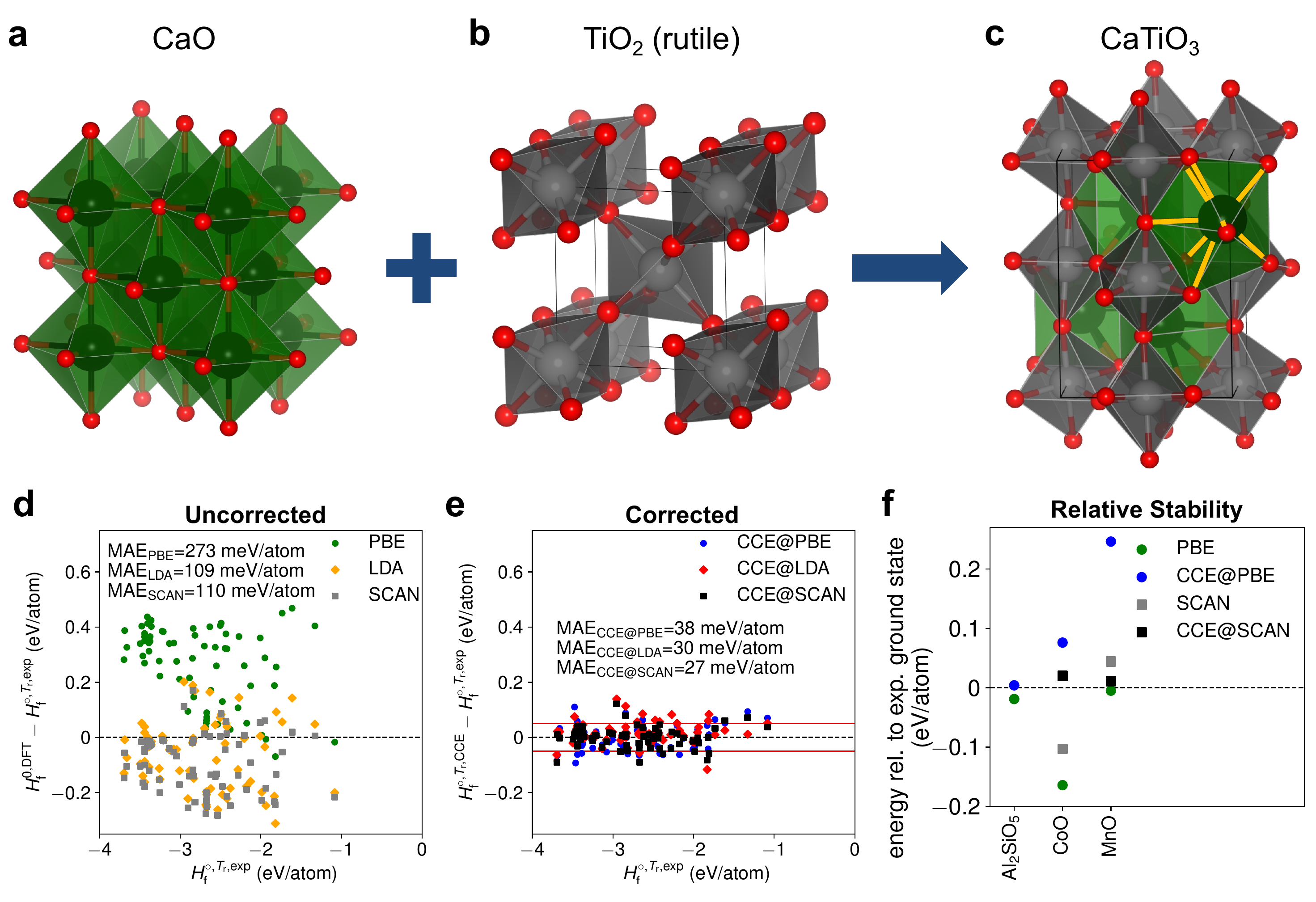}
\caption{\textbf{Motivation and validation of \CCE.}
  Crystal structures of (\textbf{a})~CaO, (\textbf{b})~rutile TiO$_2$, and (\textbf{c})~perovskite CaTiO$_3$.
	While Ti is sixfold coordinated by O anions in both rutile and perovskite, Ca changes its coordination from sixfold (CaO) to eightfold (CaTiO$_3$, Ca--O bonds highlighted in yellow).
	This indicates an important change in the number of bonds critical for the thermodynamic stability of a material.
  Deviation between (\textbf{d})~calculated $\left(H^{0,\text{DFT}}_{\text{f}}\right)$ and experimental $\left(H^{\circ,T_{\text{r}},\text{exp}}_{\text{f}}\right)$ as well as (\textbf{e})~corrected $\left(H^{\circ,T_{\text{r}},\text{CCE}}_{\text{f}}\right)$ and experimental formation enthalpies for 71 ternary oxides.
	Red lines at $\pm50$~meV/atom are visual guides representing the typical \MAE\ of previous methods~\cite{Jain_GGAU_PRB_2011,Lany_Zunger_FERE_2012}.
	({\bf f})~Correction of the relative stability of Al$_2$SiO$_5$, CoO, and MnO polymorphs.
	Color code: Ca, black; Ti, light gray; O, red~\cite{VESTA}.
}
\label{figure_CCE_motivation_validation}
\end{figure*}

\noindent\textbf{The coordination corrected enthalpies method.}
The method of \underline{c}oor\-di\-na\-tion \underline{c}orrected \underline{e}nthalpies (\CCE)~\cite{curtarolo:art150} presents a solution to this problem, improving on the established approaches
both qualitatively and quantitatively.
\CCE\ is the first scheme to leverage structural information
.
Incorrect thermodynamic behavior is avoided by construction.
The method is based on the following physical insight:
bonding is central to capturing the properties of a material, with \DFT\ revealing specific errors for each bonding type.
Thus, the number of bonds in a material is a suitable descriptor to parameterize \DFT\ errors.
The correction is hence developed per bond and per oxidation state.
The latter dependence ensures optimal transferability of the corrections since the energetic position of valence states is usually well characterized by the oxidation state.
The idea is further illustrated in the example of \Figs~\ref{figure_CCE_motivation_validation}\pan{a-c}:
in CaO, the Ca cations are sixfold (octahedrally) coordinated by oxygen anions.
Ti is also sixfold coordinated by oxygen in rutile TiO$_2$.
In perovskite CaTiO$_3$, the coordination number for Ti stays the same, but the number of Ca--O bonds changes to eight.
A variation in the number of bonds for a given cation between different materials is common and signifies that corrections to calculated DFT formation enthalpies should be obtained and applied per bond.

With binary compounds $A_{x_1}Y_{x_2}$ used as the fit set, the \CCE\ corrections $\delta H^{T_{\text{r}},A^{+\alpha}}_{A\text{--}Y}$ per cation-anion $A$--$Y$ bond and cation oxidation state $+\alpha$ are obtained from the difference between (zero-temperature and zero-pressure) \DFT\ and experimental standard room temperature formation enthalpies~\cite{curtarolo:art150,curtarolo:art172}:
\[
  H^{0,\text{DFT}}_{\text{f},A_{x_1}Y_{x_2}}-H^{\circ,T_{\text{r}},\text{exp}}_{\text{f},A_{x_1}Y_{x_2}} =x_1N_{A\text{--}Y}\delta H^{T_{\text{r}},A^{+\alpha}}_{A\text{--}Y},
\]
where $H^{0,\text{DFT}}_{\text{f},A_{x_1}Y_{x_2}}$ is the \DFT\ formation enthalpy, $H^{\circ,T_{\text{r}},\text{exp}}_{\text{f},A_{x_1}Y_{x_2}}$ is the measured standard formation enthalpy at the reference temperature {$T_{\text{r}}=$298.15~K}, and $N_{A\text{--}Y}$ is the number of nearest neighbor $A$--$Y$ bonds of element $A$ in oxidation state $+\alpha$.
Note that the \DFT\ formation enthalpies strictly include only the internal energy contributions to the enthalpies and the small pressure-volume terms are neglected~\cite{curtarolo:art150}.

The corrections can then be applied to any multinary compound $A_{x_1}B_{x_2}\dots Y_{x_n}$ to obtain the \CCE\  formation enthalpy $H^{\circ,T_{\text{r}},\text{CCE}}_{\text{f},A_{x_1}B_{x_2}\text{\dots}Y_{x_n}}$ at no additional computational cost compared to \DFT:
\begin{align*}
  H^{\circ,T_{\text{r}},\text{CCE}}_{\text{f},A_{x_1}B_{x_2}\text{\dots}Y_{x_n}} ={}& H^{0,\text{DFT}}_{\text{f},A_{x_1}B_{x_2}\text{\dots}Y_{x_n}} - \\& -\sum_{i=1}^{n-1} x_iN_{i-Y}\delta H^{T_{\text{r}},i^{+\alpha}}_{i-Y},
\end{align*}
where $N_{i-Y}$ is the number of nearest neighbor bonds between the cation $i$ and anion $Y$ species.
Temperature effects can also be parameterized on a per bond basis such that, in addition to room temperature values, also 0~K formation enthalpies are computed~\cite{curtarolo:art172}.\par

The predictive power of \CCE\ was validated on a test set of 71 ternary oxides.
\DFT\ yields, for
\PBE~\cite{PBE}, \LDA~\cite{DFT,von_Barth_JPCSS_LSDA_1972} and \SCAN~\cite{Perdew_SCAN_PRL_2015}, mean absolute errors (\MAE{}s) of the calculated enthalpies of at least 100~meV/atom (\Fig~\ref{figure_CCE_motivation_validation}\pan{d}).
The \CCE\ values show an improvement by a factor of 4--7 to 38, 30 and 27~meV/atom (\Fig~\ref{figure_CCE_motivation_validation}\pan{e}).
These mean deviations are significantly smaller than the ones of 45 and 48~meV/atom predicted by previous methods~\cite{Jain_GGAU_PRB_2011,Lany_Zunger_FERE_2012}.
The general applicability of \CCE\
was benchmarked on a set of ternary halides, achieving the same accuracy
\cite{curtarolo:art150}.
\CCE\ is also capable of correcting the relative stability of polymorphs at fixed composition --- a qualitative advantage versus all earlier schemes --- as demonstrated for several minerals and transition metal systems~\cite{curtarolo:art150}.
In \Fig~\ref{figure_CCE_motivation_validation}\pan{f} this is indicated in three examples.
While plain \PBE\ predicts the andalusite polymorph of Al$_2$SiO$_5$ to be more stable than the experimentally known ground-state kyanite (green dot), the application of \CCE\ shifts andalusite energetically above kyanite and retains the correct energetic ordering (blue dot).
Similarly, for CoO and MnO, it predicts the correct ground-state rocksalt structure, while
\DFT\ erroneously yields zincblende with only four Co/Mn-O bonds~\cite{curtarolo:art150}.
Other correction methods based on only composition cannot rectify the relative stability.
\CCE\ also gives accurate defect energies evidenced from investigating crystallographic shear compounds~\cite{curtarolo:art150}.\par

\begin{figure*}
	\centering
	\includegraphics[width=\textwidth]{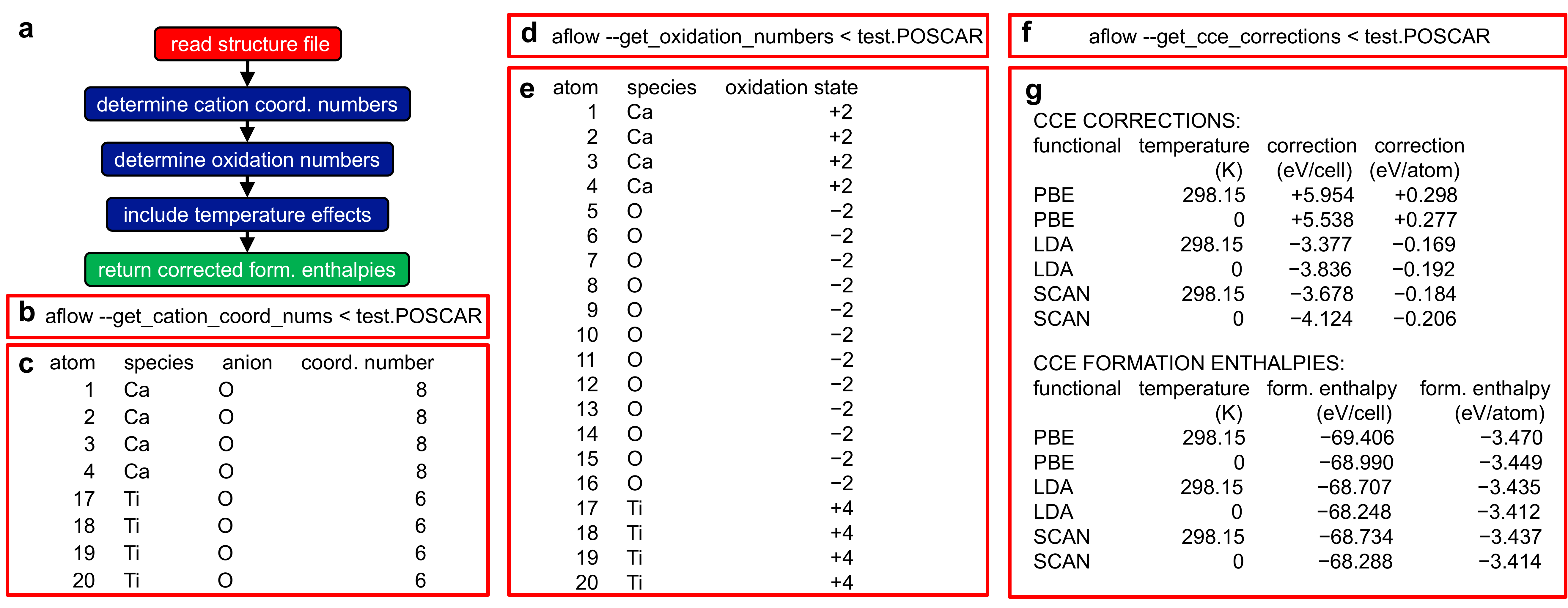}
	\caption{\textbf{\AFLOWCCE\ implementation.}
	(\textbf{a})~Workflow of the \CCE\ implementation.
  ({\bf b} and {\bf d})~Example commands for the \AFLOWCCE\ command-line tool using the input structure file \fileName{test.POSCAR} (perovskite CaTiO$_3$) to determine cation coordination numbers and oxidation states.
	({\bf c} and {\bf e})~When executed, the output is presented in tabular form.
	({\bf f})~Example command for determining \CCE\ corrections and formation enthalpies.
	Several other options to use extended functionality~\cite{curtarolo:art172} are also summarized in the main text.
	({\bf g})~When executed, the output includes the \CCE\ corrections and formation enthalpies at both 298.15 and 0~K for various functionals.
	}
	\label{figure_CCE_implementation}
\end{figure*}

\noindent\textbf{\AFLOWCCE.}
\CCE\ is integrated into and is fully interoperable with existing \AFLOW\ functionality~\cite{curtarolo:art172}.
\AFLOWCCE\ returns the correction and the corrected formation enthalpies for a given structure.
It features a command-line tool, a web interface, and a Python environment~\cite{curtarolo:art172}.
As shown in \Fig~\ref{figure_CCE_implementation}\pan{a}, the workflow analyzes the structure for cation coordination numbers, automatically determines oxidation numbers based on Allen electronegativities~\cite{Allen_electronegativity_1989,Mann_JACS_2000,Mann_JACS_2000_2}, includes temperature effects, and calculates the corrected formation enthalpies for the selected functionals.
The oxidation state determination includes the treatment of mixed-valence systems, such as Ti-O Magn\'eli phases and alkali-metal sesquioxides.
The software is also readily applicable to (su-)peroxides and multi-anion systems.
The algorithms of the implementation are detailed in \Ref~\onlinecite{curtarolo:art172}.

The command-line tool (\Figs~\ref{figure_CCE_implementation}\pan{b-g}) delivers the \CCE\ cation coordination numbers, oxidation numbers, and corrections, as well as formation enthalpies for the given structure file that can be in any format recognizable by \AFLOW, like {\VASP\ \POSCAR}~\cite{vasp}, \QUANTUMESPRESSO~\cite{quantum_espresso_2009}, \FHIAIMS~\cite{blum:fhi-aims}, \ABINIT~\cite{gonze:abinit}, \ELK~\cite{elk} and \CIF~\cite{Hall_CIF_1991}.
For {\VASP}, a \VASPFIVE\ \POSCAR\ is required, or if a \VASPFOUR\ \POSCAR\ is used, the species must be written on the right side next to the coordinates for each atom just as for the example input structure obtained from the option \inlineCommand{-{}-cce}.
If desired, oxidation numbers for all atoms can also be provided upon input.
The available options include:
\begin{lstlisting}[language=aflowBash]
aflow --cce
\end{lstlisting}
prints instructions and an example input structure.
\begin{lstlisting}[language=aflowBash]
aflow --cce=/*GEOM_FILE*/
\end{lstlisting}
prints the results of the full \CCE\ analysis, \ie\ cation coordination numbers, oxidation numbers, and \CCE\ corrections and formation enthalpies for the given structure.
\begin{lstlisting}[language=aflowBash]
aflow --get_cce_corrections < /*GEOM_FILE*/
\end{lstlisting}
determines the \CCE\ corrections and formation enthalpies for the structure.
\begin{lstlisting}[language=aflowBash]
aflow --get_oxidation_numbers < /*GEOM_FILE*/
\end{lstlisting}
determines the oxidation numbers for all atoms of the structure.
\begin{lstlisting}[language=aflowBash]
aflow --get_cation_coord_nums < /*GEOM_FILE*/
\end{lstlisting}
determines the number of anion neighbors for each cation of the structure.

\noindent
Options for \inlineCommand{-{}-cce=\textit{GEOM\_FILE}} and \inlineCommand{-{}-get\_cce\_corrections < \textit{GEOM\_FILE}}:
\begin{myitemize}
\item \inlineCommand{-{}-enthalpies\_formation\_dft=\textit{enthalpies}} :
\inlineCommand{\textit{enthalpies}} is a comma-separated list of precalculated \DFT\ formation enthalpies. They are assumed to be
negative for compounds lower in enthalpy than the elemental references and in eV/cell.
Currently, corrections are available for \PBE, \LDA, and \SCAN.
\item \inlineCommand{-{}-functionals=\textit{functionals}} :
\inlineCommand{\textit{functionals}} is a comma-separated list of functionals for which corrections should be returned.
If used together with \inlineCommand{-{}-enthalpies\_formation\_dft},
the functionals must be in the same sequence as the corresponding formation enthalpies. \
Available functionals are \PBE, \LDA, and \SCAN.
Default is \PBE\ (if only one \DFT\ formation enthalpy is provided).
\item \inlineCommand{-{}-oxidation\_numbers=\textit{oxidation\_numbers}} :
\inlineCommand{\textit{oxidation\_numbers}} is a comma-separated list of oxidation numbers.
It is assumed that one is provided for each atom of the structure and
they are in the same sequence as the corresponding atoms in the provided structure file.
\end{myitemize}

The output of these results can be returned as plain text or a \JSON\ format:
\inlineCommand{-{}-print=\textit{format}}, where \inlineCommand{\textit{format}}
can be \inlineCommand{txt} or \inlineCommand{json}, respectively.
For the full set of options and additional information, see the \AFLOWCCE\ \README:
\begin{lstlisting}[language=aflowBash]
aflow --readme=cce
\end{lstlisting}

\vspace{0.5cm}

The \AFLOWCCE\ implementation enables the enthalpy correction of an extensive library of materials as well as the accurate and quick generation of convex hull phase diagrams~\cite{curtarolo:art144}.
The tool is also readily applicable to reduced-dimensionality, \eg\ 2D systems.
\AFLOWCCE\ thus provides a valuable infrastructure for the scientific community to obtain the \CCE\ corrections and corrected formation enthalpies for a given input structure. It can also expedite various materials design applications, such as the discovery of novel 2D systems and high-entropy ceramics.

\section{Thermomechanical Analysis}

\def\sDebye{{\substack{\scalebox{0.6}{D}}}}
\def\sD{{\substack{\scalebox{0.6}{D}}}}
\def\sDFT{{\substack{\scalebox{0.6}{DFT}}}}
\def\sAGL{{\substack{\scalebox{0.6}{AGL}}}}
\def\sAEL{{\substack{\scalebox{0.6}{AEL}}}}
\def\sMP{{\substack{\scalebox{0.6}{MP}}}}
\def\sBM{{\substack{\scalebox{0.6}{BM}}}}
\def\sBCN{{\substack{\scalebox{0.6}{BCN}}}}
\def\sVIN{{\substack{\scalebox{0.6}{Vinet}}}}
\def\sVRH{{\substack{\scalebox{0.6}{VRH}}}}
\def\sVoigt{{\substack{\scalebox{0.6}{Voigt}}}}
\def\sReuss{{\substack{\scalebox{0.6}{Reuss}}}}
\def\sStatic{{\substack{\scalebox{0.6}{Static}}}}
\def\sIsothermal{{\substack{\scalebox{0.6}{Isothermal}}}}
\def\svib{{\substack{\scalebox{0.6}{vib}}}}
\def\sopt{{\substack{\scalebox{0.6}{opt}}}}
\def\sL{{\substack{\scalebox{0.6}{L}}}}
\def\sT{{\substack{\scalebox{0.6}{T}}}}
\def\sB{{\substack{\scalebox{0.6}{B}}}}
\def\sS{{\substack{\scalebox{0.6}{S}}}}
\def\sV{{\substack{\scalebox{0.6}{V}}}}
\def\sGGA{{\substack{\scalebox{0.6}{GGA}}}}
\def\sLDA{{\substack{\scalebox{0.6}{LDA}}}}

\subsection{\AFLOWAELAGL: The Automatic Elasticity and \GIBBS\ Libraries}\label{sec:aelagl}

\begin{figure}
  \includegraphics[width=0.49\textwidth]{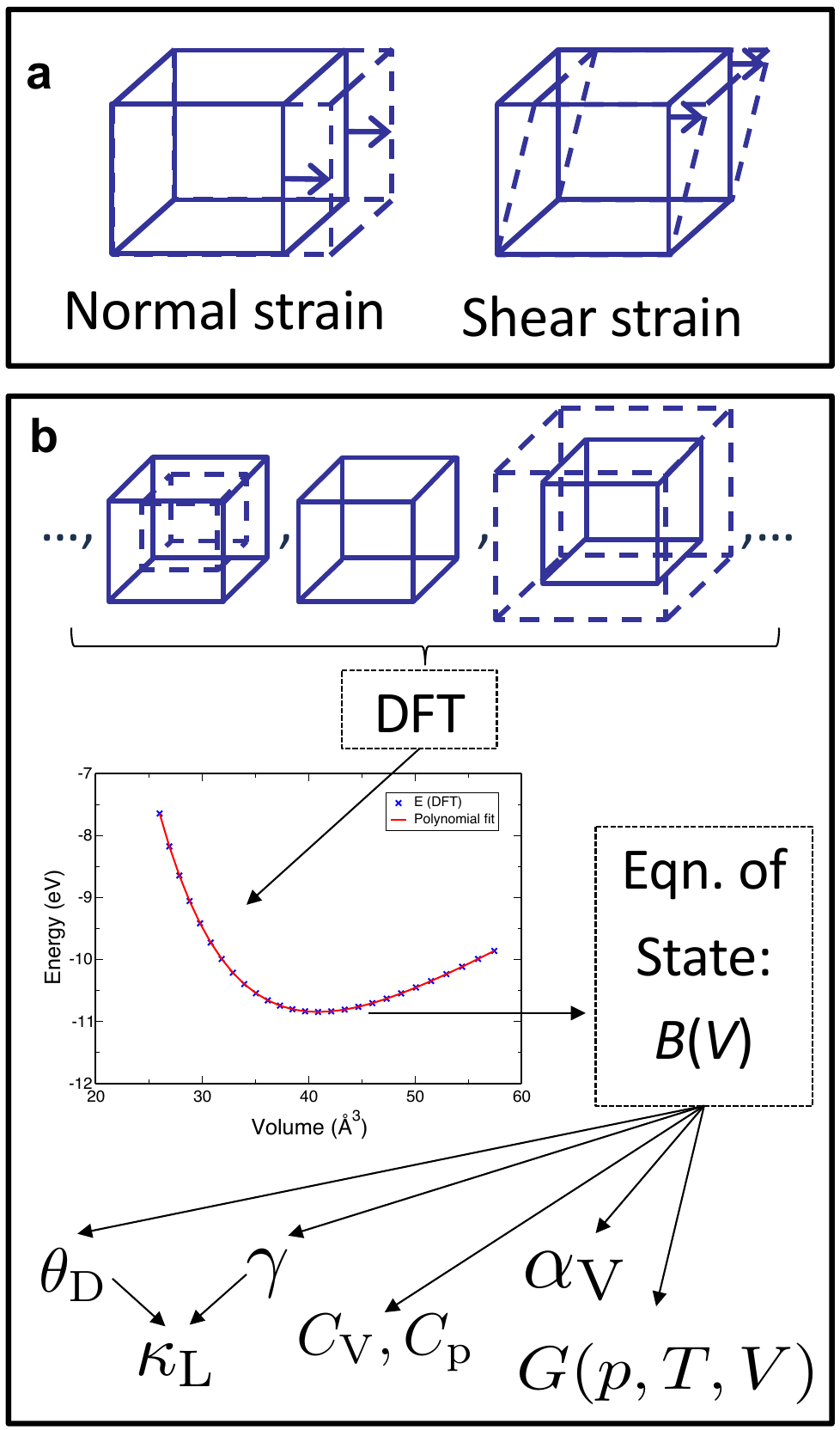}
  \vspace{-4mm}
  \caption{\small {\bf Calculations of thermomechanical properties in \AFLOW.}
   ({\bf a}) \AEL\ uses the stresses from a set of normal and shear strained cells to obtain the elastic constants.
   ({\bf b}) \AGL\ calculates the energies of a set of isotropically compressed and expanded unit cells, and fits the
   resulting $E(V)$ data by a numerical polynomial or by an empirical equation of state to obtain the bulk modulus,
   and hence other thermal and mechanical properties.}
  \label{fig:ael_agl}
\end{figure}

\noindent
\noindent\textbf{\AEL: Elastic constants.}
Thermomechanical properties of materials, such as elastic moduli, Debye temperature, heat capacity and thermal expansion, can be obtained using the \AFLOW\ \underline{A}utomatic \underline{E}lasticity \underline{L}ibrary (\AFLOW-\AEL~\cite{curtarolo:art115}) and the
\AFLOW\ \underline{A}utomatic \underline{G}IBBS \underline{L}ibrary (\AFLOW-\AGL~\cite{curtarolo:art96}) modules based on calculations of strained primitive cells.
These methods are generally computationally less costly than the \APL\ and \AAPL\ phonon calculations.
However, \APL\ and \AAPL\ generally give more quantitatively accurate results, particularly for properties where anharmonic effects are important.
\AEL\ and \AGL\ have been combined into a single automated workflow, which has been
used to calculate the thermomechanical properties of over 6000 materials in the \AFLOWorg\ repositories.

The \AEL\ module  applies a set of independent normal and shear strains to the primitive cell of a material~\cite{curtarolo:art100, curtarolo:art115}
as depicted in \Fig~\ref{fig:ael_agl}\pan{a}, and uses DFT to calculate the resulting stress tensors.
This set of strain-stress data is used to generate the elastic stiffness tensor, \ie\ the elastic constants:
  \[
\left( \begin{array}{l} \sigma_{11} \\ \sigma_{22} \\ \sigma_{33} \\ \sigma_{23} \\ \sigma_{13} \\ \sigma_{12} \end{array} \right) =
\left( \begin{array}{l l l l l l} c_{11}\ c_{12}\ c_{13}\ c_{14}\ c_{15}\ c_{16} \\
 c_{12}\ c_{22}\ c_{23}\ c_{24}\ c_{25}\ c_{26} \\
c_{13}\ c_{23}\ c_{33}\ c_{34}\ c_{35}\ c_{36} \\
c_{14}\ c_{24}\ c_{34}\ c_{44}\ c_{45}\ c_{46} \\
c_{15}\ c_{25}\ c_{35}\ c_{45}\ c_{55}\ c_{56} \\
c_{16}\ c_{26}\ c_{36}\ c_{46}\ c_{56}\ c_{66} \end{array} \right)
\left( \begin{array}{c} \epsilon_{11} \\ \epsilon_{22} \\ \epsilon_{33} \\ 2\epsilon_{23} \\ 2\epsilon_{13} \\ 2\epsilon_{12} \end{array} \right)
\]
written in the $6 \times 6$ Voigt notation using the mapping~\cite{Poirier_Earth_Interior_2000}:
$11 \mapsto 1$, $22 \mapsto 2$, $33 \mapsto 3$, $23 \mapsto 4$, $13 \mapsto 5$, $12 \mapsto 6$. The elastic constants are
combined to calculate the bulk, $B$, and shear, $G$, elastic moduli in the Voigt approximation,
which assumes a uniform strain with the stress supported by the individual grains in parallel,
giving the upper bound on the elastic moduli:
\begin{align*}
  B_{\sVoigt} ={}& \frac{1}{9} \left[ (c_{11} + c_{22} + c_{33}) + 2 (c_{12} + c_{23} + c_{13}) \right], \\
  G_{\sVoigt} ={}& \frac{1}{15} \left[ (c_{11} + c_{22} + c_{33}) -  (c_{12} + c_{23} + c_{13}) \right. + \\ &
  \left. + 3 (c_{44} + c_{55} + c_{66}) \right];
\end{align*}
the Reuss approximation, which assumes a uniform stress so that the strain is the sum of the strains of the individual grains in series,
giving the lower bound on the elastic moduli:
\begin{align*}
  B^{-1}_{\sReuss} ={}&  (s_{11} + s_{22} + s_{33}) + 2 (s_{12} + s_{23} + s_{13}), \\
  G^{-1}_{\sReuss} ={}&  \frac{1}{15} \left[ 4(s_{11} + s_{22} + s_{33}) - 4 (s_{12} + s_{23} + s_{13}) + \right. \\ &
  \left. + 3 (s_{44} + s_{55} + s_{66}) \right];
\end{align*}
and Voigt-Reuss-Hill (VRH, $B_\sVRH$ and $G_\sVRH$~\cite{Hill_elastic_average_1952}) average:
\begin{align*}
  B_{\sVRH} = \left( B_{\sVoigt} + B_{\sReuss} \right) / 2, \\
  G_{\sVRH} = \left( G_{\sVoigt} + G_{\sReuss} \right) / 2.
\end{align*}
The Poisson ratio $\nu$ is given by:
\begin{equation}
  \label{Poissonratio}
  \nu = \frac{3 B_{\sVRH} - 2 G_{\sVRH}}{6 B_{\sVRH} + 2 G_{\sVRH}}.
\end{equation}

\noindent\textbf{Performing \AEL\ calculations.}
The settings for \AEL\ calculations are provided in the \aflowin\ file.
To run \AEL, the line \inlineCommand{[{\color{codegreen} AFLOW\_AEL}]CALC} needs to be present and uncommented --- an appropriate \aflowin\ can be generated
 by including the \inlineCommand{-{}-module=ael} flag to the file generation command (\inlineCommand{-{}-aflow\_proto}).
\AEL\ can reduce the number of required strains by setting \inlineCommand{[{\color{codegreen} AFLOW\_AEL}]STRAIN\_SYMMETRY=ON}.
The number and size of the strains in each independent direction can be controlled using the parameters:
\begin{fileContentTBox}
[{\color{codegreen} AFLOW\_AEL}]NNORMAL\_STRAINS=\textit{value}\tcbbreak
[{\color{codegreen} AFLOW\_AEL}]NSHEAR\_STRAINS=\textit{value}\tcbbreak
[{\color{codegreen} AFLOW\_AEL}]NORMAL\_STRAIN\_STEP=\textit{value}\tcbbreak
[{\color{codegreen} AFLOW\_AEL}]NORMAL\_STRAIN\_STEP=\textit{value}\tcbbreak
\end{fileContentTBox}
\noindent
A full list of parameters is available in the \AEL\ \README, which can be printed using the command
\begin{lstlisting}[language=aflowBash]
aflow --readme=ael
\end{lstlisting}

\noindent\textbf{\AGL: Debye-Gr{\"u}neisen model.}
The \AGL\ module is based on the \GIBBS~\cite{Blanco_CPC_GIBBS_2004, Blanco_jmolstrthch_1996}  quasi-harmonic Debye-Gr{\"u}neisen
method, and calculates the energy as a function of volume, $E(V)$, for a set of isotropically compressed and expanded strains of the
primitive cell, as illustrated in \Fig~\ref{fig:ael_agl}\pan{b}. The $E(V)$ data are fitted by either a numerical polynomial
or an empirical equation of state to obtain the adiabatic bulk modulus $B_\sS(V)$.
The Debye temperature $\theta_\sD(V)$ as a function of volume is then calculated using the expression:
\begin{equation}
  \label{debyetemp}
  \theta_\sD = \frac{\hbar}{k_\sB}[6 \pi^2 V^{1/2} n]^{1/3} f(\nu) \sqrt{\frac{B_\sS}{M}},
\end{equation}
where $n$ is the number of atoms per unit cell, $M$ is the unit cell mass, and $f(\nu)$ is a function of the Poisson ratio $\nu$:
\[
  f(\nu) = \left\{ 3 \left[ 2 \left( \frac{2}{3} \!\cdot\! \frac{1 + \nu}{1 - 2 \nu} \right)^{3/2} \!\!\!\!\!\!\!+ \left( \frac{1}{3} \!\cdot\! \frac{1 + \nu}{1 - \nu} \right)^{3/2} \right]^{-1} \right\}^{\frac{1}{3}}\!\!\!\!,
\]
where $\nu$ can be obtained from \Eq~\ref{Poissonratio} using \AEL, or set directly by the user (\inlineCommand{[{\color{codegreen} AFLOW\_AGL}]POISSON\_RATIO=\textit{value}}). The vibrational contribution to the free energy, $F_\svib$, is given by:
\[
  F_\svib(\theta_\sDebye; T) \!=\! n k_\sB T \!\left[ \frac{9}{8} \frac{\theta_\sDebye}{T} \!+\! 3\ \mathrm{log}\!\left(1 \!-\! {\mathrm e}^{- \theta_\sDebye / T}\!\right) \!\!-\!\! D\left(\frac{\theta_\sDebye}{T}\right)\!\!\right],
\]
where $D(\theta_\sDebye / T)$ is the Debye integral:
  \[
  D \left(\theta_\sDebye/T \right) = 3 \left( \frac{T}{\theta_\sDebye} \right)^3 \int_0^{\theta_\sDebye/T} \frac{x^3}{e^x - 1} dx.
\]
The Gibbs free energy is obtained from:
  \[
  {\sf G}(V; p, T) = E_\sDFT(V) + F_\svib (\theta_\sD(V); T)  + pV.
\]
The volume which minimizes $ {\sf G}(V; p, T)$ at a given pressure $p$ and temperature $T$ is the equilibrium volume $V_{\mathrm{eq}}$, which
is used to evaluate $\theta_\sD (V_{\mathrm{eq}})$ and the Gr{\"u}neisen parameter $\gamma$, as defined by:
  \[
  \gamma = - \frac{\partial \ \mathrm{log} (\theta_\sD(V))}{\partial \ \mathrm{log} V}.
\]
Finally, $\theta_\sD$ and $\gamma$ are used to calculate other thermal properties~\cite{curtarolo:art96, Blanco_CPC_GIBBS_2004} including specific heat capacity at constant volume $C_\sV$:
  \[
C_\sV = 3 n k_\sB \left[4 D\left(\frac{\theta_\sD}{T}\right) - \frac{3 \theta_\sD / T}{\exp(\theta_\sD / T) - 1} \right];
\]
volumetric thermal expansion $\alpha_\sV$:
  \[
\alpha_\sV = \frac{\gamma C_\sV}{B_\sT V},
\]
where $B_\sT$ is the isothermal bulk modulus; specific heat capacity at constant pressure $C_{\mathrm p}$:
  \[
C_{\mathrm p} = C_\sV (1 + \gamma \alpha_\sV T);
\]
and $\kappa_{\mathrm L}$~\cite{Leibfried_formula_1954, slack, Morelli_Slack_2006}:
\begin{align*}
  \kappa_{\mathrm L} (\theta_\acoustic) ={}& \frac{0.849 \times 3 \sqrt[3]{4}}{20 \pi^3(1 - 0.514\gamma_\acoustic^{-1} + 0.228\gamma_\acoustic^{-2})}
                                         \times \\ & \times \left( \frac{k_\sB \theta_\acoustic}{\hbar} \right)^2 \frac{k_\sB m V^{\frac{1}{3}}}{\hbar \gamma_\acoustic^2},
\end{align*}
where $\theta_\acoustic$ and $\gamma_\acoustic$ is the Debye temperature and Gr{\"u}neisen parameter calculated from only the acoustic modes.

\noindent\textbf{Performing \AGL\ calculations.}
The settings for \AGL\ calculations are provided in the \aflowin\ file.
To run \AGL, the line \inlineCommand{[{\color{codegreen} AFLOW\_AGL}]CALC} needs to be present and uncommented --- an appropriate \aflowin\ can be generated
 by including the \inlineCommand{-{}-module=agl} flag to the file generation command (\inlineCommand{-{}-aflow\_proto}).
\AGL\ can run \AEL\ to obtain the Poisson ratio for use in \Eq~\ref{debyetemp}  to calculate the Debye temperature by setting \inlineCommand{[{\color{codegreen} AFLOW\_AGL}]AEL\_POISSON\_RATIO=ON}.
The number of different volumes and the size of the strain steps can be controlled using the parameters:
\begin{fileContentTBox}
[{\color{codegreen} AFLOW\_AGL}]NSTRUCTURES=\textit{value}\tcbbreak
[{\color{codegreen} AFLOW\_AGL}]STRAIN\_STEP=\textit{value}\tcbbreak
\end{fileContentTBox}
\noindent
The number of temperature and pressure points and the corresponding step sizes for the post-processing can be controlled using the parameters:
\begin{fileContentTBox}
[{\color{codegreen} AFLOW\_AGL}]NTEMP=\textit{value}\tcbbreak
[{\color{codegreen} AFLOW\_AGL}]STEMP=\textit{value}\tcbbreak
[{\color{codegreen} AFLOW\_AGL}]NPRESSURE=\textit{value}\tcbbreak
[{\color{codegreen} AFLOW\_AGL}]SPRESSURE=\textit{value}\tcbbreak
\end{fileContentTBox}
\noindent
Note that the post-processing can be run multiple times with different temperature and pressure ranges for the same set of \DFT\ calculations.
A full list of parameters is available in the \AGL\ \README, which can be printed using the command:
\begin{lstlisting}[language=aflowBash]
aflow --readme=agl
\end{lstlisting}

\subsection{\AFLOWAPL: The Automatic Phonon Library}\label{sec:apl}

\noindent\textbf{Phonons in the harmonic approximation.}
Phonons are the basis for many finite-temperature processes in solids. They can contribute to the
stabilization of a material, determine thermophysical properties such as heat capacities and thermal
expansion, and are responsible for transport phenomena such as thermal conductivity.
The \AFLOW\ \underline{A}utomatic \underline{P}honon \underline{L}ibrary~(\APL) calculates phonon modes in the harmonic approximation
by diagonalizing the dynamical matrix $D\left(\vec{q}\right)$~\cite{curtarolo:art180}:
  \[
  D\left(\vec{q}\right)\vec{e}_\lambda = \omega^2_\lambda\vec{e}_\lambda,
\]
where the phonon mode $\lambda = \left\{\vec{q}, j\right\}$ is a combined index consisting of the
reciprocal space point $\vec{q}$ and the branch index $j$. $\omega_\lambda$ and $\vec{e}_\lambda$
are the frequency and the eigenvector of the mode, respectively. The components of the dynamical
matrix are~\cite{Maradudin1968}:
\begin{equation}
\begin{split}
D_{\alpha\beta}\left(\kappa\kappap|\vec{q}\right) = {}&
    \frac{1}{\sqrt{m_\kappa m_\kappap}}
    \sum_{\lp} \IFCharm \times \\ & \times \exp\left[i\vec{q}\cdot\left(\vec{R}_\lp - \vec{R}_l\right)\right],
  \label{eq:apl_dynmat}
\end{split}
\end{equation}
with the Cartesian indices $\alpha$ and $\beta$, the atomic indices $\kappa$ and $\kappap$, and the
supercell indices $l$ and $\lp$. $m_\kappa$ is the mass of atom $\kappa$ and $\vec{R}_l$ is the
vector connecting the origin of the crystal to the origin of supercell $l$. $\IFCharm$ are the
harmonic \underline{i}nteratomic \underline{f}orce \underline{c}onstants~(\IFCs), the calculation of which is the central problem for phonon calculations.
\APL\ can obtain them either from $\Gamma$-point density functional perturbation theory or through
the finite displacement method.

The finite displacement method applies small distortions to the atomic positions inside a supercell
and calculates the forces using \DFT. For a full set of \IFCs, each symmetrically-inequivalent atom
needs to be displaced along three linearly-independent directions. To minimize the number of
calculations, \APL\ uses the following algorithm:

\begin{enumerate}[label={\textbf{\roman*.}},noitemsep]
\item Create test displacements along the unit cell axes, face diagonals, and
  body diagonal.
\item Generate  a set of displacement vectors that are orthogonal and symmetrically equivalent to
  the test displacement using the site point group of the atom~\cite{curtarolo:art135} and
  Gram-Schmidt orthogonalization.
\item Sort these sets of displacements by the number of equivalent vectors from highest to lowest.
\item Take the displacements inside the first set of this sorted list. If there are less than three,
  add the displacements in the next set of the list and use the Gram-Schmidt method and the site
  point groups to create orthogonal vectors. Repeat until three linearly-independent directions are
  found.
\end{enumerate}
This algorithm not only reduces computational requirements by minimizing the number of calculations,
using site point groups to generate the displacements also leads to supercells with the highest
possible symmetry. The calculated forces are then used to determine the \IFCs\ through finite
differences. These ``raw'' force constants do not generally fulfill the acoustic sum rule
$\sum_{\kappap} \IFCharm = 0$ nor are they commensurate with the site point groups of the atoms in
the crystal. \APL\ automatically enforces these properties when calculating the \IFCs. The dynamical
matrix can then be constructed to solve \Eq~\ref{eq:apl_dynmat}.

The resulting frequencies are used to obtain phonon dispersions and \DOS.
The \DOS\ $g(\omega)$ can be used to calculate the vibrational free energy $\Fvib$, internal energy
$\Uvib$, and entropy $\Svib$, as well as the isochoric heat capacity $C_{\mathrm V}$:
\begin{align*}
  \Fvib(T) &= {\kB}T\int_0^\infty\log\left(2\sinh\frac{\hbar\omega}{2{\kB}T}\right)g(\omega)d\omega,\\
  \Uvib(T) &= \int_0^\infty\frac{\hbar\omega}{2}\coth\left(\frac{\hbar\omega}{2{\kB}T}\right)g(\omega)d\omega,\\
  \Svib(T) &= \frac{\Uvib - \Fvib}{T},\\
  C_\textnormal{V}(T)   &= \kB\int_0^\infty\left(\frac{\hbar\omega}{2\kB T}\right)\text{csch}^2\frac{\hbar\omega}{2\kB T}g(\omega)d\omega,\\
\end{align*}
where $\kB$ and $\hbar$ are the Boltzmann and the reduced Planck constant, respectively.
Additionally, \APL\ can calculate group velocities $\vec{v}_\lambda$ and mean square atomic
displacements $\left<\left|u\right|^2\right>$ using the eigenvectors:
\begin{align*}
  \vec{v}_\lambda &= \left<\vec{e}_\lambda\left|\frac{\partial D(\vec{q})}{\partial\vec{q}}\right|\vec{e}_\lambda\right>,\\
  \left<\left|u^\alpha(\kappa, T)\right|^2\right> &= \frac{\hbar}{N_\vec{q} m_j}\sum_\lambda \omega_\lambda^{-1}\left(\frac{1}{2} + n_\lambda\right) \left|\vec{e}_\lambda^\alpha(\kappa)\right|^2,\\
\end{align*}
with $N_\vec{q}$ and $n_\lambda$ being the number of \vec{q}-points and the phonon numbers based on
the Bose-Einstein distribution, respectively.

Long-range Coulomb interactions in polar materials cause splitting between the \underline{l}ongitudinal and
\underline{t}ransversal \underline{o}ptical phonon modes (\LOTO\ splitting). This requires a corrective term $\tilde{D}$ to be added to the
dynamical matrix. \AFLOW\ uses the method by Wang \etal\ to calculate this correction~\cite{Wang_npjcm_2016}:
\begin{align*}
    \tilde{D}^{\kappa\kappap}_{\alpha\beta} = {}& \frac{4\pi e}{V\sqrt{m_\kappa m_\kappap}}
    \frac{\left[\vec{q} Z^*(i)\right]_\alpha\left[\vec{q} Z^*(i)\right]_\beta}{\vec{q}\varepsilon_\infty\vec{q}}
     \times \\ & \times \sum_{\lp} \exp\left[i\vec{q}\cdot\left(\vec{R}_\lp - \vec{R}_l\right)\right],
\end{align*}
where $e$, $Z^*$, and $\varepsilon_\infty$ are the elemental charge, the Born effective charge
tensor, and the dielectric tensor, respectively. The tensors can be directly calculated by \VASP.

\begin{figure*}
  \includegraphics[width=\textwidth]{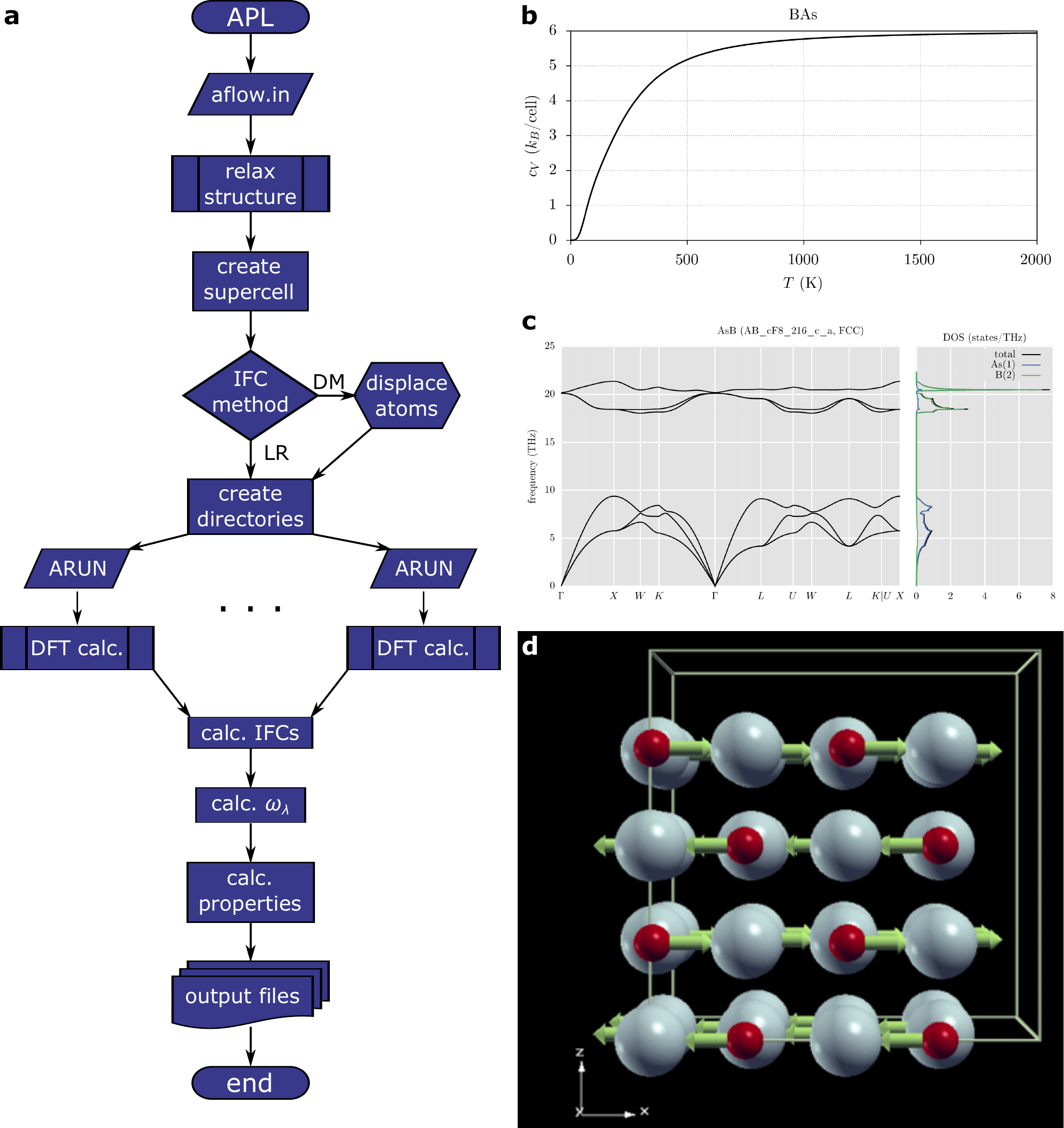}
  \caption{\textbf{Workflow and outputs of the \underline{A}utomatic \underline{P}honon \underline{L}ibrary~(\APL)}
  (\textbf{a})~Flowchart of the \APL\ workflow.
  (\textbf{b})~Isochoric heat capacity $C_V$ of zincblende BAs.
  (\textbf{c})~Combined phonon dispersion and projected \DOS\ plot for BAs.
  (\textbf{d})~Snapshot of a phonon mode visualization at the $X$ point for rocksalt MgO.
  }
  \label{fig:apl}
\end{figure*}

\noindent\textbf{Performing \APL\ calculations.}
The settings for phonon calculations are driven by the \aflowin\ file.
A full list of parameters is available in the \APL\ \README, which can be created
using the command:
\begin{lstlisting}[language=aflowBash]
aflow --readme=apl
\end{lstlisting}
To run \APL, the line \inlineCommand{[{\color{codegreen} AFLOW\_APL}]CALC} needs to be present. This can be achieved
programmatically by adding \inlineCommand{-{}-module=apl} to the input file generation command (\inlineCommand{-{}-aflow\_proto}).
The flowchart of the \APL\ workflow is shown in \Fig~\ref{fig:apl}\pan{a}.

Since phonon properties require accurate forces, the structure needs to be relaxed so that the
forces are near zero. This is done automatically by \APL\ at the beginning of the workflow. The
following parameters guide the relaxation calculations:
\begin{fileContentTBox}
[{\color{codegreen} AFLOW\_APL}]RELAX=ON\tcbbreak
[{\color{codegreen} AFLOW\_APL}]RELAX\_COMMENSURATE=ON\tcbbreak
[{\color{codegreen} AFLOW\_APL}]KPPRA=$N_k$\tcbbreak
[{\color{codegreen} AFLOW\_APL}]KPOINTS\_GRID=$k_1$x$k_2$x$k_3$\tcbbreak
\end{fileContentTBox}
\noindent
    \inlineCommand{[{\color{codegreen} AFLOW\_APL}]RELAX\_COMMENSURATE=ON} ensures that the {\bf k}-point grids of the supercell for the force
calculations and the grid of the relaxation calculations are commensurate, which increases the
accuracy of the obtained forces.
It can be toggled off with \inlineCommand{[{\color{codegreen} AFLOW\_APL}]RELAX\_COMMENSURATE=OFF}.
To determine the grid size, \inlineCommand{[{\color{codegreen} AFLOW\_APL}]KPPRA} or
\inlineCommand{[{\color{codegreen} AFLOW\_APL}]KPOINTS\_GRID}
are used, which represent the {\bf k}-points per reciprocal atom $N_k$ or the {\bf k}-point grid dimensions,
respectively, for the supercell.  Convergence criteria are set such that the forces between two
successive ionic steps are below $10^{-3}\,\text{eV/{\AA}}$ and energy differences are below
$10^{-8}\,\text{eV}$ between electronic steps.
The relaxation can be skipped entirely by setting \inlineCommand{[{\color{codegreen} AFLOW\_APL}]RELAX=OFF}.

After the relaxation, \APL\ builds the supercell using either explicit dimensions, a minimum
number of atoms, or a number of coordination shells around each atom:
\begin{lstlisting}[language=fileContentAFLOWIN]
[AFLOW_APL]SUPERCELL=3x3x3
[AFLOW_APL]MINATOMS=175
[AFLOW_APL]MINSHELL=10
\end{lstlisting}
\texttt{SUPERCELL} takes priority, followed by \texttt{MINATOMS}.
\APL\ then creates subdirectories for the force calculations using \DFT.
This process is governed by the following \aflowin\ parameters:
\begin{fileContentTBox}
[{\color{codegreen} AFLOW\_APL}]ENGINE=\textit{method}\tcbbreak
[{\color{codegreen} AFLOW\_APL}]DPM=ON\tcbbreak
[{\color{codegreen} AFLOW\_APL}]DMAG=$r_\text{displ}$\tcbbreak
[{\color{codegreen} AFLOW\_APL}]ZEROSTATE=OFF\tcbbreak
[{\color{codegreen} AFLOW\_APL}]POLAR=ON\tcbbreak
\end{fileContentTBox}
\noindent
\inlineCommand{[{\color{codegreen} AFLOW\_APL}]ENGINE} determines the method to calculate the force constants.
It can be \inlineCommand{LR} or \inlineCommand{DM}, corresponding to the linear response~($\Gamma$-point density functional perturbation theory)
and the direct method~(finite displacement), respectively. The former generates one subdirectory as
the force constants are directly calculated by \VASP.

The finite displacement method provides two ways to determine the \IFCs: the central and
the forward difference methods. \inlineCommand{[{\color{codegreen} AFLOW\_APL}]DPM=ON} always uses the central difference method,
\inlineCommand{[{\color{codegreen} AFLOW\_APL}]DPM=OFF} always uses forward differences, and
\inlineCommand{[{\color{codegreen} AFLOW\_APL}]DPM=AUTO} determines for each site
whether the forward difference is permitted by symmetry. The size of the displacement
$r_\text{displ}$~(in {\AA}ngstr\"{o}m) is set by \inlineCommand{DMAG}.
Additionally, \inlineCommand{[{\color{codegreen} AFLOW\_APL}]ZEROSTATE} can be used to calculate the forces on the atom in the unperturbed
supercell, which can be used to subtract noise for the forward difference method or to test whether
virtual forces are present in the supercells.
For both methods, \inlineCommand{[{\color{codegreen} AFLOW\_APL}]POLAR=ON} applies the non-analytical term correction, resulting in an
additional subdirectory to calculate  $Z^*$ and $\varepsilon_0$.

Each subdirectory contains an \aflowin\ file for a static \DFT\ calculation and needs to be run separately.
These calculations have a convergence criterion of $10^{-8}\,\text{eV}$ between
electronic steps and use the {\bf k}-point parameters described earlier.
Once finished, \AFLOW\ needs to be run in the parent directory again to
read the forces, determine the \IFCs, and calculate phonon dispersions, \DOS, and
thermophysical properties. Important settings for this post-processing step include:
\begin{fileContentTBox}
[{\color{codegreen} AFLOW\_APL}]DOS\_PROJECT=OFF\tcbbreak
[{\color{codegreen} AFLOW\_APL}]TPT=$T_\text{start}$:$T_\text{end}$:$T_\text{step}$\tcbbreak
\end{fileContentTBox}
\noindent
where \inlineCommand{[{\color{codegreen} AFLOW\_APL}]DOS\_PROJECT} determines whether atom-projected \DOS\ are calculated
and \inlineCommand{[{\color{codegreen} AFLOW\_APL}]TPT}
sets the start and end temperatures and the temperature step size for thermophysical properties
and atomic displacements.

\noindent\textbf{Visualization options.}
\APL\ provides several output options for its calculations.  Thermophysical properties can be
plotted and saved as an image by calling:
\begin{aflowBashTBox}
{\color{binarycolor} aflow} -{}-plotthermo
\end{aflowBashTBox}
\noindent
where \inlineCommand{-{}-plotthermo} accepts multiple optional inputs:
the \inlineCommand{\textit{directory}} of the phonon calculation~(default: current directory),
and the minimum $\left(T_\text{min}\right)$ and maximum $\left(T_\text{max}\right)$ temperatures~(default: full range of calculated temperatures);
\ie\ \inlineCommand{-{}-plotthermo=\textit{directory},$T_\text{min}$,$T_\text{max}$}.
An example plot for the isochoric
heat capacity is presented in \Fig~\ref{fig:apl}\pan{b}.
Moreover, the following additional options may be specified:
\begin{myitemize}
\item \inlineCommand{-{}-title=\textit{title}} : Title of the plot. If not used, \AFLOW\ will generate a
  generic one.
\item \inlineCommand{-{}-print=pdf} : Output format of the image. Options include:
  \inlineCommand{eps}, \inlineCommand{gif}, \inlineCommand{jpg}, \inlineCommand{png}.
\item \inlineCommand{-{}-outfile=\textit{outfile}} : Name of the output image. If unused, \AFLOW\ will
  generate a generic name.
\end{myitemize}
Phonon dispersion and phonon \DOS\ can be plotted in the same manner using
\inlineCommand{{\color{binarycolor} aflow} -{}-plotphdisp} and \inlineCommand{{\color{binarycolor} aflow} -{}-plotphdos},
respectively, or as a combined
plot with \inlineCommand{{\color{binarycolor} aflow} -{}-plotphdispdos}~(see \Fig~\ref{fig:apl}\pan{c}). For these plots,
$T_\text{min}$ and $T_\text{max}$ are replaced by the minimum and maximum frequency,
respectively. Additionally, the following options can be specified:
\begin{myitemize}
  \item \inlineCommand{-{}-unit=THz} : Unit of the frequencies. Options include:
    \inlineCommand{Hz}, \inlineCommand{eV}, \inlineCommand{meV}, \inlineCommand{rcm}, and \inlineCommand{cm-1}.
  \item \inlineCommand{-{}-projection=atoms} : Plot atom-projected \DOS.
\end{myitemize}
Individual phonon modes can be exported into animation files that can be read by the XCrysDen
software and converted into videos or \GIF{}s~\cite{xcrysden}. This can be done with the
\inlineCommand{{\color{binarycolor} aflow} -{}-visualize\_phonons} command. It has the following mandatory options:
\begin{myitemize}
  \item \inlineCommand{-{}-q=\textit{qpoints}} : the \vec{q}-points as comma-separated triplets in
    fractional coordinates, \eg\ \inlineCommand{0.5,0.5,0.5}. Multiple \vec{q}-points can be specified.
  \item \inlineCommand{-D \textit{directory}} : the directory of the phonon calculation.
\end{myitemize}
The following are optional:
\begin{myitemize}
  \item \inlineCommand{-{}-amplitude=\textit{amplitudes}} : Amplitude of the displacements.
  \item \inlineCommand{-{}-branches=\textit{branches}} : Branch numbers of the phonon mode as a
    comma-separated list.
  \item \inlineCommand{-{}-periods=\textit{periods}} : Number of oscillation periods displayed in the
    output file.
  \item \inlineCommand{-{}-steps=\textit{steps}} : Number of steps per period.
  \item \inlineCommand{-{}-scell=$s_1$x$s_2$x$s_3$} : Supercell dimensions of the structure in the animation.
\end{myitemize}
A snapshot of such a visualization \GIF\ can be found in \Fig~\ref{fig:apl}\pan{d}.

\subsection{\AFLOWQHA: The Quasi-Harmonic Approximation Library}\label{sec:qha}
\noindent\textbf{Thermomechanical properties from phonons.}
Phonon frequencies strongly depend on the volume of the structure, as shown in
\Fig~\ref{fig:qha_aapl}\pan{a}. In the harmonic approximation, they are calculated at equilibrium volume
$\Veq$, \ie\ the volume at 0\,K. At finite temperatures, the volume of the structure changes,
which also changes the phonon frequencies and thus $\Fvib$.
Knowledge about the volume-temperature curve is thus essential for a more accurate calculation of a material's free energy and thermomechanical properties.

The free energy can be calculated as:
  \[
  F (V, T) = E_0(V) + \Felec(V, T) + \Fvib(V, T),
\]
where $E_0$ is the potential energy and $\Felec$ the electronic free energy:
\begin{align*}
    \Felec(V, T) ={}& \int_0^\infty \gel\left[\fFD - \kB T s_T(\epsilon)\right]d\epsilon - \\ & - \int_0^{\EF}\gel d\epsilon.
\end{align*}
Here, $\gel$ is the electronic \DOS, $\fFD = \fFD\left(\epsilon, T\right)$ is the
Fermi-Dirac distribution, $\EF$ is the Fermi energy, and
$s_{\mathrm T}(\epsilon) = \fFD\log(\fFD) + (1 - \fFD)\log(1 - \fFD)$ is the electronic entropy of $\epsilon$.

In the \underline{q}uasi-\underline{h}armonic \underline{a}pproximation~(\QHA)~\cite{curtarolo:art114,curtarolo:art146},
$\VeqT$ is obtained by calculating $F(T)$ over a set of volumes $V_i$ and fitting the volume-dependent
free energy to an \underline{e}quation \underline{o}f \underline{s}tate~(\EOS) for each temperature. \AFLOWQHA\ provides the Murnaghan
\EOS~\cite{Murnaghan_PNAS_1944}:
  \[
  F(V) = F_\text{eq} + \frac{B \Veq}{B^\prime\left(B^\prime - 1\right)}
  \left(\frac{V}{\Veq}\right)^{1 - B^\prime} + \frac{BV}{B^\prime} - \frac{B\Veq}{B^\prime - 1},
\]
where $B$ is the bulk modulus and $B^\prime$ its pressure derivative; the $n^\text{th}$ order
Birch-Murnaghan \EOS~($2 \le n \le 4$)~\cite{Birch_PR_1947}:
  \[
  F(V) = \sum_{i = 0}^n f_i V^{-\frac{2}{3}i},
\]
where $f_i$ are parameters obtained from a polynomial fit; and the stabilized jellium
\EOS~\cite{Teter_PRB_1995,Alchagirov_PRB_2001}:
  \[
  F(V) = \sum_{i = 0}^3 f_i V^{-\frac{1}{3}i}.
\]
The fitted \EOS\ can then be used to calculate a variety of thermomechanical properties, such as
the volumetric thermal expansion coefficient $\beta$,
the bulk modulus $B$,
the isochoric and isobaric heat capacities $C_{\mathrm V}$ and $C_{\mathrm P}$,
and the average Gr\"{u}neisen parameter $\gamma$:
\begin{align*}
  \beta(T) &= \frac{d \log\Veq(T)}{dT} = \frac{1}{\Veq(T)}\frac{d\Veq(T)}{dT},\\
  B(T) &= V(T)\frac{\partial^2 F(V,T)}{\partial V^2},\\
  C_{\mathrm V}(T) &= -T \frac{\partial^2 F(V,T)}{\partial T^2},\\
  C_{\mathrm P}(T) &= C_{\mathrm V}(T) + \Veq(T) B(T) \beta^2(T) T,\\
  \gamma(T) &= \beta(T) B(T) \Veq(T) C_{\mathrm V}^{-1}(T).
\end{align*}
The Gr\"{u}neisen parameter can also be obtained as a mode-dependent quantity:
\begin{equation}
  \gamma_\lambda(T) = -\frac{V(T)}{\omega_\lambda(T)} \frac{\partial\omega_\lambda(T)}{\partial V}.
  \label{eq:qha_grueneisen_mode}
\end{equation}

\noindent\textbf{Performing \QHA\ calculations.}
\QHA\ calculations are activated by adding \inlineCommand{[{\color{codegreen} AFLOW\_QHA}]CALC} into the \aflowin\ file or by
calling the input file generation command (\inlineCommand{-{}-aflow\_proto}) with the \inlineCommand{-{}-module=qha} option.
A full list of parameters is available in the \APL\ \README, which can be created using the command:
\begin{lstlisting}[language=aflowBash]
aflow --readme=apl
\end{lstlisting}
The \AFLOWQHA\ workflow is similar to \AFLOWAPL\ in that it consists of a relaxation, a subdirectory
creation, and a post-processing step. The subdirectories, however, are different from \APL\ and
depend on the properties that are calculated. Calculating thermomechanical properties by fitting
$F(V)$ to an \EOS\ is set by \inlineCommand{[{\color{codegreen} AFLOW\_QHA}]EOS} in the \aflowin\ file:
\begin{fileContentTBox}
[{\color{codegreen} AFLOW\_QHA}]EOS=ON\tcbbreak
[{\color{codegreen} AFLOW\_QHA}]EOS\_DISTORTION\_RANGE=$V_\text{start}$:$V_\text{end}$:$V_\text{step}$\tcbbreak
[{\color{codegreen} AFLOW\_QHA}]EOS\_MODEL=\textit{EOS}\tcbbreak
[{\color{codegreen} AFLOW\_QHA}]INCLUDE\_ELEC\_CONTRIB=ON\tcbbreak
[{\color{codegreen} AFLOW\_QHA}]PDIS\_T=\textit{temperatures}\tcbbreak
\end{fileContentTBox}
\noindent
This requires phonon and static calculations at multiple volumes, which are determined by
\inlineCommand{[{\color{codegreen} AFLOW\_QHA}]EOS\_DISTORTION\_RANGE}.
\inlineCommand{[{\color{codegreen} AFLOW\_QHA}]EOS\_MODEL} selects the \EOS\ that is used to fit $F(V)$
using the \underline{M}urnaghan~(\texttt{M}), $n^\text{th}$ order \underline{B}irch-\underline{M}urnaghan~(\texttt{BM\textit{n}}),
or the \underline{s}tabilized \underline{j}ellium~(\texttt{SJ}) model.
\inlineCommand{[{\color{codegreen} AFLOW\_QHA}]INCLUDE\_ELEC\_CONTRIB} includes or
excludes $\Felec$ into the free energy. For this procedure, two subdirectories are created for each
volume, one for a static calculation to obtain $E_0$ and $\Felec$ and another one for a phonon calculation
to determine $\Fvib$. The latter is an \AFLOWAPL\ run without any prior structure relaxation. After
finishing all subdirectories, post-processing can be started by running \binary{aflow} inside the parent
directory again. \inlineCommand{[{\color{codegreen} AFLOW\_QHA}]PDIS} is a comma-separated list of temperatures for which
phonon dispersions are calculated. The temperature range for thermomechanical properties is
set via \inlineCommand{[{\color{codegreen} AFLOW\_APL}]TPT}.

It is also possible to only calculate the Gr\"{u}neisen dispersion using the following options:
\begin{fileContentTBox}
[{\color{codegreen} AFLOW\_QHA}]GP\_FINITE\_DIFF=ON\tcbbreak
[{\color{codegreen} AFLOW\_QHA}]GP\_DISTORTION=$\Delta V$\tcbbreak
\end{fileContentTBox}
\noindent
This results in three \APL\ subdirectories at $\Veq$ and $\Delta V$\% compression and expansion.

\begin{figure*}
  \includegraphics[width=\textwidth]{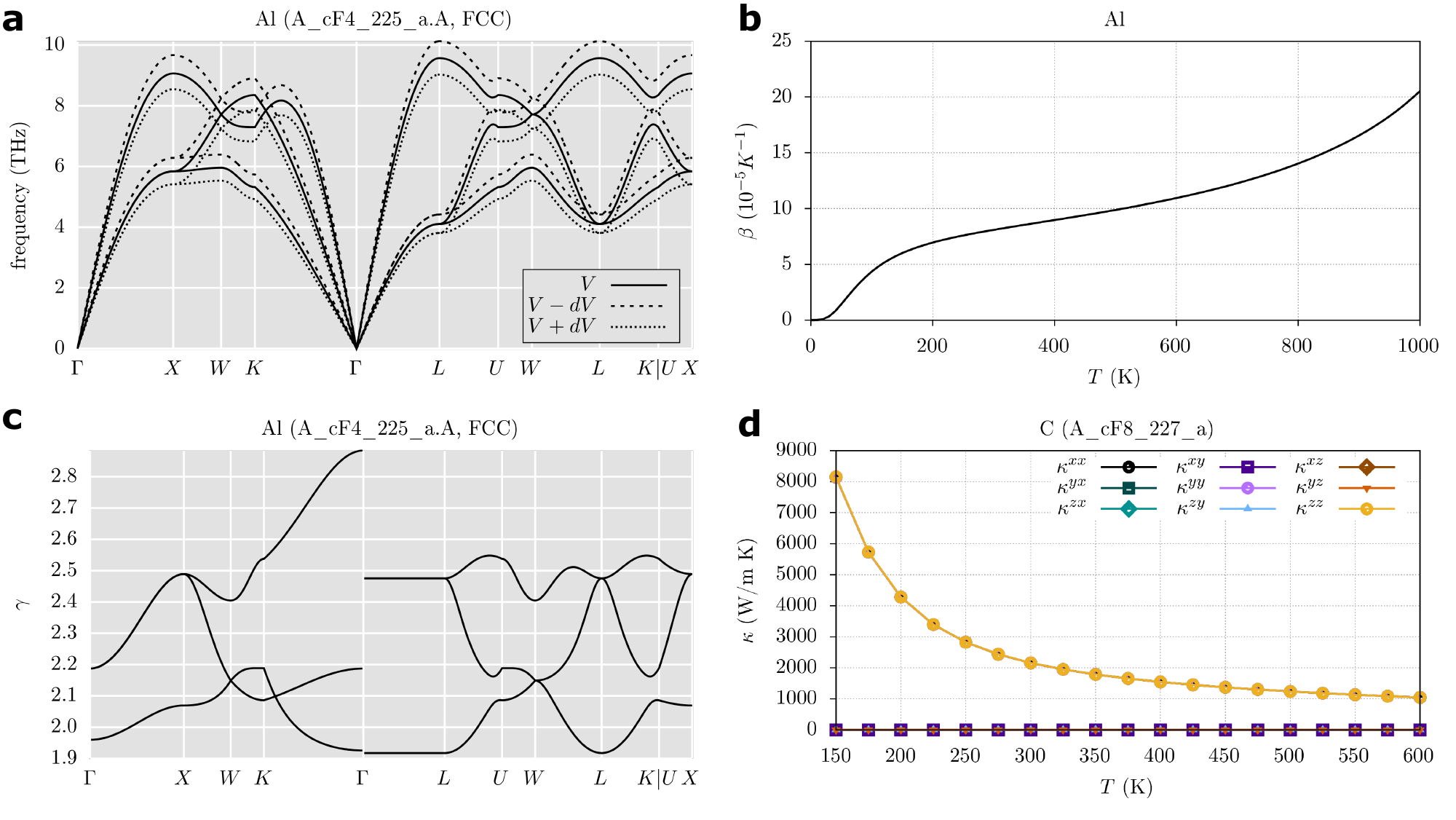}
  \caption{\textbf{Properties calculated by \AFLOWQHA\ and \AFLOWAAPL.}
  (\textbf{a})~Volume-dependent phonon dispersion,
  (\textbf{b})~volumetric thermal expansion coefficient $\beta$, and
  (\textbf{c})~dispersion of Gr\"{u}neisen parameters of Al.
  (\textbf{d})~Thermal conductivity tensor components of diamond.}
  \label{fig:qha_aapl}
\end{figure*}

\noindent\textbf{Visualization options.}
\AFLOWQHA\ provides several output options for its calculated properties. Thermophysical properties
can be plotted analogously to \inlineCommand{-{}-plotthermo} for \APL:
\begin{aflowBashTBox}
{\color{binarycolor} aflow} -{}-plotthermoqha
\end{aflowBashTBox}
\noindent
where \inlineCommand{-{}-plotthermoqha} accepts multiple optional inputs:
the \inlineCommand{\textit{directory}} of the \QHA\ calculation~(default: current directory),
and the minimum $\left(T_\text{min}\right)$ and maximum $\left(T_\text{max}\right)$ temperatures~(default: full range of calculated temperatures);
\ie\ \inlineCommand{-{}-plotthermoqha=\textit{directory},$T_\text{min}$,$T_\text{max}$}.
An example plot for the
volumetric thermal expansion coefficient of aluminum is shown in \Fig~\ref{fig:qha_aapl}\pan{b}.
The \EOS\ model can be specified using
\inlineCommand{-{}-eosmodel=M} which also accepts
\inlineCommand{BM2}, \inlineCommand{BM3}, \inlineCommand{BM4}, and \inlineCommand{SJ}.

The Gr\"{u}neisen parameter dispersion can be plotted the same way using \inlineCommand{-{}-plotgrdisp}.
For this plot, $T_\text{min}$ and $T_\text{max}$ are replaced by the minimum and maximum
Gr\"{u}neisen parameter. An example plot can be found in \Fig~\ref{fig:qha_aapl}\pan{c}.

\subsection{\AFLOWAAPL: The Automatic Anharmonic Phonon Library}\label{sec:aapl}
\noindent\textbf{Lattice thermal conductivity.}
Both \AFLOWAPL\ and \AFLOWQHA\ use harmonic phonons to describe the vibrations inside a material,
which neglects phonon-phonon scattering. The \AFLOW\ \underline{A}utomatic \underline{A}nharmonic \underline{P}honon \underline{L}ibrary~(\AAPL)
includes these interactions explicitly for three-phonon processes, allowing it to calculate the
lattice thermal conductivity of a material~\cite{curtarolo:art125}. The thermal conductivity tensor
$k$ can be calculated as:
  \[
  k^{\alpha\beta} = \frac{\hbar^2}{\kB T^2}\sum_\lambda \omega_\lambda n_\lambda\left(n_\lambda + 1\right) v_{\lambda\alpha}F_{\lambda\beta}.
\]
$\vec{F}_{\lambda}$ is the mean-free displacement of the mode and can be expanded as a first-order
perturbation:
\[
  \vec{F}_\lambda = \taul{0} \left(\vec{v}_\lambda + \vec{$\Delta$}_\lambda\right),
\]
where $\vec{$\Delta$}_\lambda$ is the perturbation and $\taul{0}$ is the relaxation time, or
inverse scattering rate, of the phonon mode. For lattice thermal conductivity, the scattering rate
consists of three components:
  \[
  \invtaul{0} = \invtaul{\text{ph-ph}} + \invtaul{\text{iso}} + \invtaul{\text{grain}}.
\]
$\invtaul{\text{grain}}$ is the grain boundary scattering rate:
  \[
  \invtaul{\text{grain}} = \frac{\left|\vec{v}_\lambda\right|}{L},
\]
with $L$ being the size of the grain. $\invtaul{\text{iso}}$ is the isotope scattering rate:
  \[
  \invtaul{\text{iso}} = \frac{1}{N_\vec{q}} \sum_\lamp \frac{\pi\omega_\lambda^2}{2}
  \sum_\kappa g(\kappa) \left|\vec{e}^*_\lambda(i)\vec{e}_\lambda(i)\right|^2 \delta(\omega_\lambda
  - \omega_\lamp),
\]
where $g(i)$ is the Pearson deviation coefficient of masses of the isotopes of atom $\kappa$.
The weights $\delta$ can be calculated using the linear tetrahedron
method~\cite{tetrahedron_method_4,Bloechl1994a}.

$\invtaul{\text{ph-ph}}$ are the phonon-phonon scattering rates and are the most
computationally expensive terms to obtain:
\begin{align*}
      \invtaul{\text{ph-ph}} = {}& \frac{1}{N_\vec{q}} \left[ \sum_{\lamp\lampp} \left(n_\lamp - n_\lampp\right)
      W^+_{\lambda\lamp\lampp} +\right. \\ & \left. +\sum_{\lamp\lampp}\frac{n_\lamp + n_\lampp + 1}{2}W^-_{\lambda\lamp\lampp}\right],
\end{align*}
where $W$ are intrinsic scattering rates. $W^+_{\lambda\lamp\lampp}$ refers to the combination of
modes $\lambda$ and $\lamp$ into $\lampp$ and $W^-_{\lambda\lamp\lampp}$ to the splitting of
$\lambda$ into $\lamp$ and $\lampp$. They can be calculated using:
\begin{align*}
    W^\pm_{\lambda\lamp\lampp} = {}& \frac{\hbar\pi}{4}
    \frac{\left|V_\pm\right|^2}{\omega_\lambda\omega_\lamp\omega_\lampp}
    \delta\left(\omega_\lambda \pm \omega_\lamp - \omega_\lampp\right) \times
    \\ & \times \delta\left(\vec{q} \pm \vec{q}^{\prime} - \vec{q}^{\prime\prime} + \vec{G}\right).
\end{align*}
The $\delta$-terms imply that only processes that conserve energy and quasi-momentum~(with a phase
of a reciprocal lattice vector $\vec{G}$) are considered. $V_\pm$ is the scattering matrix:
\begin{align*}
    V_\pm = {}& \sum \IFCanharm
    \frac{e_\lambda^\alpha(\kappa)e_{\pm\lamp}^\beta(\kappap)e_{-\lampp}^\gamma(\kappapp)}{\sqrt{m_\kappa m_\kappap m_\kappapp}} \times
    \\ & \times \exp\left[i\left(\pm\vec{q}^{\prime}\cdot\vec{R}_\lp - \vec{q}^{\prime\prime}\cdot\vec{R}_\lpp\right)\right].
\end{align*}
$\IFCanharm$ are the third-order anharmonic force constants and are the central
problem for thermal conductivity calculations. \AAPL\ uses the central difference method to
calculate them. For this purpose, two atoms need to be displaced along linearly-independent Cartesian
coordinates, resulting in many calculations that can be reduced using symmetry.

\noindent\textbf{Performing \AFLOWAAPL\ calculations.}
\AAPL\ calculations are activated by adding \inlineCommand{[{\color{codegreen} AFLOW\_AAPL}]CALC} to the \aflowin\ file. As
with \APL, this can also be achieved by adding \inlineCommand{-{}-module=aapl} to the input generation
command (\inlineCommand{-{}-aflow\_proto}). The workflow generally follows the \APL\ workflow and uses many of the same parameters, but
creates additional subdirectories to calculate forces of supercells with two atoms displaced from
their equilibrium position. It also creates the subdirectories required by \APL\ to calculate phonon
frequencies and group velocities.

As a preparatory step, \AAPL\ determines the atomic pairs that need to be displaced to calculate
anharmonic \IFCs. The cutoff distance for these pairs can be set using a radius, a number of
coordination shells, or both. For example:
\begin{fileContentTBox}
[{\color{codegreen} AFLOW\_AAPL}]CUT\_RAD=5.5\tcbbreak
[{\color{codegreen} AFLOW\_AAPL}]CUT\_SHELL=6\tcbbreak
\end{fileContentTBox}
\noindent
In this case, it will set the cutoff to at least 5.5\,{\AA} while also containing at least six
coordination shells. To use only one option, the other needs to be set to zero. \IFCs\ between
pairs with a distance beyond the cutoff are set to zero. After determining these pairs, they are
reduced by symmetry with \AFLOWSYM\ to only contain symmetrically-inequivalent
ones~\cite{curtarolo:art135}, which results in fewer calculations than in many other
packages~\cite{curtarolo:art125}.

After determining the anharmonic \IFCs, the Boltzmann Transport Equation is solved to calculate the
thermal conductivity tensor. This is a computationally expensive step and is supported by an
on-the-fly parallelization scheme inside \AFLOW. The calculation conditions are set by the following
parameters in the \aflowin\ file:
\begin{fileContentTBox}
[{\color{codegreen} AFLOW\_AAPL}]TCT=$T_\text{start}$:$T_\text{end}$:$T_\text{step}$\tcbbreak
[{\color{codegreen} AFLOW\_AAPL}]THERMALGRID=$q_1$x$q_2$x$q_3$\tcbbreak
[{\color{codegreen} AFLOW\_AAPL}]ISOTOPE=ON\tcbbreak
[{\color{codegreen} AFLOW\_AAPL}]BOUNDARY=OFF\tcbbreak
[{\color{codegreen} AFLOW\_AAPL}]NANO\_SIZE=$d_\text{grain}$\tcbbreak
\end{fileContentTBox}
\noindent
\inlineCommand{[{\color{codegreen} AFLOW\_AAPL}]TCT} sets the start and end temperatures and the temperature step size;
\inlineCommand{[{\color{codegreen} AFLOW\_AAPL}]THERMAL\_GRID} is the dimensions of the \vec{q}-point grid for phonon-phonon scattering
processes;
\inlineCommand{[{\color{codegreen} AFLOW\_AAPL}]ISOTOPE} and \inlineCommand{[{\color{codegreen} AFLOW\_AAPL}]BOUNDARY}
set whether isotope and grain boundary scattering,
respectively, are included; and \inlineCommand{[{\color{codegreen} AFLOW\_AAPL}]NANO\_SIZE} sets the size of the grains to
$d_\text{grain}$\,nm.
A full list of \aflowin\ parameters can be found in the \APL/\AAPL\ \README, which can be displayed using:
\begin{lstlisting}[language=aflowBash]
aflow --readme=aapl
\end{lstlisting}

\noindent\textbf{Visualization options.}
\AFLOWAAPL\ can plot the thermal conductivity tensor as a function of temperature. The
command is analogous to \inlineCommand{-{}-plotthermo} for \APL:
\begin{aflowBashTBox}
{\color{binarycolor} aflow} -{}-plottcond
\end{aflowBashTBox}
\noindent
where \inlineCommand{-{}-plottcond} accepts multiple optional inputs:
the \inlineCommand{\textit{directory}} of the \AAPL\ calculation~(default: current directory),
and the minimum $\left(T_\text{min}\right)$ and maximum $\left(T_\text{max}\right)$ temperatures~(default: full range of calculated temperatures);
\ie\ \inlineCommand{-{}-plottcond=\textit{directory},$T_\text{min}$,$T_\text{max}$}.
An example plot is
shown in \Fig~\ref{fig:qha_aapl}\pan{d}.
\section{Modeling Disorder}

\begin{figure*}
\includegraphics[width=\textwidth]{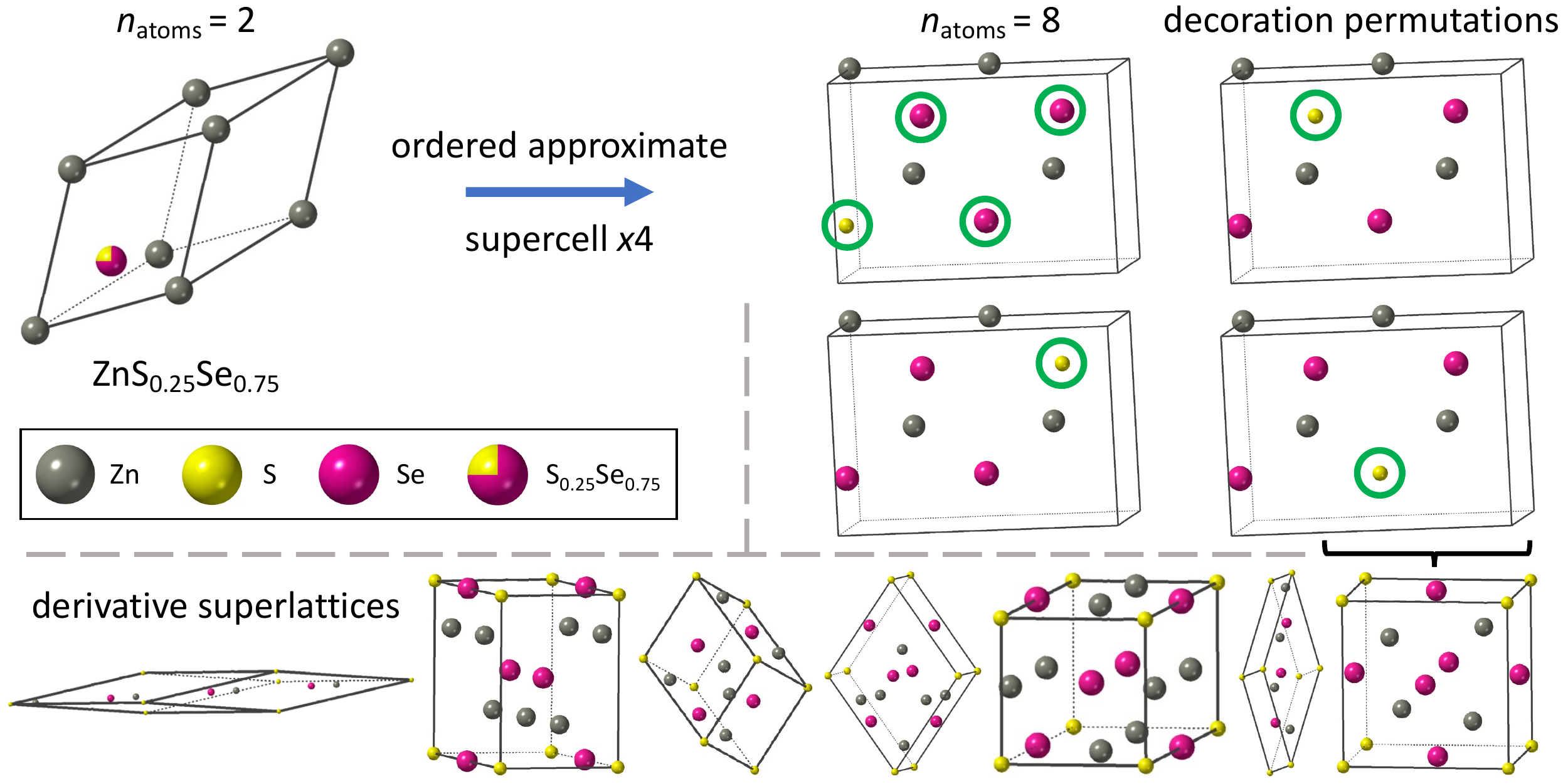}
  \caption{\textbf{Generation of ordered representatives for ZnS$_{0.25}$Se$_{0.75}$.}
The \AFLOWPOCC\ algorithm constructs the smallest supercell satisfying the stoichiometry of the disordered system.
For ZnS$_{0.25}$Se$_{0.75}$, a supercell size of four is needed given the occupancies of the shared site.
An illustration of a supercell representative is provided on the right, with the occupants of the shared site
highlighted in green.
The four decoration-permutations of this supercell are also shown, moving the sulfur (yellow) atom
to each of the allowed sites, which in this case all produce symmetrically-equivalent structures and are thus degenerate.
This derivative superlattice, a uniform expansion of the original lattice, is only one of seven
distinct possibilities producing a supercell size of four, all of which also have four decoration-permutations
to consider.
This results in 28 total representative structures, seven of which are unique.
}
\label{fig:cpp_pocc}
\end{figure*}

\subsection{\AFLOWPOCC: The Partial-Occupation Module for Chemical Disorder}\label{sec:pocc}

The \AFLOWPOCC\ module is a framework for modeling chemically disordered systems,
also known as substitutionally disordered systems or random alloys~\cite{curtarolo:art110}.
Such systems have sites with partial, probabilistic occupancies, \eg\
the high-temperature Cu$_{3}$Au phase~\cite{Nix_PR_1941,Johansson_AdP_1936} is characterized
by a single site on an fcc lattice (\href{http://www.aflow.org/prototype-encyclopedia/A_cF4_225_a.html}{A\_cF4\_225\_a})
with two occupants having probabilities matching the stoichiometry.
The properties of Cu$_{3}$Au cannot be calculated directly using standard ab-initio codes like \VASP\
relying on plane wave basis sets.
Instead, \AFLOWPOCC\ represents a random alloy as an ensemble of ordered supercells (an example is illustrated in \Fig~\ref{fig:cpp_pocc}),
the properties of which are thermally averaged to yield that of the disordered system.
The workflow is as follows: ensemble set generation (pre-processing), calculation of properties (using, \eg\ VASP),
and analysis (post-processing).
All three steps are performed by \AFLOW.

\noindent{\bf \fileName{PARTCAR} file.}
\AFLOWPOCC\ takes as input a \fileName{PARTCAR}, a modified \fileName{POSCAR} that specifies the partial occupancies of the
sites and some tolerances for the algorithm.
An example \fileName{PARTCAR} for the Ag$_{8.733}$Cd$_{3.8}$Zr$_{3.267}$ system is provided below, with
inputs specific to the \fileName{PARTCAR} file marked with a box.
\begin{fileContentTBox}
\begin{tabular}{@{}l}
\fileName{PARTCAR} of Ag$_{8.733}$Cd$_{3.8}$Zr$_{3.267}$\tabularnewline
1.0\hspace{0.4cm} \myfbox{0.001\hspace{0.4cm} 0.001}\tabularnewline
5.76\hspace{0.4cm} 5.76\hspace{0.4cm} 5.76\hspace{0.4cm} 90\hspace{0.4cm} 90\hspace{0.4cm} 90\tabularnewline
\myfbox{8{*}1+1{*}0.733\hspace{0.4cm} 3{*}1+1{*}0.8\hspace{0.4cm} 3{*}1+1{*}0.267}\tabularnewline
Direct(17)\hspace{0.4cm} \myfbox{Partial [A8.73B3.8C3.27]}\tabularnewline
0.25\hspace{0.4cm} 0.25\hspace{0.4cm} 0.25\hspace{0.4cm} Ag\hspace{0.4cm} pocc=1     \\
0.75\hspace{0.4cm} 0.75\hspace{0.4cm} 0.25\hspace{0.4cm} Ag\hspace{0.4cm} pocc=1     \\
0.75\hspace{0.4cm} 0.25\hspace{0.4cm} 0.75\hspace{0.4cm} Ag\hspace{0.4cm} pocc=1     \tabularnewline
0.25\hspace{0.4cm} 0.75\hspace{0.4cm} 0.75\hspace{0.4cm} Ag\hspace{0.4cm} pocc=1     \tabularnewline
0.25\hspace{0.4cm} 0.25\hspace{0.4cm} 0.75\hspace{0.4cm} Ag\hspace{0.4cm} pocc=1     \tabularnewline
0.75\hspace{0.4cm} 0.75\hspace{0.4cm} 0.75\hspace{0.4cm} Ag\hspace{0.4cm} pocc=1     \tabularnewline
0.25\hspace{0.4cm} 0.75\hspace{0.4cm} 0.25\hspace{0.4cm} Ag\hspace{0.4cm} pocc=1     \tabularnewline
0.75\hspace{0.4cm} 0.25\hspace{0.4cm} 0.25\hspace{0.4cm} Ag\hspace{0.4cm} pocc=1     \tabularnewline
\myfbox{0.50\hspace{0.4cm} 0.50\hspace{0.4cm} 0.50\hspace{0.4cm} Ag\hspace{0.4cm} pocc=0.733} \tabularnewline
0.00\hspace{0.4cm} 0.50\hspace{0.4cm} 0.50\hspace{0.4cm} Cd\hspace{0.4cm} pocc=1     \tabularnewline
0.50\hspace{0.4cm} 0.00\hspace{0.4cm} 0.50\hspace{0.4cm} Cd\hspace{0.4cm} pocc=1     \tabularnewline
0.50\hspace{0.4cm} 0.50\hspace{0.4cm} 0.00\hspace{0.4cm} Cd\hspace{0.4cm} pocc=1     \tabularnewline
\myfbox{0.00\hspace{0.4cm} 0.00\hspace{0.4cm} 0.00\hspace{0.4cm} Cd\hspace{0.4cm} pocc=0.8}   \tabularnewline
0.50\hspace{0.4cm} 0.00\hspace{0.4cm} 0.00\hspace{0.4cm} Zr\hspace{0.4cm} pocc=1     \tabularnewline
0.00\hspace{0.4cm} 0.50\hspace{0.4cm} 0.00\hspace{0.4cm} Zr\hspace{0.4cm} pocc=1     \tabularnewline
0.00\hspace{0.4cm} 0.00\hspace{0.4cm} 0.50\hspace{0.4cm} Zr\hspace{0.4cm} pocc=1     \tabularnewline
\myfbox{0.50\hspace{0.4cm} 0.50\hspace{0.4cm} 0.50\hspace{0.4cm} Zr\hspace{0.4cm} pocc=0.267} \tabularnewline
\end{tabular}
\end{fileContentTBox}
\noindent
The first box is next to the scaling factor and contains the site and stoichiometry tolerances for the supercell size algorithm.
The default tolerance is 0.001 for both, which can be changed in the \fileName{aflow.rc} file.
The second box marks the line containing the number of each type of atom, which now accepts fractional occupancies.
The `+' sign separates occupants of the same type with different occupancy values.
The third box is next to the `Direct' coordinates designation and signals to \AFLOW\ that the sites that follow
will have `Partial' occupancy.
Only the `P' in `Partial' is required, similar to the `D' in `Direct'.
The following three boxes mark the two partially occupied sites:
one shared between Ag ($x=0.733$) and Zr ($x=0.267$) and the other between Cd ($x=0.8$) and a vacancy ($x=0.2$).
Take note that the order of the sites matches that specified by the number-of-each-type line with
the species listed in alphabetic order.

\begin{table*}
\caption{
Evolution of the algorithm to determine the optimal supercells size for Ag$_{8.733}$Cd$_{3.8}$Zr$_{3.267}$.
$i$ is the iteration step of the algorithm, $\delta$ is the site occupancy resolution,
$f$ is the iteration's choice fraction for the site,
$\Delta$ is the error in site occupancy,
$\Delta_{x}$ is the error in the stoichiometry, and
$n$ is the supercell size.
}
{\footnotesize
\begin{tabular}{ccccccccccc}
\toprule
  \multirow{2}{*}{$i$} & \multirow{2}{*}{$\delta$} & \multicolumn{2}{c}{Ag$_{0.733}$} & \multicolumn{2}{c}{Zr$_{0.267}$} & \multicolumn{2}{c}{Cd$_{0.8}$} & \multirow{2}{*}{$\max\left(\Delta\right)$} & \multirow{2}{*}{$\max\left(\Delta_{x}\right)$} & \multirow{2}{*}{$n$}\tabularnewline
\cmidrule{3-8}
 &  & $f$  & $\Delta$  & $f$  & $\Delta$  & $f$  & $\Delta$ &  &  &  \tabularnewline
\midrule
1  & 1.000 & 1/1   & 0.267  & 0/1  & 0.267  & 1/1   & 0.200  & 0.267 & 0.019 & 1\tabularnewline
2  & 0.500 & 1/2   & 0.233  & 1/2  & 0.233  & 2/2   & 0.200  & 0.233 & 0.021 & 2\tabularnewline
3  & 0.333 & 2/3   & 0.066  & 1/3  & 0.066  & 2/3   & 0.133  & 0.133 & 0.006 & 3\tabularnewline
4  & 0.250 & 3/4   & 0.017  & 1/4  & 0.017  & 3/4   & 0.050  & 0.050 & 0.003 & 4\tabularnewline
5  & 0.200 & 4/5   & 0.067  & 1/5  & 0.067  & 4/5   & 0.000  & 0.067 & 0.004 & 5\tabularnewline
6  & 0.167 & 4/6   & 0.066  & 2/6  & 0.066  & 5/6   & 0.033  & 0.066 & 0.005 & 6\tabularnewline
7  & 0.143 & 5/7   & 0.019  & 2/7  & 0.019  & 6/7   & 0.057  & 0.057 & 0.003 & 7\tabularnewline
8  & 0.125 & 6/8   & 0.017  & 2/8  & 0.017  & 6/8   & 0.050  & 0.050 & 0.003 & 4\tabularnewline
9  & 0.111 & 7/9   & 0.045  & 2/9  & 0.045  & 7/9   & 0.022  & 0.045 & 0.004 & 9\tabularnewline
10 & 0.100 & 7/10  & 0.033  & 3/10 & 0.033  & 8/10  & 0.000  & 0.033 & 0.002 & 10\tabularnewline
11 & 0.091 & 8/11  & 0.006  & 3/11 & 0.006  & 9/11  & 0.018  & 0.018 & 0.001 & 11\tabularnewline
12 & 0.083 & 9/12  & 0.017  & 3/12 & 0.017  & 10/12 & 0.033  & 0.033 & 0.002 & 12\tabularnewline
13 & 0.077 & 10/13 & 0.036  & 3/13 & 0.036  & 10/13 & 0.031  & 0.036 & 0.003 & 13\tabularnewline
14 & 0.071 & 10/14 & 0.019  & 4/14 & 0.019  & 11/14 & 0.014  & 0.019 & 0.001 & 14\tabularnewline
15 & 0.067 & 11/15 & 0.000  & 4/15 & 0.000  & 12/15 & 0.000  & 0.000 & 0.000 & 15\tabularnewline
\bottomrule
\end{tabular}
}
\label{tab:pocc_hnf}
\end{table*}

A \fileName{PARTCAR} file can be constructed automatically by \AFLOW\
using the aforementioned \inlineCommand{-{}-proto} and \inlineCommand{-{}-aflow\_proto} commands,
which leverage the extensive prototyping suite~\cite{curtarolo:art170} and
library of naturally-occurring compounds~\cite{curtarolo:art121,curtarolo:art145,aflowANRL3}
to construct the parent structure.
The first step is to construct the \inlineCommand{-{}-proto} command for the base structure.
For example,
\begin{lstlisting}[language=aflowBash]
aflow --proto=AB_hP4_186_b_b-001:S:Zn --abccar\end{lstlisting}
generates the \fileName{POSCAR} for the high-temperature (wurtzite) phase of ZnS~\cite{Kisi_wurtziteZnSZnO_1989}.
\begin{fileContentTBox}
\begin{tabular}{l}
\fileName{POSCAR} of SZn/AB\_hP4\_186\_b\_b-001.AB\tabularnewline
1.0\tabularnewline
3.5155\hspace{0.4cm} 3.5155\hspace{0.4cm} 3.5155\hspace{0.4cm} 90\hspace{0.4cm} 90\hspace{0.4cm} 120\tabularnewline
2\hspace{0.4cm} 2\tabularnewline
Direct(4) [A2B2]\tabularnewline
0.333333\hspace{0.4cm} 0.666667\hspace{0.4cm} 0.374800\hspace{0.4cm} S\tabularnewline
0.666667\hspace{0.4cm} 0.333333\hspace{0.4cm} 0.874800\hspace{0.4cm} S\tabularnewline
0.333333\hspace{0.4cm} 0.666667\hspace{0.4cm} 0.000000\hspace{0.4cm} Zn\tabularnewline
0.666667\hspace{0.4cm} 0.333333\hspace{0.4cm} 0.500000\hspace{0.4cm} Zn\tabularnewline
\end{tabular}
\end{fileContentTBox}
\noindent
The \inlineCommand{-{}-abccar} flag prints the lattice as
$\left|\mathbf{a}\right|$,
$\left|\mathbf{b}\right|$,
$\left|\mathbf{c}\right|$,
$\alpha$,
$\beta$, and
$\gamma$
instead of the usual matrix format.
Take note that this structure has four sites and two species.
To construct the ZnS$_{0.5}$Se$_{0.5}$ structure, where both sulfur sites are
partially occupied with selenium, the \inlineCommand{-{}-proto} command is modified as such:
\begin{lstlisting}[language=aflowBash]
aflow --proto=AB_hP4_186_b_b-001:S:Se:Zn --abccar --pocc_params=P0-0.5xA-0.5xB_P1-0.5xA-0.5xB_P2-1xC_P3-1xC [--pocc_tol=0.001:0.001]
\end{lstlisting}
Notice that the changes from the original \inlineCommand{-{}-proto} command include
adding selenium to the colon-separated species list (in alphabetic order) and
the \inlineCommand{-{}-pocc\_params}/\inlineCommand{-{}-pocc\_tol} options.
The \inlineCommand{-{}-pocc\_params} option is a list of underline-separated values for each site (\underline{P}osition)
with the first site of the parent structure indicated by \inlineCommand{P0}.
Following the site specification is a list of comma-separated values of the occupancies for that site,
with \inlineCommand{A}, \inlineCommand{B}, and \inlineCommand{C} referring to the colon-separated, alphabetically-ordered set of species
provided in the \inlineCommand{-{}-proto} option.
All sites must be specified with the \inlineCommand{-{}-pocc\_params} options.
The optional \inlineCommand{-{}-pocc\_tol} takes as input a colon-separated list of site and stoichiometry tolerances
for the supercell size algorithm.
For the site tolerance, a negative integer can be accepted instead, which fixes
the desired supercell size and allows the generation of larger ensemble sets (better sampling) beyond what is
required to satisfy the stoichiometry exactly.
The command generates the following \fileName{PARTCAR}:
\begin{fileContentTBox}
\begin{tabular}{l}
\fileName{PARTCAR} of S$_{0.5}$Se$_{0.5}$Zn/AB\_hP4\_186\_b\_b-001.AB\tabularnewline
1.0\hspace{0.4cm} 0.001\hspace{0.4cm} 0.001\tabularnewline
3.6017\hspace{0.4cm} 3.6017\hspace{0.4cm} 5.8988\hspace{0.4cm} 90\hspace{0.4cm} 90\hspace{0.4cm} 120\tabularnewline
2*0.5\hspace{0.4cm} 2*0.5\hspace{0.4cm} 2*1\tabularnewline
Direct(6)\hspace{0.4cm} Partial [A1B1C2]\tabularnewline
0.333333\hspace{0.4cm} 0.666667\hspace{0.4cm} 0.374800\hspace{0.4cm} S \hspace{0.4cm} pocc=0.5\tabularnewline
0.666667\hspace{0.4cm} 0.333333\hspace{0.4cm} 0.874800\hspace{0.4cm} S \hspace{0.4cm} pocc=0.5\tabularnewline
0.333333\hspace{0.4cm} 0.666667\hspace{0.4cm} 0.374800\hspace{0.4cm} Se\hspace{0.4cm} pocc=0.5\tabularnewline
0.666667\hspace{0.4cm} 0.333333\hspace{0.4cm} 0.874800\hspace{0.4cm} Se\hspace{0.4cm} pocc=0.5\tabularnewline
0.333333\hspace{0.4cm} 0.666667\hspace{0.4cm} 0.000000\hspace{0.4cm} Zn\hspace{0.4cm} pocc=1\tabularnewline
0.666667\hspace{0.4cm} 0.333333\hspace{0.4cm} 0.500000\hspace{0.4cm} Zn\hspace{0.4cm} pocc=1\tabularnewline
\end{tabular}
\end{fileContentTBox}
\noindent
The same \fileName{PARTCAR} can be generated with a shorter command:
\begin{lstlisting}[language=aflowBash]
aflow --proto=AB_hP4_186_b_b-001:S:Se:Zn --abccar --pocc_params=S0-0.5xA-0.5xB_S1-1xC
\end{lstlisting}
where the partial occupancy values are specified by \underline{S}pecies,
\ie\ the group of sites having sulfur atoms, \inlineCommand{S0}, (zinc atoms, \inlineCommand{S1}) in the parent structure.
This specification can truncate the input size substantially for larger structures.
The two specifications, \inlineCommand{P} and \inlineCommand{S}, can be mixed, but should be avoided in practice as
it can be challenging to keep track of the two sets of indices simultaneously.

A \fileName{PARTCAR} can also be converted to a \CIF\ file with partially occupied sites and vice versa.
For example:
\begin{lstlisting}[language=aflowBash]
aflow --proto=AB_hP4_186_b_b-001:S:Se:Zn --cif --pocc_params=S0-0.5xA-0.5xB_S1-1xC
\end{lstlisting}
generates the following output:
\begin{lstlisting}[language=fileContentAFLOWIN]
# AFLOW.org Repositories
# SSeZn/AB_hP4_186_b_b-001.AB:POCC_S0-0.5xA-0.5xB_S1-1xC
data_SSeZn
_pd_phase_name AB_hP4_186_b_b-001.AB:POCC_S0-0.5xA-0.5xB_S1-1xC
_cell_length_a  3.6017
_cell_length_b  3.6017
_cell_length_c  5.8988
_cell_angle_alpha  90.0000
_cell_angle_beta  90.0000
_cell_angle_gamma  120.0000
_symmetry_space_group_name_H-M  `P1'
_symmetry_Int_Tables_Number  1
loop_
_symmetry_equiv_pos_site_id
_symmetry_equiv_pos_as_xyz
  1  x,y,z
loop_
_atom_site_label
_atom_site_occupancy
_atom_site_fract_x
_atom_site_fract_y
_atom_site_fract_z
_atom_site_type_symbol
S1  0.5000 0.3333 0.6667 0.3748 S
S2  0.5000 0.6667 0.3333 0.8748 S
Se3 0.5000 0.3333 0.6667 0.3748 Se
Se4 0.5000 0.6667 0.3333 0.8748 Se
Zn5 1.0000 0.3333 0.6667 0.0000 Zn
Zn6 1.0000 0.6667 0.3333 0.5000 Zn
\end{lstlisting}
A \CIF\ file can also be converted to a \fileName{PARTCAR} with the \inlineCommand{-{}-vasp} flag:
\begin{lstlisting}[language=aflowBash]
aflow --proto=AB_hP4_186_b_b-001:S:Se:Zn --cif --pocc_params=S0-0.5xA-0.5xB_S1-1xC | aflow --vasp
\end{lstlisting}

By switching the \inlineCommand{-{}-proto} to \inlineCommand{-{}-aflow\_proto}, \AFLOW\ will generate
an \aflowin\ file containing the \fileName{PARTCAR} inside.
The directory structure is largely the same as that specified in Section~\ref{sec:aflowin},
except
in this case, the third (structure) layer is a mix of the parent structure prototype \inlineCommand{AB\_hP4\_186\_b\_b-001.AB}
and the \mbox{\inlineCommand{-{}-pocc\_params}.}
The directory tree also protects from the generation of duplicates, although it does not prevent the ones
produced from mixing \inlineCommand{P} and \inlineCommand{S} specifications.
The \AFLOWPOCC\ \aflowin\ contains \inlineCommand{[{\color{codegreen} AFLOW\_POCC}]CALC} activating the module
and the \fileName{PARTCAR} wrapped with \inlineCommand{[{\color{codegreen} POCC\_MODE\_EXPLICIT}]START.POCC\_STRUCTURE}
and \inlineCommand{[{\color{codegreen} POCC\_MODE\_EXPLICIT}]STOP.POCC\_STRUCTURE}.
Running \inlineCommand{{\color{binarycolor} aflow} -{}-run}
in the directory with this \aflowin\ will,
in a single step, determine the optimal supercell size, generate the ensemble of
ordered representatives, and create a subdirectory for each (denoted with \inlineCommand{ARUN.POCC}).
Each directory contains a child \aflowin\ setting up that particular calculation.
The subdirectory structure enables parallelization across the ordered representatives.
Note that the characterization of the ordered representatives is not strictly limited to
the usual relaxation/static workflows, but can also include the analysis of thermomechanical
properties, themselves creating additional parent/child layers which ultimately resolve the
properties of the disordered system.
These advanced workflows will be covered in subsequent sections.

\noindent{\bf Creating the right supercell size.}
Following the creation of a \fileName{PARTCAR},
the \AFLOWPOCC\ algorithm determines the smallest supercell size that satisfies the site and stoichiometry tolerances specified.
The algorithm iterates through incrementally larger supercell sizes, identifying the optimal occupation $\left(f\right)$
and the associated site $\left(\Delta\right)$ and stoichiometry $\left(\Delta_{x}\right)$ errors.
An example is provided in Table~\ref{tab:pocc_hnf} for Ag$_{8.733}$Cd$_{3.8}$Zr$_{3.267}$.
The stop condition is achieved when both the maximum of the site and stoichiometry errors are below
the tolerance specified in the \fileName{PARTCAR}.
The default tolerances of 0.001 should find the supercell size that satisfies the stoichiometry exactly.
To optimize either the site or stoichiometry errors individually, set the other tolerance to 1.

One challenge that the algorithm overcomes is the filling of sites for supercell sizes that may be too small.
For example, consider the first iteration in Table~\ref{tab:pocc_hnf}.
For this case, there is only one site and two possible occupants.
The best choice (the one reducing the error) is to fill the site with the occupant having the higher concentration (Ag).
If both occupants had equal (50\%) concentrations, the algorithm would leave the site unfilled.
While filling the site with one of the occupants would yield the same error,
the algorithm has no way of deciding which occupant to pick.
The issue becomes more problematic with more occupants having equal concentrations.
This requires grouping these same-site, equi-concentration occupants and incrementing their fillings simultaneously or not at all.
To avoid the generation of vacancies, the input occupancy values should be altered to bias the components of interest.

\noindent{\bf Ensemble set generation.}
After the algorithm determines the optimal supercell size,
it proceeds to construct all of the possible derivative supercells.
An illustration of the two steps --- determining the unique derivative superlattices (geometries)
and all of the corresponding decorations (colorings) --- is shown in \Fig~\ref{fig:cpp_pocc}.
The Hermite normal form matrices~\cite{Santoro_HNF1,Santoro_HNF2} generate the unique set of derivative superlattices.
For the ZnS$_{0.25}$Se$_{0.75}$ system shown in \Fig~\ref{fig:cpp_pocc}, there are seven unique derivative superlattices,
each of which has four decoration-permutations, generating 28 total representative structures.
\AFLOWPOCC\ then determines which of these supercells are unique,
employing the Universal Force Field method~\cite{Rappe_1992_JCAS_UFF} to calculate an approximate energy for each structure
which can be quickly compared to resolve whether two structures are identical.
For ZnS$_{0.25}$Se$_{0.75}$,  seven unique supercells will be passed along
to the next workflow stage: \DFT\ calculation with \VASP.
The algorithm has been optimized for speed and reduction of memory footprint,
critical as the number of total derivative structures to consider
can easily run into the billions.
It is important to mention that \AFLOWPOCC\ considers all of the possible ordered representatives,
and not just a subset limited to a particular superlattice.
This ensures proper sampling for the calculation of spectral descriptors like the
\underline{e}ntropy-\underline{f}orming-\underline{a}bility (\EFA) discussed in detail in the next section.

\noindent{\bf Post-processing analysis.}
Upon completion of the \inlineCommand{ARUN.POCC} subdirectory calculations,
running \inlineCommand{{\color{binarycolor} aflow} -{}-run}
again in the parent directory will initiate the
post-processing analysis.
For the usual relaxation/static runs of the ordered representatives,
\AFLOWPOCC\ will resolve ensemble average \DOS,
bandgaps, and magnetic moments~\cite{curtarolo:art110}.
A Boltzmann-weight is calculated for the averaging:
the $i$-structure's probability depends on its
degeneracy count $\left(g_{i}\right)$,
amount of disorder relative to the minimum (ground-state) structure in the set $\left(H_i-H_{\text{gs}}\right)$,
and a tuning parameter mimicking temperature $\left(T_\sPOCC\right)$:
  \[
  P_i=\frac{g_i e^{-\left(H_i-H_{\text{gs}}\right)/k_{\text{B}}T_\sPOCC}}{\sum_{i} g_i e^{-\left(H_i-H_{\text{gs}}\right)/k_{\text{B}}T_\sPOCC}}.
\]
The temperature parameter dictates how much disorder to incorporate
in the analysis, and roughly correlates with the synthesis temperature of the material.
The ensemble average \DOS\ is written to \fileName{DOSCAR}-type
files at various temperature snapshots, \eg\ \fileName{DOSCAR.pocc\_T0300K.xz}.
The snapshots can be set from the command line upon creation of the
parent \aflowin\ (in combination with the \inlineCommand{-{}-aflow\_proto} command)
with \inlineCommand{-{}-temperature=0:2400:300},
or by hand inside the parent \aflowin\
by adding the line
\inlineCommand{[{\color{codegreen} AFLOW\_POCC}]TEMPERATURE=0:2400:300},
both of which will perform temperature snapshots between 0--2400~K
in 300~K increments.
The default temperature snapshots are set inside the \fileName{aflow.rc} file.
The other properties are written to the \inlineCommand{aflow.pocc.out} file
containing a summary of the full analysis, including
the degeneracy count for each unique structure and associated
temperature-specific properties, including the Boltzmann-weighted properties of each structure.
The properties are organized by temperature snapshots.

\AFLOWPOCC\ also calculates the \EFA, which has led to the discovery of 10 high-entropy carbides~\cite{curtarolo:art140,curtarolo:art164}.
The descriptor is the inverse of the standard deviation of the energy spectrum
of the ordered representatives.
Being a spectral descriptor, the \EFA\ analysis is reliant on proper sampling,
requiring full exploration of possible ordered representatives.

\AFLOWPOCC\ has been validated for a number of systems and properties~\cite{curtarolo:art110,curtarolo:art140,curtarolo:art164,curtarolo:art179,curtarolo:art180},
including electronic, magnetic, thermodynamic, and thermomechanical properties.
It has also been demonstrated that the accuracy improves with larger supercell sizes~\cite{curtarolo:art110},
providing better sampling and capturing longer-range effects.
Convergence with supercell size should be checked for each property and balanced with
the feasibility of larger supercell calculations.

\noindent{\bf Comparison to other disordered system models.}
\AFLOWPOCC\ is a multiple-supercells approach, which is often compared with special quasirandom structures (\SQS),
a supercell approach~\cite{sqs}.
\SQS\ is, for a given supercell size, the singular supercell structure that minimizes the site correlations to emulate the random alloy,
and is thus among the set of structures enumerated by \AFLOWPOCC.
Using a single supercell to model the disordered system represents the strictly infinite-temperature solution.
\AFLOWPOCC\ allows finite-temperature modeling, including near the transition temperature,
with the introduction of an ensemble set (degeneracies) and tuning parameters into the framework.
To best model the properties of the disordered system, the needed \SQS\ is often quite large and very low-symmetry,
making it a challenging calculation to converge.
Alternatively, \AFLOWPOCC\ approaches disordered-systems-modeling from the other end,
generating many smaller cell representatives, which can be easily parallelized for high-throughput workflows.

\noindent\textbf{Elasticity and GIBBS analysis.}
The \AFLOW\ Elasticity Library (\AEL) and \AFLOW\ GIBBS Library (\AGL) modules
(see Section~\ref{sec:aelagl})
can be used to calculate the thermal and elastic properties of ordered materials.
These methods have now been integrated with the \AFLOWPOCC\ module to enable calculation
of the thermal and elastic properties of configurationally disordered materials.

For every derivative structure or ``{\it tile}'' generated by \POCC, full \AEL\ and/or \AGL\ calculations
are performed to obtain their thermoelastic properties, including the elastic constants,
bulk and shear moduli, Debye temperature, specific heat capacity at constant volume and at constant
pressure, and coefficient of thermal expansion.
The results are then thermally averaged using the \POCC\ Boltzmann probabilities $P_i$,
a function of the \POCC\ temperature $T_\sPOCC$.
For example, the thermally averaged {\small VRH} bulk modulus for a configurationally disordered material would be calculated as
  \[
  B_\sVRH^\textnormal{avg}\left(T_\sPOCC\right) = \sum_i P_i\left(T_\sPOCC\right) {B_\sVRH}_{, i},
\]
where ${B_\sVRH}_{, i}$ is the  {\small VRH} bulk modulus for tile $i$.
Since the equations for the bulk and shear moduli are linear in the elastic constants, using the thermally averaged elastic
constants to calculate the bulk and shear moduli would give the same result as averaging the bulk and shear
moduli for each tile.

Note that in the case of \AGL\ calculations, there are two sets of temperatures: the \POCC\ temperature,
which determines the distribution of the ``tiles'' present in the material and is often equivalent to
a synthesis or annealing temperature; and the \AGL\ temperature, which corresponds to the instantaneous
temperature determining the vibrational properties of the material.
Both sets of temperatures can be set separately during post-processing, so the same set of \DFT\ calculations can
be used to investigate a range of different temperature regimes and combinations.

Performing \AGL\ calculations in the \POCC\ framework requires a completed \POCC\ run with all
``tiles'' relaxed and all enthalpies calculated, so that the probabilities
needed to ensemble-average the thermal elastic properties are available.
To run an \AFLOWPOCC+\AEL/\AGL\ calculation, the line \inlineCommand{[{\color{codegreen} AFLOW\_AGL}]CALC}
or \inlineCommand{[{\color{codegreen} AFLOW\_AEL}]CALC}
should be present in the \aflowin\ file of the parent directory.
It is recommended to copy the original \aflowin\ file into a new file (\eg\ \fileName{aflow\_agl.in}) before adding/uncommenting the line,
since \aflowin\ files of the same name will be created in the subdirectories.
As in the case of ordered materials, combined \AELAGL\ workflows can also be run,
\eg\ by setting \inlineCommand{[{\color{codegreen} AFLOW\_AGL}]AEL\_POISSON\_RATIO=ON}.
Other \AEL\ and \AGL\ parameters can be set for all tiles by including the appropriate options
(described in Section~\ref{sec:aelagl})
in the \aflowin\ file of the parent directory.

\noindent\textbf{Phonon analysis.}
The \AFLOW\ Automatic Phonon Library~(\APL) discussed earlier is limited to calculating phonon
properties for ordered materials. Since the \POCC\ method uses a statistical ensemble of ordered
structures, \APL\ can be expanded to determine integrated vibrational properties of disordered
materials as well~\cite{curtarolo:art180}.

These properties include the vibrational free energy $\Fvib$, internal energy $\Uvib$, and vibrational
entropy $\Svib$. They are calculated from the phonon \DOS\ as shown in the \APL\ section. This opens
two avenues to calculate these quantities as ensemble properties: determining them for each derivative
structure and ensemble-averaging, or using the ensemble-averaged \DOS\ and integrating it.
Due to the linear relationship between the phonon \DOS\ and $\Fvib$, $\Uvib$, and
$\Svib$, the results are independent of that choice:
\begin{align}
  \Fvib^\textnormal{avg}(T)& = \sum_i P_i(T_\sPOCC) {\Fvib}_{, i}\nonumber\\
                           & = \sum_i P_i(T_\sPOCC) \int_0^\infty f(T,\omega) g_i(\omega)d\omega\nonumber\\
                           & = \int_0^\infty f(T,\omega) \sum_i P_i(T_\sPOCC)g_i(\omega)d\omega\nonumber\\
                           & = \int_0^\infty f(T,\omega) g^\textnormal{avg}(\omega)d\omega\nonumber.
\end{align}
Here, $\Fvib^\textnormal{avg}$ and $g^\textnormal{avg}$ are the ensemble-averaged vibrational free
energy and phonon \DOS, respectively, and
$f(T,\omega) = \kB T \log\left(2\sinh\frac{\hbar\omega}{2{\kB}T}\right)$. The same relationship
can be shown for $\Uvib$ and $\Svib$. The \POCCAPL\ workflow ensemble averages the \DOS, which also
provides access to the phonon \DOS\ of the disordered material.

Performing \APL\ calculations in the \POCC\ framework requires a completed \POCC\ run with all
derivative structures relaxed and all enthalpies calculated. This ensures that the probabilities
needed to ensemble-average the phonon \DOS\ are available. To start the \AFLOWPOCC\ calculation,
the line \inlineCommand{[{\color{codegreen} AFLOW\_APL}]CALC} needs to be present in the \aflowin\ file of the parent
directory. It is recommended to copy the original \aflowin\ file into a new file (\eg\ \fileName{aflow\_apl.in}) before
adding/uncommenting the line since \aflowin\ files of the same name will be created in the
subdirectories.

Running \binary{aflow} in this directory will create \aflowin\ files for the \APL\ calculations for each
derivative structure. All \APL\ options set in the parent input file will be propagated into
the subdirectories. The initial structure is taken from the last relaxed run of the ordered
representative. The next step is to run \inlineCommand{{\color{binarycolor} aflow} -{}-run} inside each subdirectory, which creates the
additional layer of subdirectories needed to calculate force constants and the non-analytical correction, if requested.
After finishing all \DFT\ calculations in these directories, \inlineCommand{{\color{binarycolor} aflow} -{}-run} needs to be run again inside
the \POCC\ parent directory. It is not required to execute the \APL\ post-processing runs in the
individual subdirectories --- \AFLOWPOCC\ will calculate the force constants if not present.

The force constants are then determined and used to calculate the phonon frequencies and phonon
\DOS\ for each ordered representative. The \DOS\ are ensemble-averaged to calculate thermophysical
properties for each \POCC\ temperature. Dynamically unstable derivative structures,
\ie\ structures with imaginary frequencies in the phonon dispersion, are automatically discarded
from the ensemble. This behavior can be turned off via the \aflowin\ option
\inlineCommand{[{\color{codegreen} AFLOW\_POCC}]EXCLUDE\_UNSTABLE=OFF}.

\AFLOWPOCC\ outputs the phonon \DOS\ for each \POCC\ temperature~(in the \VASP\ \fileName{DOSCAR} format),
and the vibrational properties calculated from the ensemble-averaged \DOS. Both can be plotted the same
way as in \APL: the \DOS\ plots for all temperatures and the thermophysical properties are generated
using \inlineCommand{-{}-plotphdos} and \inlineCommand{-{}-plotthermo} commands, respectively.

\subsection{\AFLOWQCA: The Quasi-Chemical Approximation Module for Chemical Disorder}\label{sec:qca}

\noindent\textbf{Solid solution phase transition.}
Forming solid solution alloys can offer enhanced
thermodynamic, chemical and mechanical properties~\cite{HEAapp1,HEAapp2}.
The \AFLOWQCA\ module provides a high-throughput~\cite{curtarolo:art81} ab-initio method~\cite{curtarolo:art139} to predict the temperature at which phase-separated multi-component alloys undergo a  transition to become highly disordered solid solutions~\cite{hea1,hea2}.
This method rests on calculating an order parameter, within the \underline{g}eneralized \underline{q}uasi-\underline{c}hemical \underline{a}pproximation (\GQCA)~\cite{GQCA1,GQCA2}, whose maximal change, with respect to temperature, defines the order-disorder phase transition at equi-concentration.
Then, tracing the locus of the equi-concentration relative entropy~\cite{RE2,RE3,RE4}, the phase transition temperature for the whole concentration spectrum is found.
In the subsequent sections, the predictive capabilities of this method are demonstrated and corroborated by Monte Carlo simulations~\cite{Massalski}, \CALPHAD\ calculations~\cite{Andersson_CALPHAD_2001_THERMOCALC_DICRA}, and experimental data~\cite{hea2,GaoAlman_CoFeMnNi_2013,familyfcc,CoCrFeNi,GaoAlman_CoFeMnNi_2013,AlNbTiV,HfNbTiZr,MoNbTaW,MoNbTaW2,NbTaTiV,CrNbVTiZr,MoNbTiVZr,AlCrMoTiW,HfNbTaTiZr,MoNbTaTiV,HfNbTiVZr}.

\noindent\textbf{\GQCA\ model.}
The \GQCA\ model factorizes a parent lattice of $N$ sites and $K$ species into an ensemble of non-overlapping clusters, which are statistically and energetically independent of the surrounding medium.
Let an alloy with macroscopic concentration $\{X_1,X_2,\ldots,X_K\}$ be characterized by an ensemble of $J+1$ clusters, with each cluster containing $n$ atoms.
Here, each $j$-type cluster has a distinct energy $\varepsilon_j$ and concentration $\{x_{1j},x_{2j},\ldots,x_{Kj}\}$.
Then the mixing energy for a given set of clusters is determined by~\cite{chen1995alloys}:
\begin{equation}
\Delta E(\mathbf{X},T) = \sum_{k=1}^KX_k\varepsilon_k^0 +
\sum_{j=0}^JP_j(\mathbf{X},T)\Delta_j,
\nonumber
\end{equation}
where $\varepsilon_k^0$ are the energies of the pure cluster only containing the $k^{\text{th}}$ species, $P_j$ are the probabilities of the $j$-type cluster in the alloy, and $\Delta_j$ are the reduced excess energies defined as:
  \[
    \Delta_j = \varepsilon_j-\sum_{k=1}^Kx_{kj}\varepsilon_k^0 .
  \]
Likewise, the mixing entropy is given by:
\begin{align*}
    		\Delta S(\mathbf{X},T) = {}& -k_B\left [n\sum_{k=1}^KX_k\log{X_k} \right. + \\ & \left. + \sum_{j=0}^JP_j(\mathbf{X},T)\log{\left (\frac{P_j(\mathbf{X},T)}{P_j^0(\mathbf{X})}\right )}\right],
\end{align*}
where $P_j^0$ are the probabilities to find the $j$-type cluster for the ideal solution model~\cite{Rutherford_EnumDS_1992,gus_enum} and the last term is known as the Kullback–Leibler divergence $D_{KL}$ or relative entropy~\cite{RE1}.

Finding the cluster probabilities $P_j$ at equilibrium involves minimizing the mixing free energy $\Delta F = \Delta E - T\Delta S$ by solving the set of equations $\partial \Delta F/\partial \mathbf{P}=\mathbf{0}$ with $K$ constraints:
\begin{align*}
   	& 	\sum_{j=0}^JP_j = 1, \,\,\,\,\, \,\,\,\,\,
      \sum_{j=0}^JP_jx_{1j} = X_{1}, \\
    &  \sum_{j=0}^JP_jx_{2j} = X_{2},  \,\,\, \cdots  \,\,\, \sum_{j=0}^JP_jx_{K-1j} = X_{K-1},
\end{align*}
leaving $J+1-K$ coupled equations.
Using the method of Lagrangian multipliers, the solution to this set of equations yields~\cite{chen1995alloys,curtarolo:art139}:
\begin{equation*}
    P_j(\mathbf{X},T) = \frac{P_j^0(\mathbf{X})e^{n\beta\left[\sum_{k=1}^{K-1}x_{kj}\lambda_k(\mathbf{X},T)-\Delta_j\right ]}}{\sum_{j=0}^JP_j^0(\mathbf{X})e^{n\beta\left[\sum_{k=1}^{K-1}x_{kj}\lambda_k(\mathbf{X},T)-\Delta_j\right ]}},
    \label{qca_eqn:equil_prob}
\end{equation*}
where $\beta=1/k_BT$ and $\lambda_k$ are the Lagrangian multipliers.
The energies used in evaluating the previous equation are calculated
with the
Cluster Expansion technique,
as implemented in \Ref~\onlinecite{atat1,atat2} using the \AFLOWorg\ repositories~\cite{aflowlib,curtarolo:art92,curtarolo:art104,curtarolo:art128}.

\noindent\textbf{Phase transition order parameter.}
The order-disorder transition of an alloy can be determined by performing a common tangent construction of the mixing free energy~\cite{kittel1980thermal}.
However, in our model, where the clusters are uncorrelated, this technique is invalid due to the absence of coherency effects~\cite{Cahn_CoherentEquil_Actametal_1984}.
To overcome this problem, an order parameter is defined by:
  \[
    \alpha(\mathbf{X},T) = \frac{\mathbf{P}\cdot\mathbf{P}^0}{\lVert\mathbf{P}\rVert\lVert\mathbf{P}^0\rVert},
  \]
which measures the deviation of the probability distribution from the high-$T$ limit~\cite{guggenheimmixtures1952}.
Furthermore, $\alpha$ is only evaluated at the equi-concentration $\mathbf{X}^{(\textrm{ec})}$, where the correlation effects are minimal~\cite{Taggart_Correlations_1973}.
Next, the transition temperature at equi-concentration $T_c^{(\textrm{ec})}$ is defined where the temperature gradient of the order parameter is maximum, similar to Monte Carlo simulations~\cite{axel_MC}, as shown in \Fig~\ref{fig:qca}\pan{a}.
Finally, an assumption is made: the relative entropy at the transition temperature $T_c$ is independent of the macroscopic concentration, such that:
  \[
    D^{(\textrm{ec})}_{KL}\equiv D_{KL}(\mathbf{X}^{(\textrm{ec})},T_c^{(\textrm{ec})})\approx D_{KL}(\mathbf{X},T_c)
  \]
yields the transition temperature for the whole concentration spectrum.

\begin{figure}[hbt]
\centering
\includegraphics[width=0.9\linewidth]{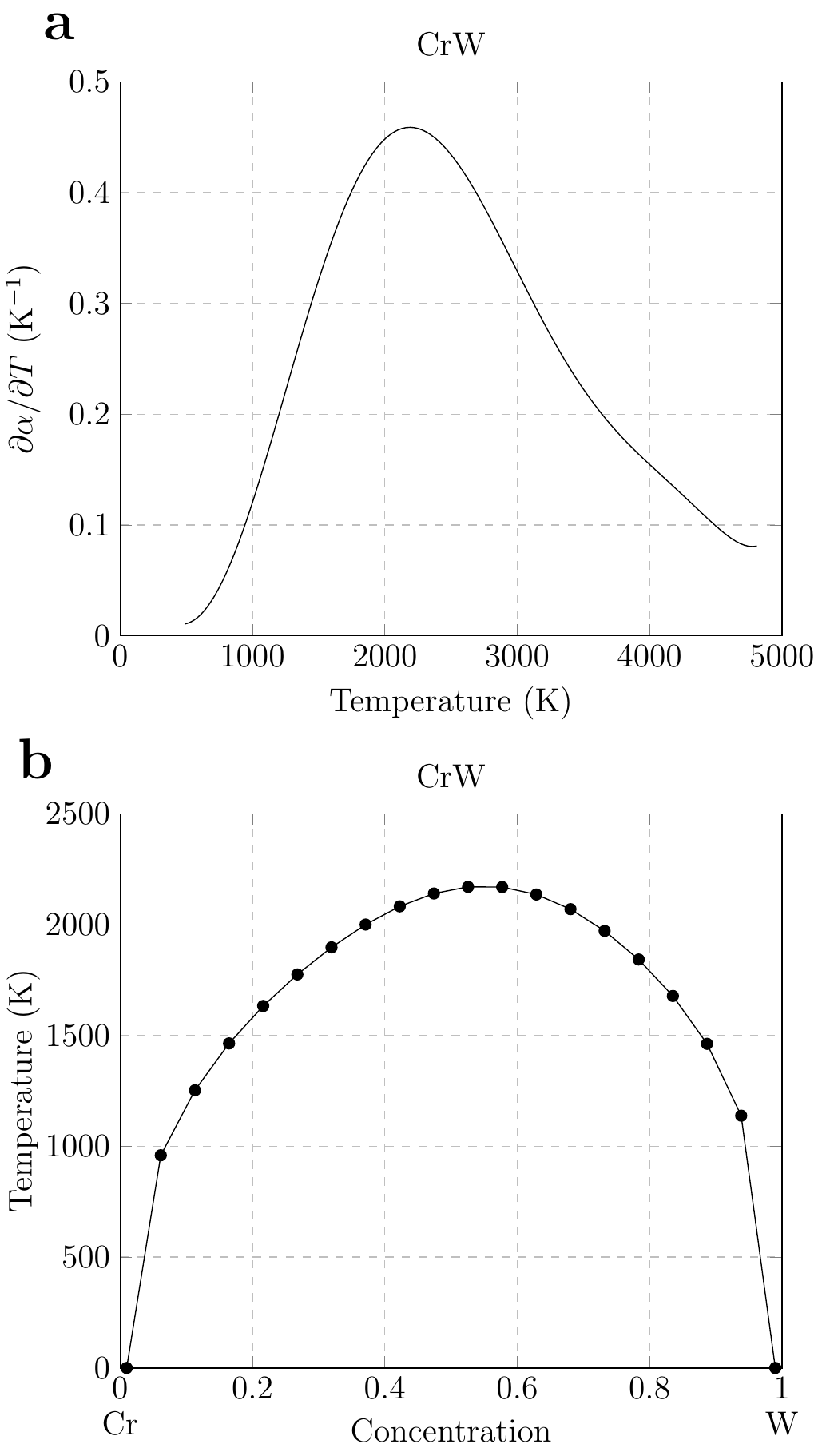}
\caption{\textbf{Properties calculated by \AFLOWQCA.}
(\textbf{a})~Partial derivative of the order parameter with respect to temperature and
(\textbf{b})~the binodal curve, for the CrW alloy, respectively.}
\label{fig:qca}
\end{figure}

\noindent\textbf{Usage.}
The following command loads the alloy data from the \AFLOWorg\ repositories, constructs the binodal curve, as shown in \Fig~\ref{fig:qca}\pan{b}, and returns the output in plain text (\inlineCommand{txt}) format:

\begin{lstlisting}[language=aflowBash]
aflow --qca --plattice=fcc --elements=Au,Pt
\end{lstlisting}

\begin{myitemize}
\item \inlineCommand{-{}-qca} : Necessary argument, enters the mode to calculate the phase equilibria properties.
\item \inlineCommand{-{}-plattice=\textit{lattice}} : Necessary argument, specifies the parent lattice of the alloy (\eg\ \inlineCommand{fcc}).
\item \inlineCommand{-{}-elements=\textit{elements}} : Necessary argument, \inlineCommand{\textit{elements}} is a comma-separated list of components present in the alloy.
\item \inlineCommand{-{}-directory=\textit{directory}} : Optional argument, specifies the directory where to run the calculation. Default is `\inlineCommand{./}'.
\end{myitemize}
A full list of parameters is available by invoking the following command:
\begin{lstlisting}[language=aflowBash]
aflow --qca --usage
\end{lstlisting}

\subsection{\AFLOWGFA: The Glass-Forming-Ability Module for Structural Disorder}\label{sec:gfa}

\noindent
Metallic glasses are a unique class of materials without the crystalline order typically found in metals~\cite{Kruzic_aem_BMGstruct_2016, schroers2013bulk}.
Finding new metallic glasses is constrained by the vast combinatorial space~\cite{Li_acscombsci_numMG_2017} and time-consuming experiments.
\AFLOW\ can accelerate the discovery of new suitable candidates through the  \underline{g}lass-\underline{f}orming \underline{a}bility (\GFA) prediction module.

The first iteration of \GFA\ prediction was included in \AFLOW\ by Perim~\etal\ in 2016~\cite{curtarolo:art112}.
A spectral descriptor was constructed to capture the structural confusion
during vitrification into a glass upon cooling of the melt, as described by Greer~\cite{greer1993confusion}.
It was calculated based on the structures of competing crystalline phases available
at a specific stoichiometry in the \AFLOWorg\ repositories.
This first work focused on binary alloy systems, where a discrete exploration along the concentration axis already revealed a good insight into the possible glass-forming structures.

\begin{figure}[thb]
\centering
\includegraphics[width=0.95\linewidth]{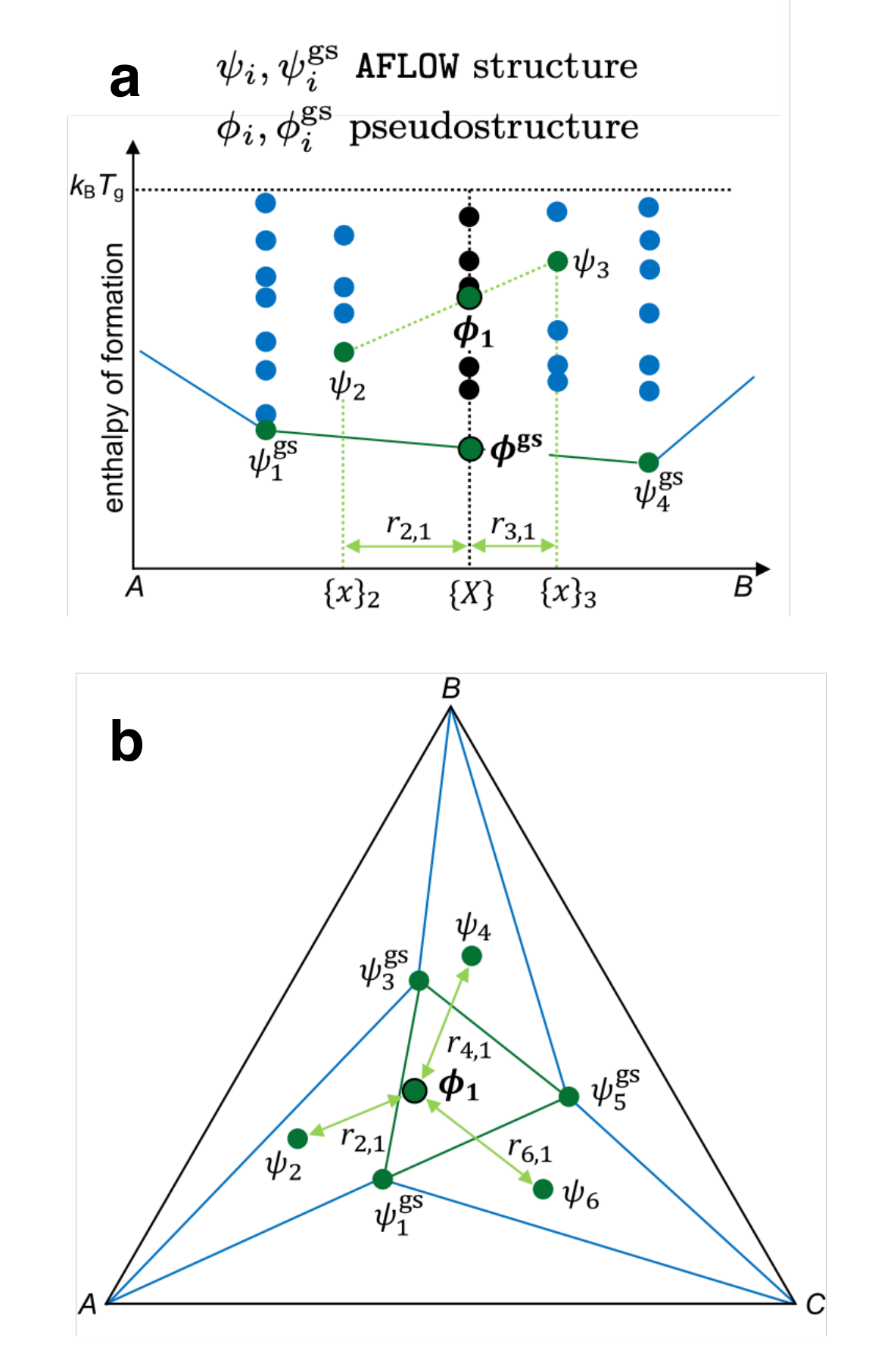}
\caption{
{\bf Schematic of contributions to the \GFA\ for binary (a) and ternary (b) systems.}
Structures are represented as circles: black circles are at the global stoichiometry,
green circles are pairs and triplets that contribute to the \GFA\ at the global stoichiometry, and blue circles are at other stoichiometries.
The reference state is the \underline{g}round \underline{s}tate ($\mathrm{gs}$) defined by the convex hull.
In the ternary schematic, $\phi^{\mathrm{gs}}$ is located directly beneath $\phi_1$.
}
\label{fig:gfa}
\end{figure}

In 2019, the \GFA\ module was subsequently expanded to better capture multi-component alloys by Ford~\etal~\cite{curtarolo:art154}.
Employing a new approach of global stoichiometry $\{X\}$,
the \GFA\ prediction now covers the whole concentration space of an alloy
which can comprise of combinations of phases at different compositions from the nominal one
and is no longer limited to points with multiple entries in the database.
To improve the descriptions of the reference states $\psi_{i}^{\mathrm{gs}}$ and competing structures $\psi_i$,
we include pseudo-structures $\phi_i$ from neighbors around the target composition.
\Fig~\ref{fig:gfa} shows the formation of pseudo-structures based on entries in the \AFLOWorg\ repositories.
Different entries (green circles) are weighted to represent the target composition.
Using \AFLOWCHULL~\cite{curtarolo:art144} (see Section~\ref{sec:chull}}) the ground states for the investigated concentrations are identified.
Based on their energetic distance to the ground state, the analysis will be limited to combinations that could occur at a typical glass transition temperature $T_\mathrm{g}$.

The current implementation relies on two primary sources to predict a GFA score for a specific composition.
The first one is the structural similarity of the competing entries.
To capture the structural component, the \underline{a}tomic \underline{e}nvironments (\AE) are constructed.
The neighbors around each atom in an entry's unit-cell are utilized to form a collection of \AE{}s.
The categorization of neighbors is based on a distance histogram, as suggested by Brunner and Schwarzenbach~\cite{brunner:environments}, and later applied to \AE{}s by Daams~\etal~\cite{AtomEnviron}.
A categorization scheme is used as the constructed \AE\ are not directly comparable.
Each \AE\ is described by a polyhedron code created from the number of vertexes connected to a specific mix of facets~\cite{daams_villars:environments_2000}.
An example is shown in \Fig~\ref{fig:gfa.ae}.

\begin{figure}
\centering
\includegraphics[width=0.9\linewidth]{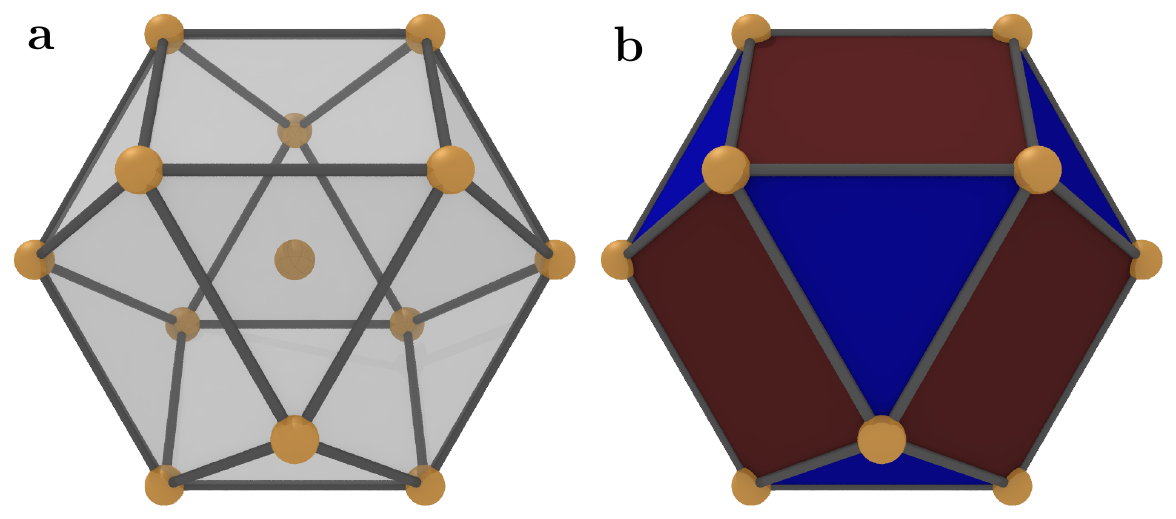}
\caption{
{\bf The Atomic environment of Copper (fcc, cF4).} The polyhedron code of this cuboctahedron ({\bf a}) is $12^{2.2}$, as it has 12 vertexes, each connected to two triangles and two quadrangles as highlighted in ({\bf b}).
}
\label{fig:gfa.ae}
\end{figure}

This classification describes the structural information of an entry on a high level, and details such as distortions or size differences are not captured.
Based on this description, the structural compatibility of entries is expressed by comparing the composition of the \AE\ ensemble for each element type.
The descriptor is zero if the structures have equal \AE{}s, and reaches a maximum when they have no \AE\ in common.
All possible combinations in a limited area around a given stoichiometry will be combined to create the \underline{s}tructural \underline{s}imilarity $\overline{\mathrm{ss}}$ factor.
Additionally, the function $ f(|\phi_i\rangle)$, describing the direct structural difference between an entry and the ground state, is used in the final representation of the \GFA.
Both structural descriptors include Gaussian distributed weights $w_i$ assigned to each entry depending on the dimensionless distance in stoichiometry to the global stoichiometry.

The second source of information to predict a GFA is the formation enthalpy, which is captured by the exponential function $g(H_i)$.
This function tends towards zero as the difference between the formation enthalpies of a considered entry and the ground state increases.
Through $g(H_i)$, entries near the ground state have a bigger impact on the overall result.

Overall the \GFA\ at a global stoichiometry $\{X\}$ is calculated as:
\[
\chi_{\mathrm{GFA}}\left(\left\{X\right\}\right) = \dfrac{100\,\overline{\mathrm{ss}}^2\sum_i f(|\phi_i\rangle)\,g(H_i)}{\sum_i w_i}\, ,
\]
where $100$ is an arbitrary scaling factor.
A detailed definition of the different segments is presented in \Ref~\onlinecite{curtarolo:art154}.

The \GFA\ module in \AFLOW\ can be invoked following this pattern:
\begin{lstlisting}[language=aflowBash]
aflow --gfa --alloy=CaCu
\end{lstlisting}
\begin{myitemize}
\item \inlineCommand{-{}-alloy=\textit{alloy}} : Sorted, case-sensitive string of the alloy system (\eg\ \inlineCommand{CaCu}).
\item \inlineCommand{-{}-ae\_file=\textit{file}} : Optional argument, file containing pre-calculated atomic environments (\eg\ \inlineCommand{AE\_input.dat}).
\item \inlineCommand{-{}-cutoff\_energy=\textit{cutoff}} : Optional argument, is the formation enthalpy cutoff in eV/atom. Default is \inlineCommand{0.05} (eV~$\sim$~580~K).
\end{myitemize}
This creates the following outputs:
\begin{myitemize}
\item \fileName{standard output} : Information about the current calculation process.
\item \fileName{GFA\_entries.dat} : Contains the stoichiometries and formation enthalpies of the structures used in the \GFA\ calculation.
\item \fileName{GFA\_\textit{alloy}.dat} : Contains the calculated \GFA\ for each point on the stoichiometry grid.
\item \fileName{All\_atomic\_environments.dat} : Contains the atomic environments (one per species) for each entry used in the \GFA\ calculation.
\end{myitemize}
For the full set of options and additional information, see the \AFLOWGFA\ \README:
\begin{lstlisting}[language=aflowBash]
aflow --readme=gfa
\end{lstlisting}
\section{\AFLOWAPE: The \AFLOW\ Python Environment}\label{sec:ape}

\noindent
\AFLOW\ is a powerful tool with various fully-automated workflows that can easily be integrated into custom code environments.
Python is particularly important to support due to its
popularity and because many machine learning frameworks are implemented in this language.
To facilitate the usage of \binary{aflow} with Python, the \AFLOW\ \underline{P}ython \underline{E}nviornment~(\AFLOWAPE), has been
developed and includes wrappers for
\AFLOWSYM~\cite{curtarolo:art135}, \AFLOWCHULL~\cite{curtarolo:art144},
\AFLOWXTALFINDER~\cite{curtarolo:art170}, and \AFLOWCCE~\cite{curtarolo:art172}.

\AFLOW\ Python modules can be installed using the command
\begin{aflowBashTBox}
{\color{binarycolor} aflow} -{}-python\_modules[=\textit{modules}] -D \textit{directory}
\end{aflowBashTBox}
\noindent
where \inlineCommand{\textit{modules}} are a comma-separated list and \inlineCommand{\textit{directory}} is
where the modules are installed, \eg\ the site-packages directory of the Python
installation.
Square brackets \inlineCommand{[...]} indicate optional arguments; the brackets themselves are not part of the command.
If no modules are given, \binary{aflow} will install all available ones. A complete
installation of \binary{aflow} is required for them to run. They can also be installed into a
virtual environment during the installation of \binary{aflow} with the \binary{install-aflow.sh} script by
adding the \inlineCommand{-{}-venv} option. Some of these modules require additional Python packages.
A list of dependencies can be accessed via
\begin{lstlisting}[language=aflowBash]
install-aflow.sh --pip_modules
\end{lstlisting}
They are automatically installed when using \inlineCommand{-{}-venv} with \binary{install-aflow.sh}.

The \AFLOWSYM\ wrapper~(module name: \inlineCommand{aflow\_sym}) provides functionality to calculate all
\AFLOW\ symmetry groups~(lattice point group, reciprocal lattice point group, crystallographic point
group, the dual of the crystallographic point group, Patterson point group, factor group, space
group, and atom-site point group), extended crystallographic data~(\inlineCommand{edata}), and space group
data~(\inlineCommand{sgdata}). It supports magnetic moments as well.

\AFLOWCHULL\ consists of two different modules. \inlineCommand{aflow\_chull} contains the \inlineCommand{CHull}
class, which calculates convex hulls, hull energies, distances to the convex hull, and stability
criteria. \inlineCommand{aflow\_chull\_plotter} provides a \inlineCommand{Plotter} class that can create Jupyter
notebooks with convex hull plots.

\AFLOWXTALFINDER\ is implemented via the \inlineCommand{XtalFinder} class inside the
\inlineCommand{aflow\_xtal\_finder} module. It can take a list of files or a directory as input and provide
the structure comparison output. Single input files can also be compared to the \AFLOWorg\ repositories and
prototype encyclopedia. Unique atom decorations can be obtained as well.

The \AFLOWCCE\ module~(\inlineCommand{aflow\_cce}) contains the \inlineCommand{CCE} class and
provides the same features as the \texttt{C++} version of \AFLOW, \ie\ corrections to formation enthalpies,
oxidation numbers, and coordination numbers around cations.

With these wrappers, \AFLOW\ can be seamlessly integrated into other Python workflows. Most
functions return dictionaries, a basic built-in data type in Python. This allows the output to be
further processed, \eg\ to populate feature vectors in machine learning applications.

\section{Summary}
This article describes \AFLOW, an interconnected collection of
algorithms and workflows, written \texttt{C++}, that have been developed to address the
challenge of accelerated and autonomous materials' calculation and identifications.
The article highlights the upgrades that have been developed since
the original \AFLOW\ report ~\cite{curtarolo:art65}, and
demonstrate their interoperability within the overall environment.
The various modules and tools included in the standard distribution
(version \AFLOWVERSION, Fall 2022) are listed in Table
\ref{tableacronyms}.
The code, download/installation instructions and operation manuals describing
all the features, are freely available at \AFLOWorg.
Through ongoing innovation and implementation of robust descriptors and workflows,
\AFLOW\ continues to deliver valuable solutions as well as
playing a role in accelerating the pace of automation in the materials community.

\vspace{1cm}

{\small

  \section*{Declaration of competing interest}
  \noindent
  The authors declare that they have no known competing financial
  interests or personal relationships that could have appeared to
  influence the work reported in this paper.

  \section*{Data Availability}
  \noindent
  The \AFLOW\ software suite is freely available through the \AFLOWorg\
  website. Instructions for download and installations are provided in this article.

  \section*{Acknowledgments}
  \noindent
  The authors thank
  Adam~Zettel, Asa~Guest, Doug~Wolfe, Don~Brenner, Jon-Paul~Maria, Bill~Fhrenholtz,
  Douglas~Hofmann, Christian~Carbogno, Luca~Ghiringhelli, Gus~Hart,
  Eric~Gossett, Cheryl~Li, Harry~Wang,
  Frisco~Rose, Mana~Rose,
  Max Brenner, William~Schmitt, and Stuart~Ki for fruitful discussions.
  Research sponsored by DOD-ONR
  (N00014-21-1-2132, N00014-20-1-2525, N00014-20-1-2299,
  N00014-20-1-2200)
  and by NSF (NRT-HDR DGE-2022040).
  R.F. acknowledges support from the Alexander von Humboldt Foundation
  under the Feodor Lynen research fellowship.
  A.vdW acknowledges support from DOD-ONR (N00014-20-1-2225).
  C.T. acknowledges support from NSF (DMR-2219788).
}

\newcommand{\Ozolins}{Ozoli{\c{n}}{\v{s}}}

\end{document}